\documentclass[twocolumn]{aastex631}

\usepackage[caption=false]{subfig}
\usepackage{array}
\newcolumntype{H}{>{\setbox0=\hbox\bgroup}c<{\egroup}@{}}

\usepackage{comment}
\usepackage{color}
\usepackage{url}

\usepackage{amsmath}	
\usepackage{amssymb}	

\usepackage[percent]{overpic}

\newcommand{\Lya}{Ly$\alpha$}
\newcommand{\OII}{[O\,{\sc ii}]}
\newcommand{\OIII}{[O\,{\sc iii}]}

\newcommand{\Ha}{H$\alpha$}
\newcommand{\Hb}{H$\beta$}
\newcommand{\Hg}{H$\gamma$}
\newcommand{\Hd}{H$\delta$}
\newcommand{\CIII}{C{\sc iii}]}

\newcommand{\HeII}{He{\sc ii}}
\newcommand{\NIII}{N{\sc iii}]}
\newcommand{\NIV}{N{\sc iv}]}

\newcommand{\NeIII}{[Ne\,{\sc iii}]}
\newcommand{\NeIV}{[Ne\,{\sc iv}]}
\newcommand{\NeV}{[Ne\,{\sc v}]}
\newcommand{\NII}{[N\,{\sc ii}]}

\newcommand{\xiion}{$\xi_{\rm ion}$}

\newcommand{\kms}{km\,s$^{-1}$}
\newcommand{\Oabundance}{$12+\log ({\rm O/H})$}
\newcommand{\Zsun}{$\rm Z_{\odot}$}

\newcommand{\ebv}{E(B$-$V)}

\newcommand{\Msunyr}{$M_{\odot}\,{\rm yr}^{-1}$}
\newcommand{\Msun}{$M_{\odot}$}
\newcommand{\Mstar}{$M_{\star}$}

\newcommand{\Muv}{$M_{\rm UV}$}

\submitjournal{ApJ Supplement on Jan 30, 2023 / Accepted on May 8, 2023}

\shorttitle{
Mass-Metallicity Star-Formation Relations at $z=4-10$
}
\shortauthors{Nakajima et al.}

\begin{document}

\title{
JWST Census for the Mass-Metallicity Star-Formation Relations at $z=4-10$ with\\
the Self-Consistent Flux Calibration and the Proper Metallicity Calibrators
}

\correspondingauthor{Kimihiko Nakajima}
\email{kimihiko.nakajima@nao.ac.jp}

\author[0000-0003-2965-5070]{Kimihiko Nakajima}
\affiliation{National Astronomical Observatory of Japan, 2-21-1 Osawa, Mitaka, Tokyo 181-8588, Japan}

\author[0000-0002-1049-6658]{Masami Ouchi}
\affiliation{National Astronomical Observatory of Japan, 2-21-1 Osawa, Mitaka, Tokyo 181-8588, Japan}
\affiliation{Institute for Cosmic Ray Research, The University of Tokyo, 5-1-5 Kashiwanoha, Kashiwa, Chiba 277-8582, Japan}
\affiliation{Kavli Institute for the Physics and Mathematics of the Universe (WPI), University of Tokyo, Kashiwa, Chiba 277-8583, Japan}

\author[0000-0001-7730-8634]{Yuki Isobe}
\affiliation{Institute for Cosmic Ray Research, The University of Tokyo, 5-1-5 Kashiwanoha, Kashiwa, Chiba 277-8582, Japan}
\affiliation{Department of Physics, Graduate School of Science, The University of Tokyo, 7-3-1 Hongo, Bunkyo, Tokyo 113-0033, Japan}

\author[0000-0002-6047-430X]{Yuichi Harikane}
\affiliation{Institute for Cosmic Ray Research, The University of Tokyo, 5-1-5 Kashiwanoha, Kashiwa, Chiba 277-8582, Japan}

\author[0000-0003-3817-8739]{Yechi Zhang}
\affiliation{Institute for Cosmic Ray Research, The University of Tokyo, 5-1-5 Kashiwanoha, Kashiwa, Chiba 277-8582, Japan}
\affiliation{Department of Physics, Graduate School of Science, The University of Tokyo, 7-3-1 Hongo, Bunkyo, Tokyo 113-0033, Japan}

\author[0000-0001-9011-7605]{Yoshiaki Ono}
\affiliation{Institute for Cosmic Ray Research, The University of Tokyo, 5-1-5 Kashiwanoha, Kashiwa, Chiba 277-8582, Japan}

\author{Hiroya Umeda}
\affiliation{Institute for Cosmic Ray Research, The University of Tokyo, 5-1-5 Kashiwanoha, Kashiwa, Chiba 277-8582, Japan}
\affiliation{Department of Physics, Graduate School of Science, The University of Tokyo, 7-3-1 Hongo, Bunkyo, Tokyo 113-0033, Japan}

\author[0000-0003-3484-399X]{Masamune Oguri}
\affiliation{Center for Frontier Science, Chiba University, 1-33 Yayoi-cho, Inage-ku, Chiba 263-8522, Japan}
\affiliation{Department of Physics, Graduate School of Science, Chiba University, 1-33 Yayoi-Cho, Inage-Ku, Chiba 263-8522, Japan}



\begin{abstract}
We present the evolution of the mass-metallicity (MZ) relations at $z=4-10$ derived with 135 galaxies identified in the JWST/NIRSpec data taken from the three major public spectroscopy programs of ERO, GLASS, and CEERS. Because there are many discrepancies between flux measurements reported by early ERO studies, we first establish our NIRSpec data reduction procedure for reliable emission-line flux measurements and errors successfully explaining Balmer decrements with no statistical tensions via thorough comparisons of the early ERO studies. Applying the reduction procedure to the 135 galaxies, we obtain emission-line fluxes for physical property measurements. We confirm that 10 out of the 135 galaxies with \OIII$\lambda 4363$-lines have electron temperatures of $\simeq (1.1-2.3)\times 10^4$\,K, similar to lower-$z$ star-forming galaxies, that can be explained by heating of young massive stars. We derive metallicities of the 10 galaxies by the direct method and the rest of the galaxies with strong lines by the metallicity calibrations of \citet{nakajima2022_empressV} applicable for these low-mass metal-poor galaxies, anchoring the metallicities with the direct-method measurements. We thus obtain MZ relations and star-formation rate (SFR)-MZ relations over $z=4-10$. We find that there is a small evolution of the MZ relation from $z\sim 2-3$ to $z=4-10$, while interestingly that the SFR-MZ relation shows no evolution up to $z\sim 8$ but a significant decrease at $z>8$ beyond the error. This SFR-MZ relation decrease at $z>8$ may suggest a break of the metallicity equilibrium state via star-formation, inflow, and outflow, while further statistical and local-baseline studies are needed for a conclusion.
\end{abstract}

\keywords{Chemical abundances(224) --- Galaxy chemical evolution(580) --- Galaxy evolution(594) --- High-redshift galaxies(734) --- James Webb Space Telescope(2291)}

\section{Introduction} \label{sec:introduction}

The James Webb Space Telescope (JWST) has dramatically advanced the exploration of the high redshift universe. One important advancement for the high redshift community is provided by the high sensitivity of the Near Infrared Spectrograph (NIRSpec; \citealt{jakobsen2022}) at $\lambda \simeq 2-5\,\mu$m, allowing us to directly address the key questions relating to the physical conditions of inter-stellar medium (ISM) and the nature of the ionizing spectrum for galaxies in the early universe. 
The properties of the hot ISM, especially the gas-phase metallicity determined by the oxygen abundance (\Oabundance), can be accurately constrained using rest-frame optical emission lines. These emission lines have been well calibrated and understood over decades, as discussed in a recent review by \citet{MM2019}.
Gas-phase metallicities provide a crucial experimental tool for studying the early chemical enrichment and feedback processes in galaxy evolution. These measurements can be used in conjunction with cosmological simulations (e.g., \citealt{FD2008,ma2016_MZR, torrey2019, langan2020, ucci2021}) and chemical evolution models (e.g., \citealt{dave2012,lilly2013,DF2018}) to gain insights into galaxy formation and evolution.
Moreover, the efficiency of ionizing photon production, as quantified by \xiion\ (the production rate of hydrogen ionizing photons per unit luminosity in the UV-continuum), can be used to study the nature of the ionizing spectrum. The Hydrogen Balmer emission via recombination physics provides the best constraints on \xiion, while high ionization lines such as \HeII\ offer key spectroscopic tools to investigate the hardness of the ionizing spectrum. These approaches are crucial for understanding the stellar population of early galaxies, diagnosing the presence of accreting black holes in the system, and detecting galaxies hosting the very first generation of stars in the early universe (e.g., \citealt{schaerer2003,inoue2011_metal_poor,kewley2013_theory,bouwens2016_xiion,NM2022,katz2022_popIII,trussler2022_popIII}).

Determinations of these properties ideally require key optical emission lines such as \OIII$\lambda\lambda5007,4959$, \OIII$\lambda 4363$, \OII$\lambda\lambda 3726,3729$%
\footnote{%
We use the notation \OII$\lambda 3727$ as the sum of the doublet.
}, and the Balmer lines. In particular, \OIII$\lambda 4363$ is crucial for gas-phase metallicity studies based on electron temperature ($T_e$).
However, these lines are limited in their applicability to galaxies up to $z\lesssim 3$ in the pre-JWST era, with the detection of \OIII$\lambda 4363$ from sources at $z\gtrsim 1$ still challenging due to faintness (e.g., \citealt{christensen2012_laes,jones2015_directZ,sanders2020,sanders2021}). These lines are beyond the reach of ground-based telescopes at higher redshift. 
Now JWST/NIRSpec has become online to directly examine these key emission lines in detail for high redshift sources.

Using the Early Release Observations (ERO) of NIRSpec within one month after the data release, several studies report a detection of \OIII$\lambda 4363$ from three objects at $z=7.6-8.5$ which are magnified by the strong lensing galaxy cluster SMACS J0723.3-7327 \citep{schaerer2022_ero,curti2023_ero,trump2022_ero,rhoads2023_ero,arellano-cordova2022_ero,brinchmann2022_ero}.
These objects have provided the first reliable metallicity determinations at high redshift based on the direct $T_e$ measurement. 
The metallicity values from the different teams are overall consistent to each other for each source, resulting in a mass-metallicity relation which is scattered around the extrapolation of the $z\simeq 2-3$ relation towards the lower mass end (\Mstar\ $= 10^{7.5}-10^{9}$\,\Msun).
\citet{curti2023_ero} also examine the mass-metallicity-star formation rate (SFR) relation as indicated to be a fundamental relation from $z=0$ to $z\lesssim 2-3$ (\citealt{mannucci2010, MM2019, sanders2021}, and the references therein),
suggesting that no clear evolution is seen from $z=0-3$ to $z=7.6-8.5$, particularly excepting for the $z=8.5$ object (see below).
However, the sample size is obviously too small to conclude the evolution on the mass-metallicity relation and its SFR dependence.

Moreover, several pieces of evidence in the previous ERO studies imply that the early release of the NIRSpec products contain some issues. 
At the early stage of the ERO data release, the reductions including the flux calibration are partly based on the predicted pre-flight data such as the throughputs of the spectrograph, the optical telescope element, and the NIRSpec fore optics.
A sensitivity function, whose shape is strongly dependent on the wavelength, is necessary for appropriate flux measurements as adopted in \citet{curti2023_ero} using standard stars taken during the commissioning.
Even after the flux-calibration, there remains some tensions for faint emission lines. A notable issue is seen in the Balmer decrements such as \Hg/\Hb\ and \Hd/\Hb\ which are (not always but sometimes) unexplainable in a consistent way within the uncertainties according to the Case B recombination. Such tensions, especially seen in the faint emission lines, may be caused by background residuals and/or hot pixels that persist in the released final spectra. 
Interestingly, the $z=8.5$ object (ID:04590) is suggested to present a high \OIII$\lambda\lambda4363/5007$ line ratio, and accordingly a very high $T_e \gtrsim 25000$\,K, which is hard to be explained by heating of young massive stars alone \citep{katz2023_jwst}. The low metallicity indicates it falls significantly below the mass-metallicity-SFR relation \citep{curti2023_ero} and can be in an early stage of galaxy evolution.
Such a high electron temperature and low metallicity need to be confirmed by a more carefully reduced spectrum free from any tensions.
The NIRSpec data reduction procedure will be revisited as one of the key parts of this paper for reliable emission-line flux measurements and errors.

Despite these possible uncertainties, recent NIRSpec observations are now providing increasing numbers of spectra from high-redshift objects, and the reports of the identifications of strong emission lines such as \OIII$\lambda\lambda5007, 4959+$\Hb\ from objects at $z>7-8$ follows. 
Notable results has come from the Early Release Science (ERS) observations of GLASS \citep{treu2022_glass,morishita2022_glass,mascia2023_glass} and CEERS \citep{finkelstein2022_ceers,bagley2022_ceers,sanders2023_jwst,tang2023_ceers,fujimoto2023_ceers,shapley2023_ceers}, two Director's Discretionary Time (DDT) programs \citep{williams2022_jwst,heintz2022_jwst,wang2022_jwst,langeroodi2022_ddt}, and the Guaranteed Time Observations of JADES \citep{robertson2022_jades,curtis-lake2022_jades,bunker2023_jades,cameron2023_jades,saxena2023_jades,curti2023_jades}.
Regarding the ISM properties, 
\citet{sanders2023_jwst} examine the ionization properties of $\sim 160$ galaxies at $z=2-9$ from CEERS by using high-to-low ionization emission line ratios such as \OIII$\lambda 5007$$/$\OII$\lambda 3727$, and suggest that galaxies tend to present hard ionizing spectra at high-redshift. The metallicities at $z>6.5$ are suggested to be sub-solar on average 
(see also \citealt{tang2023_ceers,shapley2023_ceers}). 
A similarly high ionization state and modest metallicity is also suggested in $29$ galaxies at $z=4.8-8$ from GLASS \citep{mascia2023_glass} and in $26$ galaxies at $z=5.5-9.5$ from JADES (\citealt{cameron2023_jades}, see also \citealt{saxena2023_jades}).
\citet{fujimoto2023_ceers} explicitly derive metallicities of $\sim 10$ $z=8-9$ CEERS galaxies using a metallicity indicator of \OIII$\lambda 5007$/\Hb, suggesting that these high-redshift galaxies have a lower metallicity than $z=0-3$ galaxies for a given stellar mass. 
More extremely, \citet{williams2022_jwst} report a $z=9.5$ galaxy with \OIII$+$\Hb, revealing a very high \OIII$/$\OII\ ratio and a modestly low-metallicity, \Oabundance\ $=7.48\pm 0.08$, falling below the $z=0$ mass-metallicity relation but still consistent within $2\sigma$. Including the $z=9.5$ object, \citet{heintz2022_jwst} analyzed five objects at $z>7.8$ from the DDT programs and suggest a systematic decrease of metallicity at $z>7.8$ on the mass-metallicity-SFR relation found in lower-redshift, as seen in the ERO $z=8.5$ object. 
A similar conclusion is also derived by \citet{langeroodi2022_ddt}.
\citet{boyett2023_jwst} report another $z=9$ galaxy from GLASS whose \NeIII$\lambda 3869$/\OII\ ratio implies a sub-solar metallicity.
Finally, \citet{bunker2023_jades} report the astonishing spectrum of GN-z11 at $z=10.6$ using the deep JADES observations. 
The author use several metal lines such as \NeIII, \OII, as well as the UV lines of \NIV$\lambda 1486$, \NIII$\lambda 1748$, and \CIII$\lambda 1909$ and imply an unusually high nitrogen-to-oxygen abundance ratio for its modest ($\sim$ sub-solar) oxygen abundance (see also \citealt{cameron2023_nitrogen}). 
We note that these metallicities are based on the strong emission line ratios which are empirical indicators and calibrated mostly in the local universe (e.g., \citealt{MM2019}). The applicability of these methods at high-redshift need to be carefully confirmed (e.g., \citealt{curti2023_ero}), particularly given the indication of different degrees of ionization in the ISM of galaxies between $z=0$ and high-redshift for a fixed metallicity \citep{sanders2023_jwst}.

In this paper, we provide a summary of three major public NIRSpec observation programs of ERO, GLASS, and CEERS, aimed at characterizing the mass-metallicity relation at $z=4-10$ and examining its evolution over cosmic time. 
We establish a reliable data reduction procedure for emission-line flux measurements and errors using NIRSpec data, which is applied to construct a large sample of galaxies at $z=4-10$ as detailed in Section \ref{sec:data}.
In Section \ref{sec:results}, we utilize key emission line ratios to derive metallicities for the JWST objects with improved NIRSpec spectra. We then investigate the mass-metallicity relation and its dependence on SFR at high redshift.
We begin with the 10 objects in which \OIII$\lambda 4363$ is identified and metallicity is determined using the direct $T_e$ method. The empirical metallicity indicators validated with these 10 objects are then applied to estimate metallicities for the remaining JWST objects.
We analyze the mass-metallicity and mass-metallicity-SFR relations at $z=4-10$ using the metallicity measurements obtained for the JWST objects. The results and implications of these metallicity relationships are discussed in Section \ref{sec:discussion}, and we provide a summary of our conclusions.
Throughout the paper we adopt a standard $\Lambda$CDM cosmology with $\Omega_{\Lambda}=0.7$, $\Omega_{m}=0.3$, and $H_0=70$\,\kms\,Mpc$^{-1}$.

\begin{figure*}
  \centering
    \begin{tabular}{c}
      \begin{minipage}{0.475\hsize}
        \begin{center}
         \includegraphics[bb=0 0 546 515, width=0.9\columnwidth]{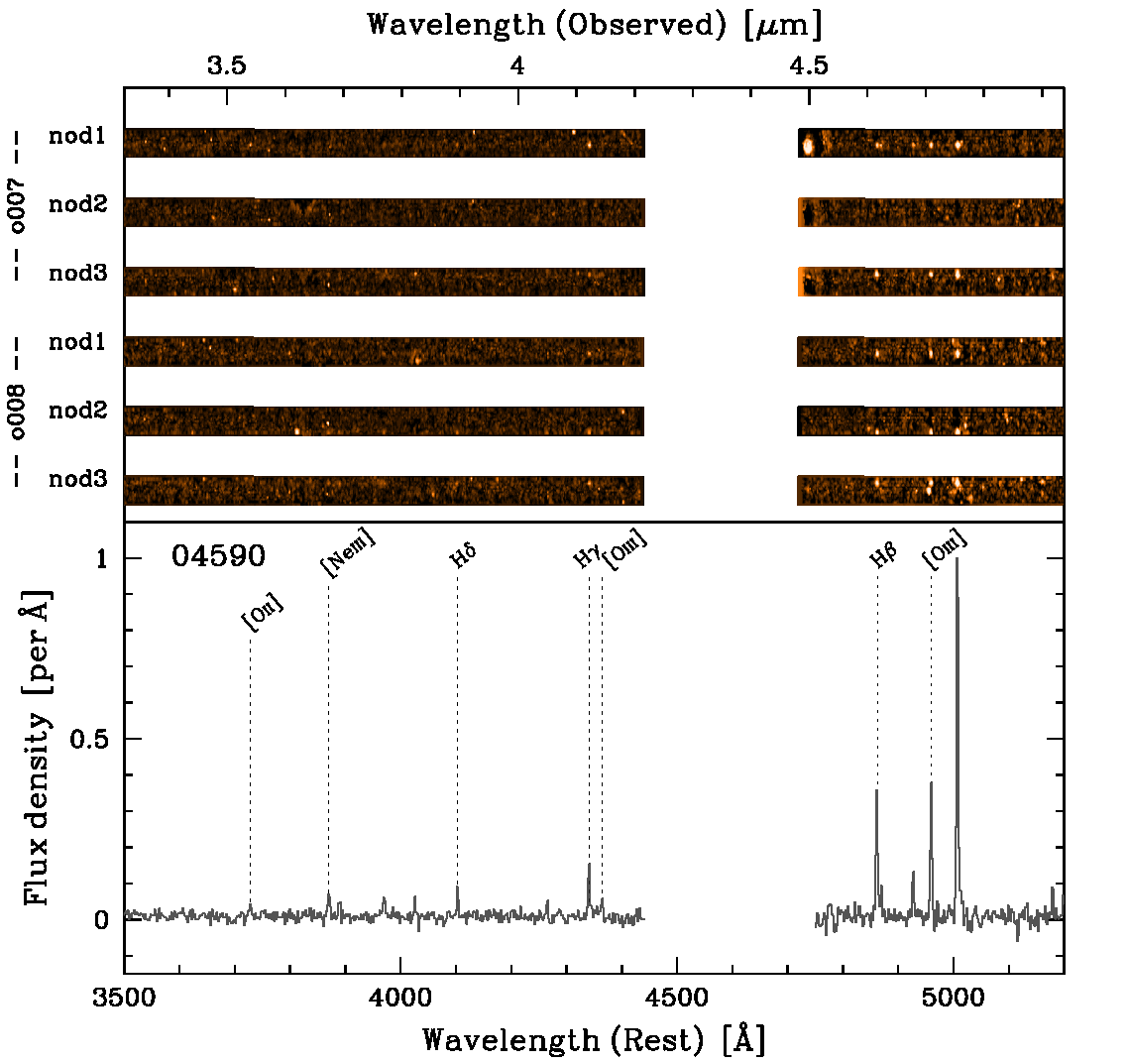}
        \end{center}
      \end{minipage}
      \begin{minipage}{0.475\hsize}
        \begin{center}
         \includegraphics[bb=0 0 546 515, width=0.9\columnwidth]{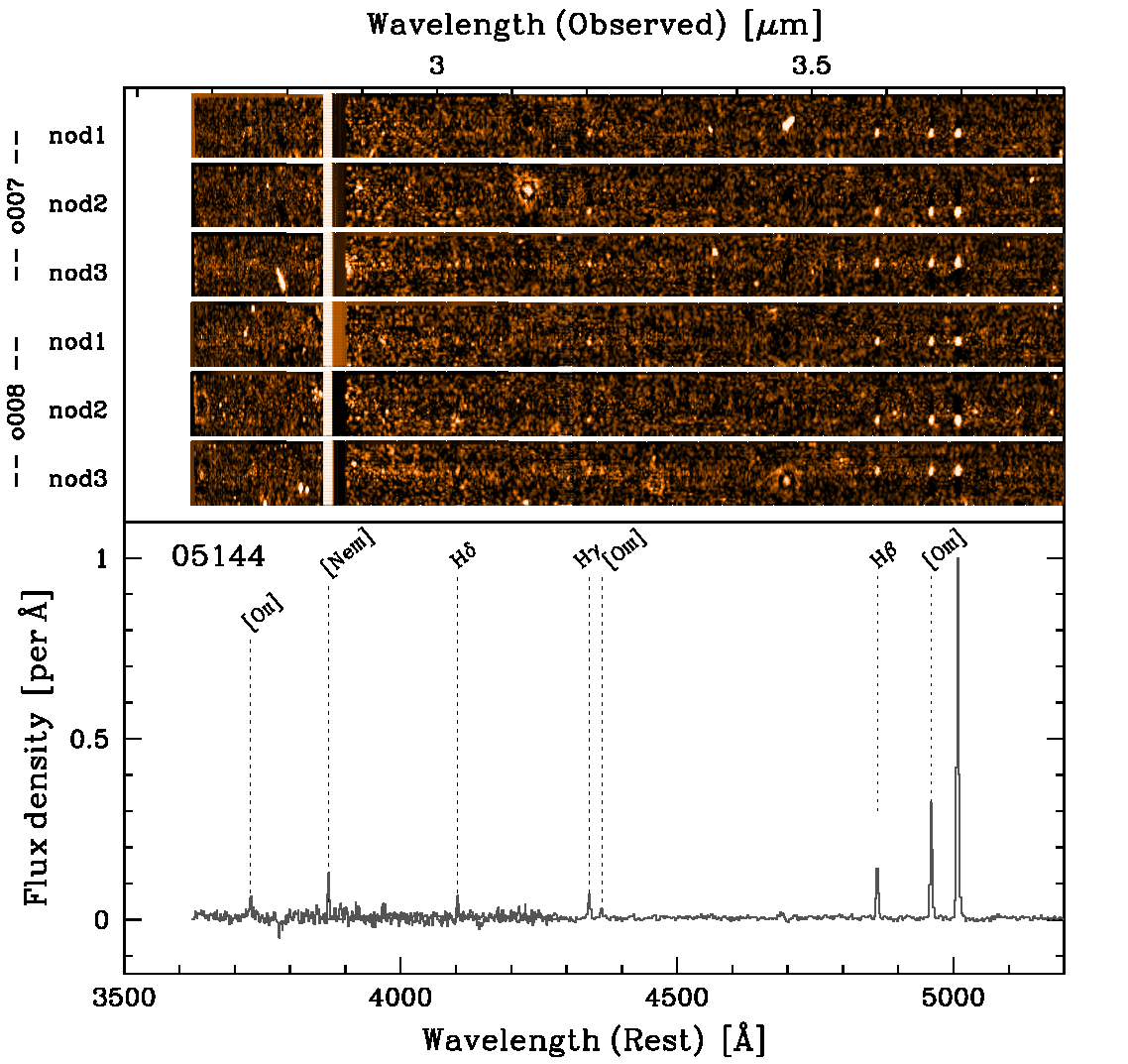}
        \end{center}
      \end{minipage}
      \\\\
      \begin{minipage}{0.475\hsize}
        \begin{center}
         \includegraphics[bb=0 0 546 515, width=0.9\columnwidth]{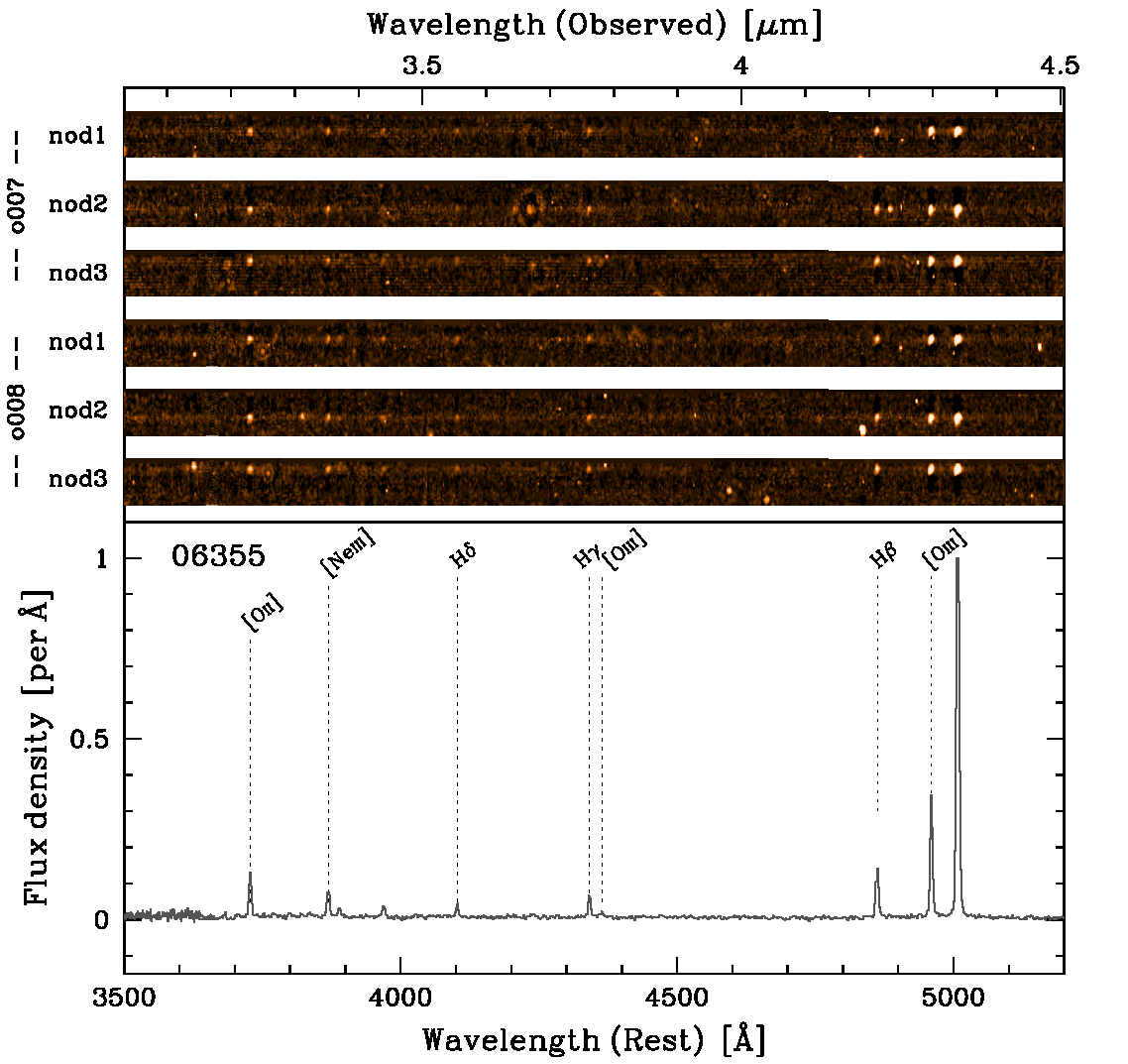}
        \end{center}
      \end{minipage}
      \begin{minipage}{0.475\hsize}
        \begin{center}
         \includegraphics[bb=0 0 546 515, width=0.9\columnwidth]{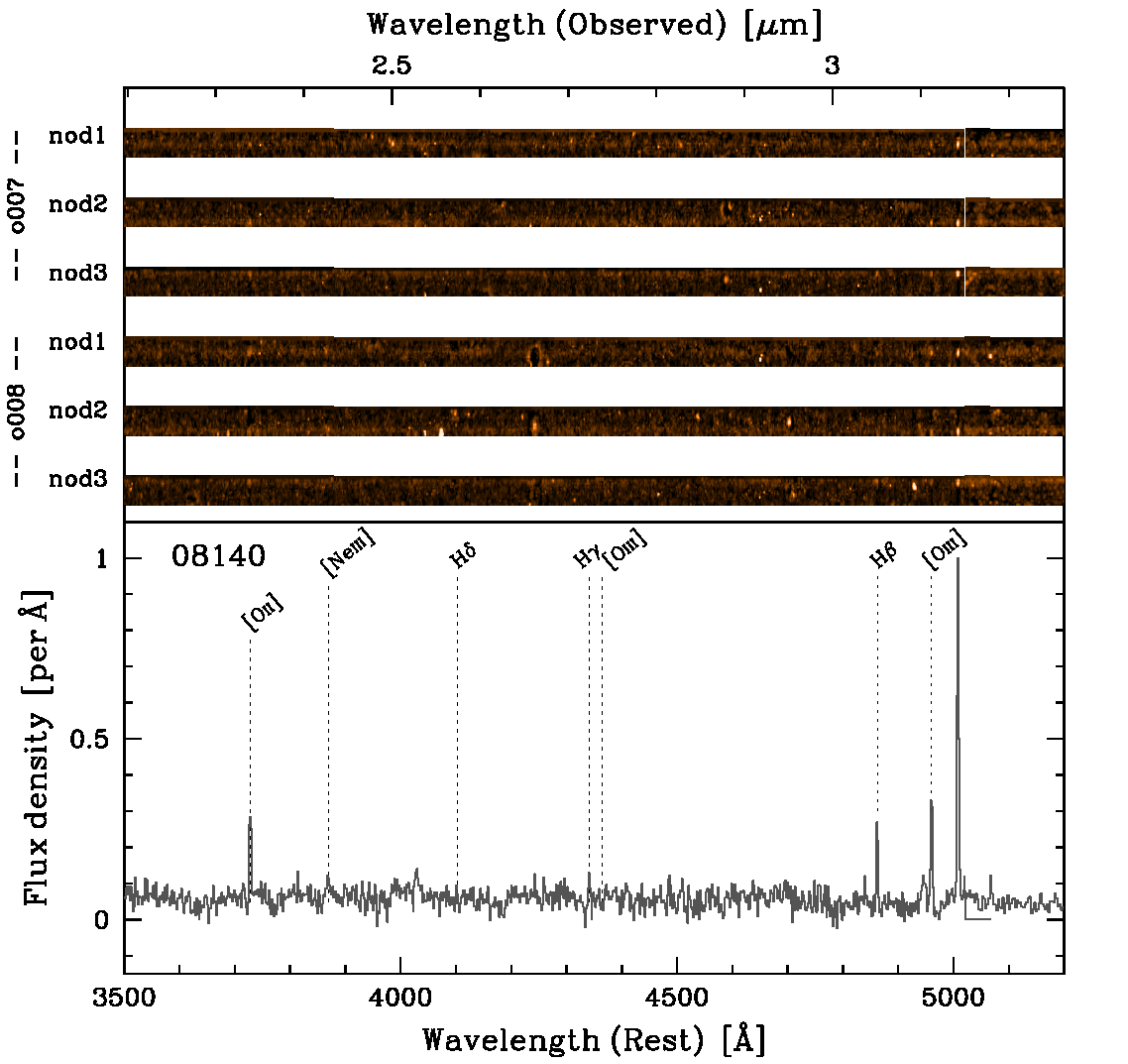}
        \end{center}
      \end{minipage}
      \\\\
      \begin{minipage}[b]{0.475\hsize}
        \begin{center}
         \includegraphics[bb=0 0 546 515, width=0.9\columnwidth]{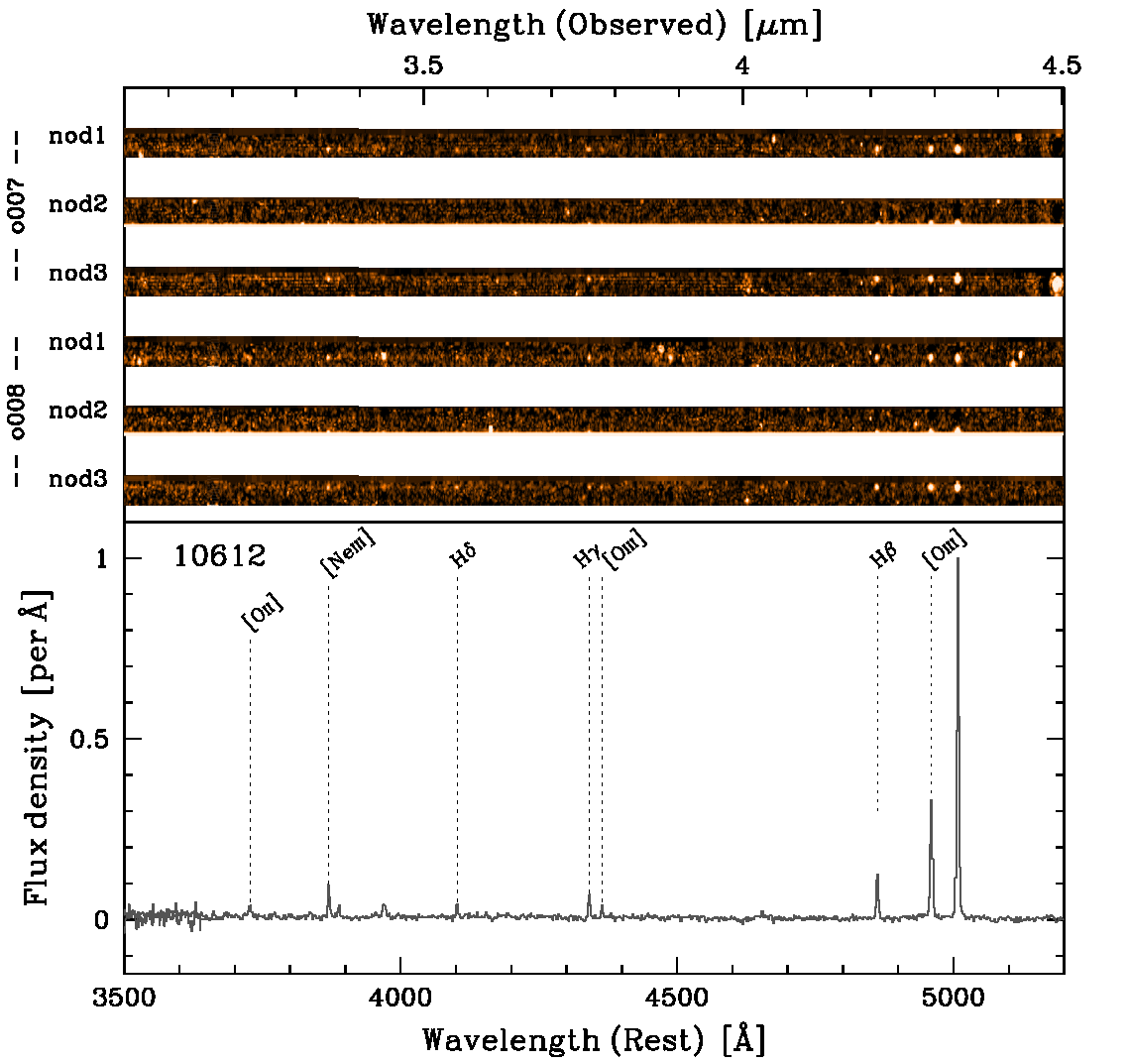}
        \end{center}
      \end{minipage}
      \begin{minipage}[b]{0.475\hsize}
        \begin{flushleft}
         \caption{%
            Observed NIRSpec spectra for the five $z=5-8.5$ ERO objects. For each object: (top) 2D spectra for the six individual exposures (three nods each in o007 and o008). (bottom) 1D composite spectrum. 
            We generate the 1D composite spectrum by first median-stacking the individual 2D spectra and then extracting the 1D spectrum with a small aperture (see text). This procedure allows us to minimize the possible impacts of hot pixels and to rescue some individual 2D frames where the spectrum is partly out of the slit (e.g., two nod2 frames of 10612).
         }
        \label{fig:spectra_ero}
        \end{flushleft}
      \end{minipage}
    \end{tabular}
\end{figure*}


\section{Data} \label{sec:data}

We assemble the publicly available NIRSpec data of ERO and ERS, re-reduce and analyze the spectra, and identify high-redshift galaxies with the rest-frame optical emission lines to discuss their nebular properties. In the following subsections, we detail our improved reduction procedures, particularly for the ERO data to demonstrate the improvements by comparing with the early results.

\subsection{ERO Data} \label{ssec:data_ero}

\subsubsection{NIRSpec observations and Data reduction} \label{sssec:data_ero_observations}

The JWST/NIRSpec ERO observations were undertaken on UT 2022 June 30,
targeting the SMACS 0723 lensing cluster field (\citealt{pontoppidan2022}; Proposal ID: 2736).
Multi-slit spectroscopy was taken using the micro-shutter assembly (MSA), 
with the medium resolution ($R\sim 1000$) gratings/filters of G235M/F170LP and G395M/F290LP
sampling the wavelength range of $1.7-3.1$ and $2.9-5.1\,\mu$m, respectively.
Two observing blocks were carried out, o007 and o008, 
each of which consisted of three exposures of 2918\,sec in G395M and F235M each, with 
a 3-shutter slitlet nod pattern, i.e.,
slightly nodded positions, nod1, nod2, and nod3, in different shutters within the slitlets.
The total integration time was $4.86$\,hr in G235M and G395M each.

By visually checking all of the 1D and 2D composite spectra that were publicly available, 
we identified 5 objects whose \OIII$\lambda\lambda 5007,4959+$\Hb\ were clearly detected and 
that were useful for the investigation of the nebular properties at high-redshift $z>5$, 
with IDs 04590, 05144, 06322, 08140, and 10612.
Although there were additional two sources in the ERO sample that were tentatively reported at $z=5-6$ \citep{brinchmann2022_ero,mahler2022},
they were not included in the following analysis because of their lines' poorer confidence levels.

For the five sources, data reduction was re-performed for a more careful analysis and flux calibration.
Starting with the Level1 products provided by the observatory, 
we carried out the Spec2 and Spec3 pipelines per nod and per observing block
using the Python library for JWST observations by STScI (ver.1.8.5).
We used the latest reference files stored in a pmap file of either 1028 or 1027, where
notably the flat files were created from in-flight data for all MOS data taken after the launch,
resolving the MOS flux calibration issues reported in the early report of the NIRSpec ERO data
(see below for more details about the flux calibration).
According to the original reduction procedure for NIRSpec MSA data provided by the CEERS team%
\footnote{
\url{https://web.corral.tacc.utexas.edu/ceersdata/SDR2/NIRSpec}
}, during the Spec2 pipeline, a single science image for a particular nodding position
was background-subtracted by using the other nods' images.
However, we realized some hot pixels including uncleaned cosmic rays in one of the background images 
fell on the spatial position of the science objects and badly influenced the final 1D spectrum.
To avoid such additional noises, we carefully checked the background-subtracted science images
and removed one of the other nods' images, if necessary, for the background images and 
re-ran the Spec2 pipeline in an iterative manner. 
Furthermore, we subtracted the local residual background at each spectral pixel for each science image 
by masking the spatial position of the object and taking the median value of the pixels in the range 
[$-10$\,pix:$+10$\,pix] along the spectral direction%
\footnote{
We have checked that the choice of the box size does not significantly affect the final spectra,
by changing it from $\pm5$ to $\pm15$\,pix.
The pixel size is $1.06\times 10^{-3}$ and $1.79\times 10^{-3}\,\mu$m in G235M and G395M, respectively.
}.
We performed the Spec2 pipeline assuming that all of the five sources were point sources, as they all looked compact in the NIRCam images with the half-light radii smaller than the PSF (e.g., \citealt{harikane2022_jwst,tacchella2022_ero}; see also \citealt{ono2022_jwst} for the size evolution at high-redshift).
This assumption affected how the pathloss corrections, which accounted for various types of signal loss in NIRSpec data, were calculated and applied.
The correction depended on the position of the source within the configured MSA shutter, the aperture used for the observations, the wavelength, as well as the type of the object.
We used the original MSA metadata files to refer to the relative placement of the sources within the MSA shutters assuming the pointing uncertainties during the observations were negligible, but manually changed the values in the column \verb+STELLARITY+ to force the objects to be treated as point sources.
Figure \ref{fig:spectra_ero} summarizes the six 2D outcomes (3 nods for 2 observing blocks) for each of the ERO sources.

Checking the six single exposures in both 1D and 2D for each of the sources,
we confirmed no signal was detected in the nod2/o007 observing block for 04590, 
as already pointed out by the earlier studies (e.g., \citealt{curti2023_ero,arellano-cordova2022_ero}).
This exposure obviously needed to be removed from the final composition.
Moreover, we noticed the spectra of nod2/o008 of 05144 were badly affected especially around the \Hg+\OIII$\lambda 4363$ wavelength regime.
To be conservative, we decided to remove the visit for 05144 to make a composite. 
In short, the integration time we used for the composite spectra of 04590 and 05144 was $4.05$\,hr, and $4.86$\,hr for 06355, 08140, and 10612.

The composition was done by median-stacking the available 5--6 2D spectra.
The individual 2D spectra were residual background-subtracted, and shifted to have common spatial and spectral ranges for the composition. All the shifts were integer movements.
We adopted the 2D median-stacking procedure to further alleviate the possible impacts of hot pixels that persisted in the final spectra, which were produced by the standard procedure by the Spec3 pipeline.
Moreover, the 2D median-stacking method allowed us to rescue the frames where the spectrum was partly out of the slit. 
Examples were found in the two nod2 positions for 10612 whose losses of the fluxes were $>3\sigma$ level if extracted in individual frames and compared with the other nods' 1D. Our method treated the outer regions of the slit as NaN and calculated the median value without NaN at each pixel in 2D. 
Using the 2D composite, one-dimensional spectrum was produced for each source and grating via the summation of 3 pixels  ($=0\farcs3 \simeq 2\times$\,FWHM) along the spatial direction centered on the spatial peak position. We adopted the narrow extraction aperture as compared to the previous studies to minimize the effects coming from the noisy regions close to the edge, particularly for 04590, 08140, and 10612 that were observed with a short MSA slitlet.

The flux calibration was one key procedure to be improved for the early NIRSpec studies,
as some of the reference files needed for the successful flux calibration, such as the flat frames for the spectrograph and the fore optics, were based on the predicted pre-flight throughputs. 
As of the pmap file of 1022, these flat files have been updated and created from in-flight data for all MOS data. 
The MOS flux calibration accuracy is now estimated at approximately 5 percent or less over the full wavelength range%
\footnote{%
\url{https://jwst-crds.stsci.edu/display_all_contexts/}
}.
We therefore used the flux solutions provided by the spec2pipeline together with the latest reference files.
Finally, we scaled the spectrum to match the broadband photometry of JWST/NIRCam to combine the spectroscopic and photometric measures based on a common aperture to derive the quantities such as equivalent widths of the optical emission lines and \xiion. We used the NIRCam broadband that covered the \OIII+\Hb\ emission for the scaling. 
The final composite 1D spectrum is presented in the bottom panel for each of the sources in Figure \ref{fig:spectra_ero}.

To evaluate the noise level, we used the 2D noise frames of read-out noise (\verb+VAR_RNOISE+) and Poisson noise (\verb+VAR_POISSON+) outputted in the Spec2 procedure for each object and nod.
In addition, we added the standard deviation of the residual background at each spectral bin in quadrature. 
We then averaged the noise frames of the available nods and extracted the 1D noise spectra in quadrature by using the same spectral and spatial ranges as adopted for the science spectra.
Finally the scaling was performed in the same manner as done for the science spectra.

\begin{figure}
    \begin{center}
    \includegraphics[bb=21 153 588 705, width=0.98\columnwidth]{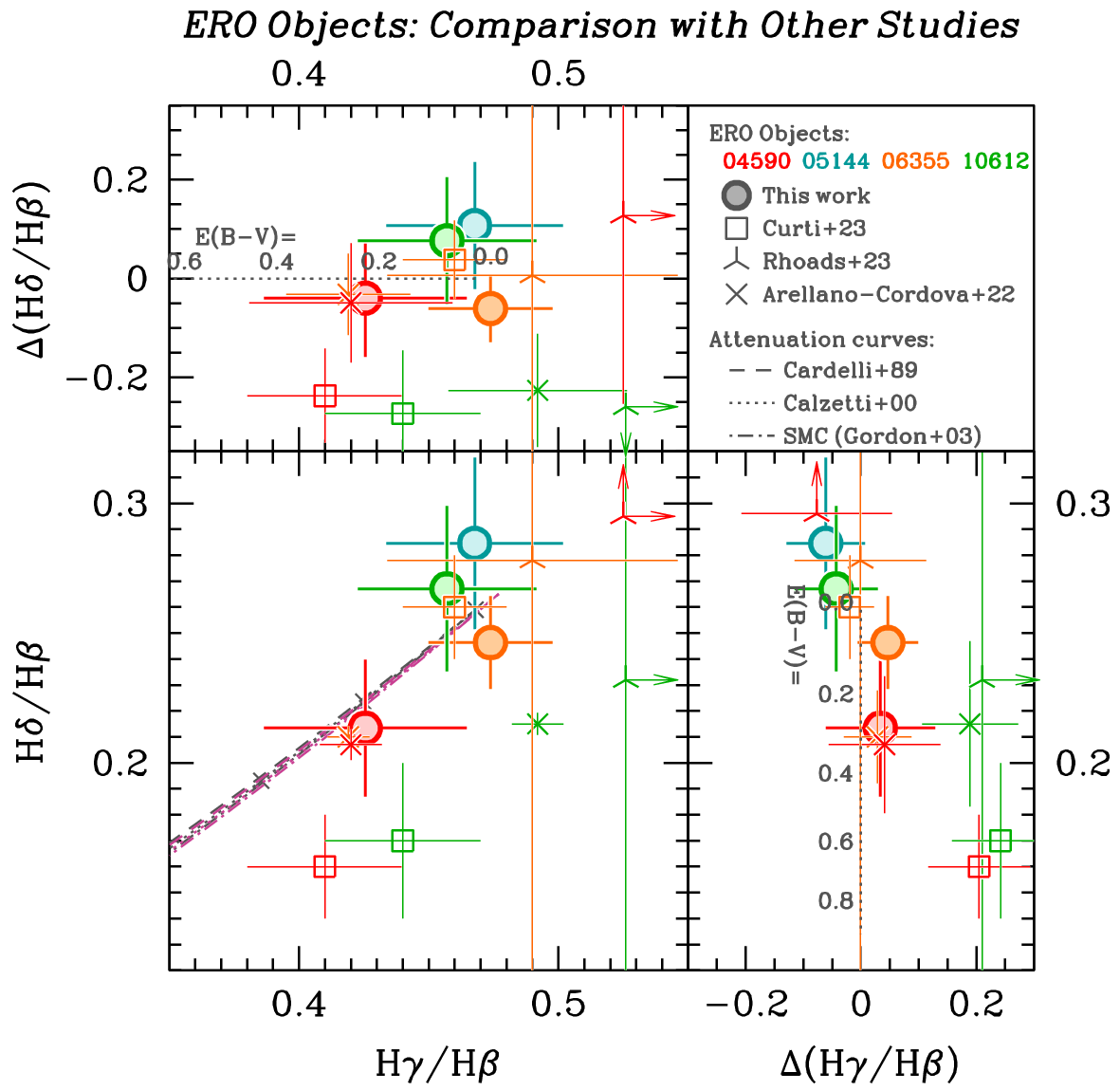}
    \caption{%
        Balmer emission line ratios of \Hg/\Hb\ and \Hd/\Hb\ for the ERO objects (04590 in red, 05144 in emerald green, 06355 in orange, and 10612 in green) in the main panel. The colored filled circles represent the new emission line flux measurements and errors with our improved reduction, while the other symbols show the measurements based on the early data release, as shown in the legend \citep{curti2023_ero,rhoads2023_ero,arellano-cordova2022_ero}. For the measurements of \citet{arellano-cordova2022_ero}, we adopt the average values based on the two spectra (o007 and o008) for 06355 and 10612, but refer only to the o008 value for 04590. For those with line ratios outside the ranges of the plot, we place them close to the edge of the plot with an arrow to indicate the direction toward the actually reported line ratio.
        The lines show expected Balmer decrements as a function of \ebv\ for the three different attenuation curves (\citealt{cardelli1989} with dashed, \citealt{calzetti2000} with dotted, and SMC (\citealt{gordon2003}) with dot-dashed) and the two different electron temperatures ($T_e=10000$\,K in gray and $25000$\,K in magenta), although the different attenuation curve and $T_e$ have a small impact on this diagram. The deviation of the \Hd/\Hb\ ratio from the theoretically expected value based on the \Hg/\Hb\ ratio ($\Delta$ \Hd/\Hb\ = ((\Hd/\Hb)${\rm obs}$ $-$ (\Hd/\Hb)${\rm expected}$) normalized by (\Hd/\Hb)$_{\rm expected}$) is highlighted in the upper panel, and that of the \Hg/\Hb\ ratio based on the \Hd/\Hb\ ratio in the left panel, for each symbol as in the main panel. The corresponding \ebv\ values are printed in these sub-panels, where the \citealt{calzetti2000} attenuation curve and $T_e=17500$\,K are adopted for reference. Our measurements all follow the sequence of the expected Balmer decrements within the uncertainties, while quite a few of the measurements from the earlier studies show the ratios that are not physically and/or consistently explainable.
    }
    \label{fig:balmer_lines}
    \end{center} 
\end{figure}

\subsubsection{Emission line flux measurements} \label{sssec:data_ero_measurements}

We measured the key optical emission lines of hydrogen Balmer lines, 
\OII$\lambda 3727$,
\NeIII$\lambda 3869$,
\OIII$\lambda 4363$, and 
\OIII$\lambda\lambda 5007, 4959$
by performing a Gaussian profile fitting to each line, and used the noise spectrum to weight the fit.
In the fitting procedure for the relatively faint objects of 04590, 05144, and 08140, 
the redshift of the 
\OIII$\lambda 5007$ (i.e., the strongest emission line) was
adopted as Gaussians for the other lines.
We perform all wavelength measurements on the vacuum wavelength scale, although we follow a longstanding convention of specifying the emission lines by their wavelength in air.
The error of each flux measurement was calculated by adding the noise levels of the spectral bins in quadrature within the wavelength range of $\pm$ FWHM centered on the Gaussian peak.
Using the flux measurements and the error evaluations, we confirm that four of the ERO objects, 04590, 05144, 06355, and 10612 present a $>3\sigma$ detection of \OIII$\lambda 4363$.
We also derived equivalent widths of the optical emission lines by combining the fluxes with the continuum level provided from a best-fit SED to the appropriate rest-frame optical broad-band photometry (Section \ref{sssec:data_ero_sedfit}).
We summarize the key emission line measurements in Table \ref{tbl:objects_directTe} for the \OIII$\lambda 4363$-detected sources to discuss the metallicity indicators (Section \ref{ssec:results_metallicity}). 
For the full sample, we list the metallicity diagnostic emission line ratios in Table \ref{tbl:objects_all} 
in Appendix \ref{sec_app:tbl_all}
together with the other physical quantities.

To test our new flux measurements with the improved reduction and calibration of the NIRSpec spectra, we show in Figure \ref{fig:balmer_lines} the Balmer decrements of \Hg/\Hb\ and \Hd/\Hb\ for the four ERO objects with a significant detection of the three lines, and compare them with the prediction of Case B recombination.
Observed Balmer line ratios should follow the sequence of the expected Balmer decrements as a function of dust attenuation, as shown with the curves adopting different reddening laws \citep{cardelli1989,calzetti2000,gordon2003} and electron temperatures ($T_e=10000$ and $25000$\,K).
The four ERO objects all follow the sequence and present physically-reasonable Balmer decrements within the uncertainties. The other object (08140) is not detected with \Hg\ and \Hd\ but the non-detections as well as the \Ha/\Hb\ line ratio are consistent with the theoretically expected line ratios at the $\lesssim 3\sigma$ level.
In contrast, previous studies claim some tensions of the observed Balmer decrements based on the early release products which show inconsistent with the expected line ratios, as referenced with the open and the cross symbols in Figure \ref{fig:balmer_lines}. 
We speculate such inconsistencies have been improved by implementing the residual background subtractions and the 2-dimensional stacking procedure, where we can ease the potential impacts of hot pixels that could have persisted in the final spectra in the early release. 
We note that the error-bars of our measurements are slightly larger than those in the previous studies \citep{curti2023_ero,arellano-cordova2022_ero}, mostly due to the addition of the uncertainty of the subtracted residual background into the noise levels.

Prior to quantitative analysis, it is necessary to consider corrections for dust reddening. We use the observed Balmer decrement of \Hg/\Hb\ for the four objects to estimate the dust attenuation assuming the \citet{calzetti2000} attenuation curve and a Case B recombination with $T_e=17500$\,K.
Adding a lower signal-to-noise ratio emission of \Hd\ does not change the estimate as indicated in Figure \ref{fig:balmer_lines}.
For 08140, we use the \Ha/\Hb\ ratio instead. 
We adopt \ebv\ $=0.22$ for 04590, $=0.03$ for 05144, $=0.08$ for 10612, and $=0.00$ for 08140 and 06355, and do not propagate the uncertainty in \ebv\ in the following analysis.

\subsubsection{NIRCam photometry and SED fit} \label{sssec:data_ero_sedfit}

All five ERO objects have been observed with JWST/NIRCam in the F090W, F150W, F200W, F277W, F356W, and F444W filters. These photometric data are crucial for deriving stellar properties such as stellar mass, and for evaluating the continuum level to correct for slit-loss in the NIRSpec observations. The derived corrections are important for obtaining key properties, such as the total SFR via recombination physics, \xiion, and EWs of optical emission lines such as EW(\Hb).

We refer to the catalog provided by \citet{harikane2022_jwst} to extract the necessary photometric data.
We adopt the total magnitudes that are estimated from the $0\farcs3$-diameter aperture magnitudes with the aperture corrections.
The uncertainties of the magnitudes include the photometric errors from the images as well as the 10\,\%\ error floor to account for systematic uncertainties arising from e.g., the zero-point corrections.

Using the total magnitudes and the errors of NIRCam, we perform the Bayesian spectral energy distribution (SED) fitting with \verb+Prospector+ \citep{johnson2021} to derive the stellar population properties.
The procedure of the SED fitting is the same as that of \citet{harikane2022_jwst}, except for the fixed redshifts based on the NIRSpec emission line measurements.
We use the stellar population synthesis package, Flexible Stellar Population Synthesis (FSPS; \citealt{conroy2009,conroy2010}) for stellar SEDs, and include nebular emission from the photoionization models of Cloudy \citep{byler2017}.
We assume the stellar initial mass function (IMF) of \citet{chabrier2003}, the intergalactic medium (IGM) attenuation model of \citet{madau1995}, the \citet{calzetti2000} dust attenuation law, and a fixed metallicity of $0.2$\,\Zsun.
We choose a flexible star formation history as adopted in \citet{harikane2022_jwst}.

The stellar masses we obtain from the SED fitting are the apparent values, i.e., after experiencing the strong lensing magnifications for the ERO objects. To correct for the lensing effects, we refer to the mass model of SMACS J0723 constructed with \verb+glafic+ \citep{oguri2010, oguri2021} updated with the new JWST ERO data, as adopted in \citet{harikane2022_jwst}.
The adopted magnification factors as well as the stellar masses after the lensing magnification correction are summarized in Table \ref{tbl:objects_all}.
The errors of the best-fit parameters from the SED fitting correspond to the $1\sigma$ confidence interval ($\Delta\chi^2<1$) for each parameters.

\begin{figure}[t]
    \begin{center}
    \includegraphics[bb=18 175 516 541, width=0.95\columnwidth]{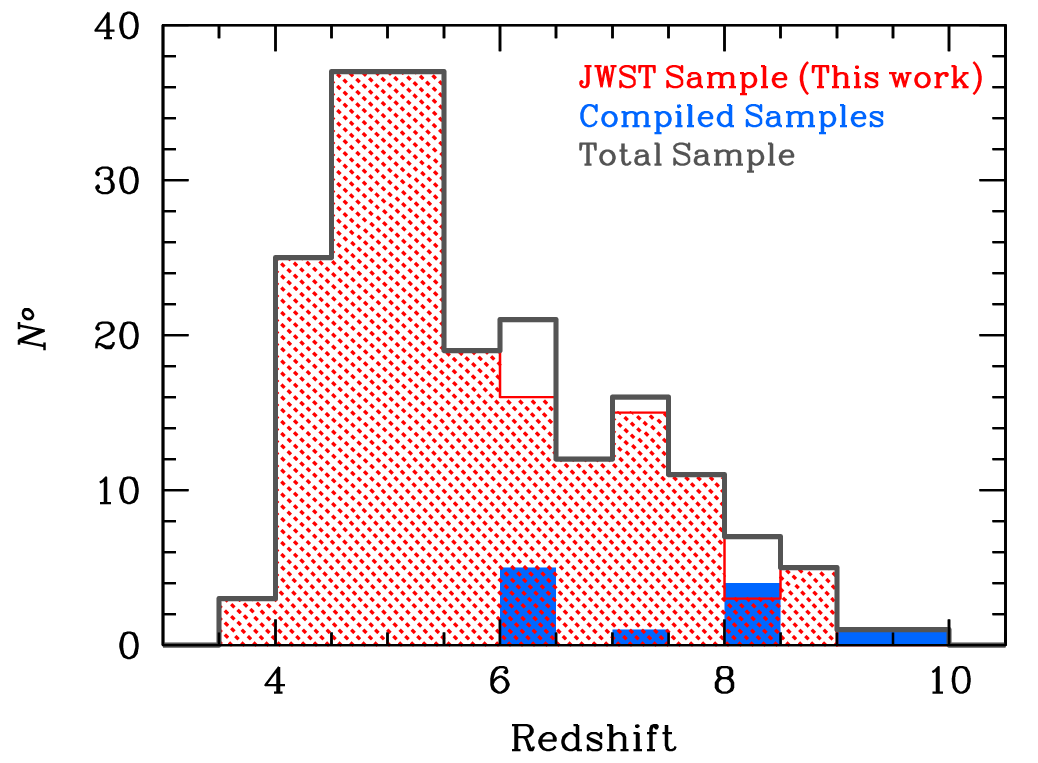}
    \caption{%
        Redshift distribution of the high-redshift galaxy samples used in this work. The red histogram presents the JWST sample constructed in this study using our analyzed ERO, GLASS, and CEERS spectra. The blue histogram shows the samples compiled from the literature (see Section \ref{sssec:results_metallicity_literature}). The sum of these samples, represented by the gray histogram, is used to investigate the mass-metallicity relationships at $z=4-10$ in this work.
    }
    \label{fig:hist_redshift}
    \end{center} 
\end{figure}

\subsection{ERS GLASS Data} \label{ssec:data_glass}

The ERS GLASS NIRSpec MSA observations were carried out on UT 2022 Nov 10-11
for objects behind the galaxy cluster Abell 2744 (Proposal ID: 1324; \citealt{treu2022_glass}).
They adopted the $R\sim 2700$ high resolution gratings/filters of G140H/F100LP, G235H/F170LP, and G395H/F290LP
sampling the wavelength range of $1.0-1.9$, $1.7-3.2$, and $2.8-5.3\,\mu$m, respectively.
Twelve exposures of $1473$\,sec were conducted in one pointing of NIRSpec MSA for each spectral configuration with the 3-shutter slitlet nod pattern, totaling the on-source integration of $4.9$\,hr in each of G140H, G235H, and G395H.

Through the initial screening of all of the public 1D and 2D composite spectra, we identified $15$ objects at $z>4$ with \OIII$+$\Hb%
\footnote{
There were six other objects whose \OIII$+$\Hb\ appeared at $\lambda=2.5-5.2\,\mu$m in the spectrum, with IDs 340899, 341905, 340055, 341715, 341712, and 80121. After our careful investigations, they were concluded as fake detections and hence not high-redshift objects, since they were located along the same spatial (cross-dispersion) position of the other high-redshift objects on the MSA mask, and their spectra were contaminated by the others. Indeed, the fake \OIII$+$\Hb\ detections showed inconsistent redshifts, indicating the spectrum was coming from a different slit.
}.
The sample contains the members of the protocluster at $z=7.89$ as presented by \citet{morishita2022_glass}. 
We re-reduced the spectra of the 15 objects with our custom procedure as described in Section \ref{sssec:data_ero_observations}.
We carefully checked all of the 12 nods' spectra for each objects to determine the extraction aperture and any nods to be removed from the composition. 
For 10005, the light from another member of the $z=7.89$ protocluster partly came into the same slit at a spatially-offsetted position. We changed the background frames during the spec2 procedure so that the signal from 10005 was not over-subtracted by the neighboring object in different nods.

The emission line flux measurements were performed in the same manner as explained in Section \ref{sssec:data_ero_measurements}. Among the 15 objects, we identified 5 with \OIII$\lambda 4363$. One of the \OIII$\lambda 4363$-detected source, 40066, lacked the \OII$\lambda 3727$ emission which fell in the detector gap. The other 4 objects, whose flux measurements are given in Table \ref{tbl:objects_directTe}, are thus used for the metallicity measurements with the direct $T_e$ method (Section \ref{ssec:results_metallicity}).
The nebular dust attenuation was evaluated using \Ha, \Hb, and/or \Hg.

The NIRCam multi-band photometric catalog was generated in the same manner as in Section \ref{sssec:data_ero_sedfit} based on the images publicly available from the GLASS and UNCOVER (Proposal ID: 2561) programs.
We use the data of F115W, F150W, F200W, F277W, F356W, and F444W for the subsequent SED fitting analysis.
Note that one of 15 objects (160122) is not covered by NIRCam, and its stellar mass are not determined.

All of the GLASS objects are strongly lensed. To correct for the magnification factors, we use the \verb+glafic+ \citep{oguri2010,oguri2021} mass model of \citet{kawamata2018} with a minor update based on the Multi Unit Spectroscopic Explorer (MUSE) follow-up spectroscopy \citep{bergamini2022}. The updated mass model uses 45 multiple image systems, 27 of which have spectroscopic redshifts. This model reproduces the observed multiple image positions with an accuracy of $0 \farcs 42$. The derived magnification factors, the stellar masses after the lensing magnification correction, as well as the other key quantities for the entire GLASS sample are summarized in Table \ref{tbl:objects_all}
in Appendix \ref{sec_app:tbl_all}.

\begin{figure*}[t]
    \begin{center}
        \subfloat{
            \includegraphics[bb=0 0 520 393, height=0.35\textheight]{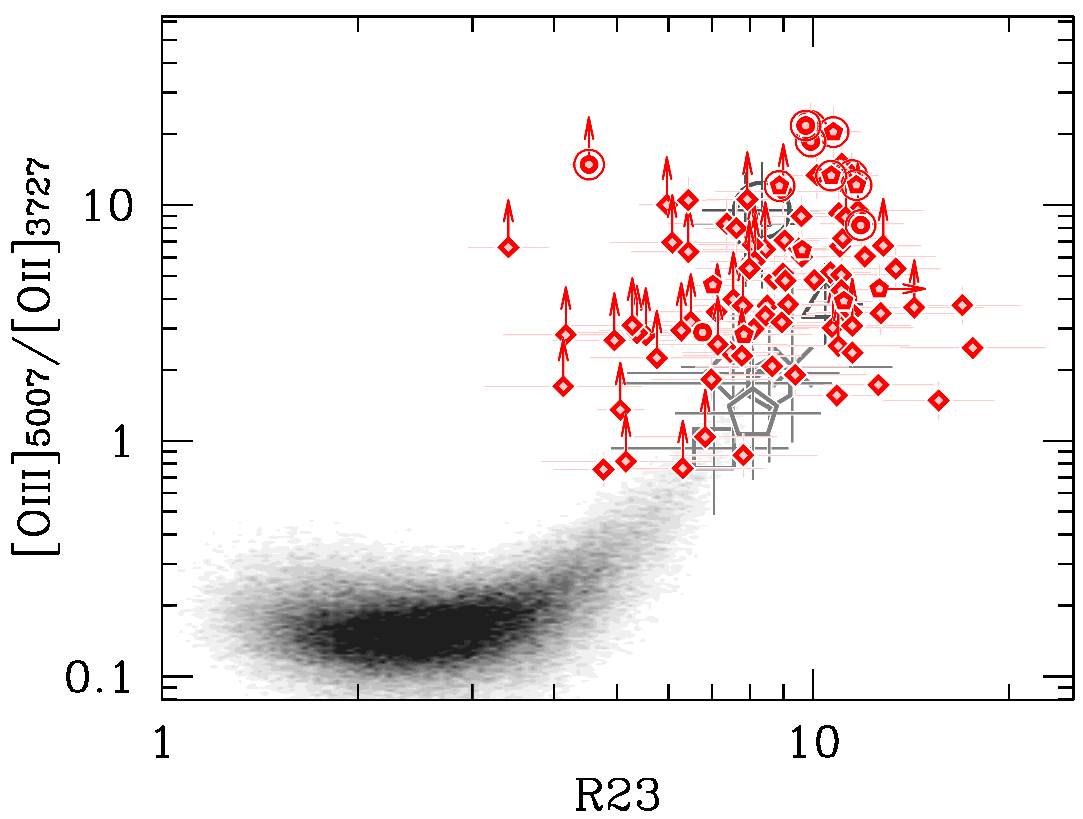}
        }
        \subfloat{
            \includegraphics[bb=29 115 549 502, height=0.2\textheight]{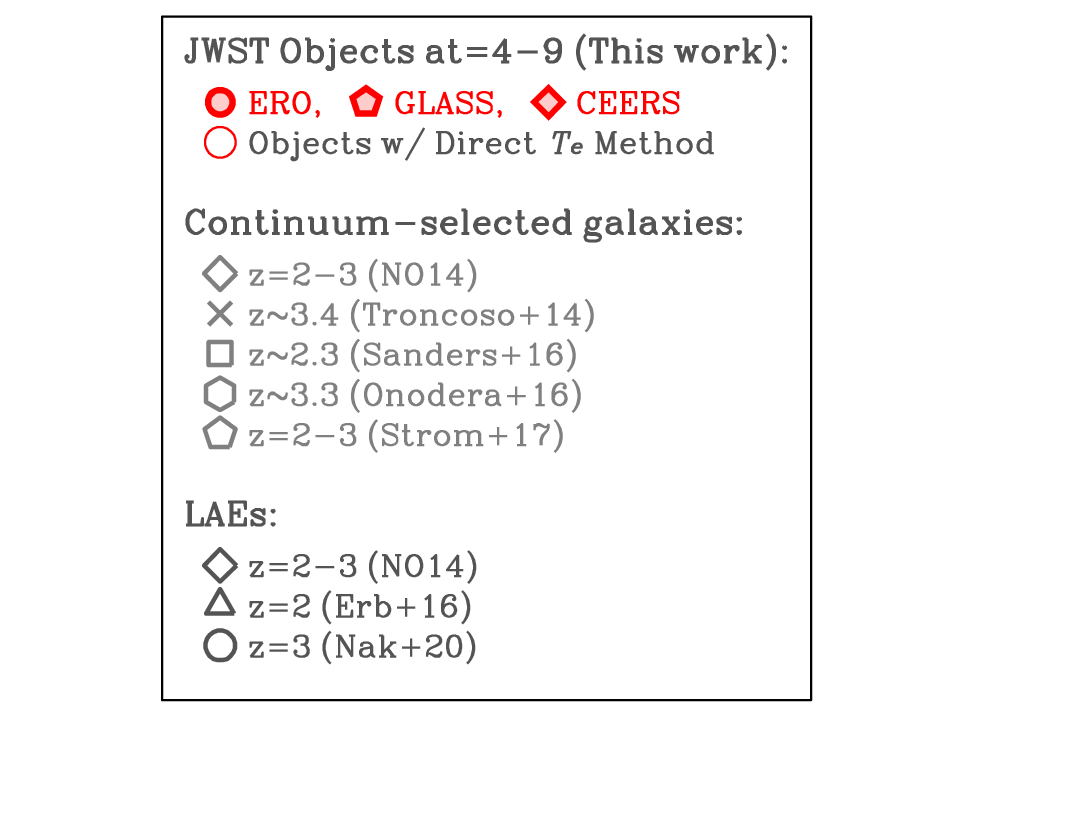}
        }
        \caption{%
            O32 vs. R23 diagram comparing the JWST sample at $z=4-9$, analyzed in this work, with lower redshift samples at $z=2-3$. The JWST objects are shown as red filled symbols, with circles representing ERO, pentagons representing GLASS, and diamonds representing CEERS. Objects with metallicity measurements obtained using the direct $T_e$ method are indicated with red open circles. The gray open symbols represent the averages of lower-redshift samples of continuum-selected galaxies \citep{NO2014,troncoso2014,sanders2016,onodera2016,strom2017} and \Lya\ emitting galaxies (LAEs; \citealt{NO2014,erb2016,nakajima2020}), as indicated in the legend. Arrows indicate $3\sigma$ lower limits. Gray shading illustrates the equivalent distribution for nearby SDSS galaxies.
        }
        \label{fig:o3o2r23}
    \end{center} 
\end{figure*}

\subsection{ERS CEERS Data} \label{ssec:data_ceers}

The ERS CEERS NIRSpec MSA observations were carried out on UT 2022 Dec 20-22, and 24
in the blank field of EGS (\citealt{finkelstein2022_ceers,fujimoto2023_ceers}; Proposal ID: 1345).
They used both of the Prism ($R\sim 100$) and the medium resolution ($R\sim 1000$) gratings to cover from $1.0$ to $5\,\mu$m in 6 pointings (P4, P5, P7, P8, P9, and P10) with some overlaps of objects. 
Note that two of the Prism pointings, P9 and P10, were not fully completed. 
To compensate the two Prism pointings, additional observations were performed on UT 2023 Feb 9-10 (P11 and P12; see also \citealt{arrabal-haro2023_ceers}). These additional data were also analyzed and presented in this paper to provide a complete spectroscopic sample of CEERS to the community.
Three exposures of $1036$\,sec were conducted for each spectral configuration with the 3-shutter slitlet nod pattern, totaling the on-source integration of $0.9$\,hr in each Prism, G140M, G235M, and G395M, excepting for P9 and P11 where another set of three exposures of $1036$\,sec were repeated in the Prism observations.

At redshifts $z>4$, we detected a total of 147 Prism spectra and 77 medium grating spectra with \OIII$+$\Hb\ emission lines. Among these, three objects (IDs 00618, 00717, and 01027) were observed with two different Prism and medium grating pointings each, resulting in four duplicate observations. Additionally, one object (ID 00603) was observed with two different Prism pointings and one medium grating pointing, resulting in three duplicate observations. Furthermore, 50 objects were observed with both Prism and medium grating, resulting in two duplicate observations with one pointing each.
In summary, a total of 163 objects were identified through the ERS CEERS NIRSpec observations, as summarized in Table \ref{tbl:objects_all}
in Appendix \ref{sec_app:tbl_all}.
For the 163 objects, we performed our re-reduction procedure as described in Section \ref{sssec:data_ero_observations}.

The same emission line flux measurements were performed as detailed in Section \ref{sssec:data_ero_measurements}. 
We identified two out of the 163 objects with \OIII$\lambda 4363$ among the CEERS sample.
Their flux measurements are added in Table \ref{tbl:objects_directTe}.
The nebular dust attenuation was evaluated using \Ha, \Hb, and/or \Hg\ depending on the detections and the redshift.

The CEERS NIRCam observations consists of 10 pointings, four of which were taken in Jun 2022 and the rest in Dec 2022 \citep{bagley2022_ceers}. The early observations of 4 pointings are presented in \citet{harikane2022_jwst}. We reduce the latter observations' data in the same manner, and generate the NIRCam full photometry catalog, including the seven bands of F115W, F150W, F200W, F277W, F356W, F410M, and F444W (Y. Harikane et al. in prep.).
One caveat is that not all of the NIRSpec CEERS objects are covered by NIRCam. We find 91 of the 163 CEERS objects are in the NIRCam coverage. For the 91 objects, we perform the SED fitting as in Section \ref{sssec:data_ero_sedfit} to derive the stellar masses, and scale the NIRSpec spectra to be consistent with the NIRCam total magnitude photometry (i.e., slit-loss correction).
For the others except for a single object (ID 01115), we rely on the Hubble Space Telescope (HST) WFC3 photometry \citep{stefanon2017}, as detailed in \citet{isobe2023_jwst}. Briefly, we derive the UV absolute magnitudes (\Muv) from the HST/WFC3 photometry, use the empirical relationships between stellar mass and \Muv\ at $z=4-8$ \citep{song2016} to estimate the stellar masses, and rescale them to the \citet{chabrier2003} IMF. 
We exclude the single object with ID 01115 from the following analysis as its HST photometry is not given.
For those not detected in the HST bands, we translate the corresponding $5\sigma$ limiting magnitudes to the $5\sigma$ upper-limits of stellar mass. 
While acknowledging that the empirical method utilizing \Muv\ for estimating stellar mass may be less reliable compared to the method involving an SED fit to NIRCam photometry, we demonstrate in the subsequent sections (Sections \ref{ssec:results_MZ} and \ref{ssec:results_FMR}) that our main findings remain unchanged even when excluding objects without NIRCam photometry. This is partly because the fraction of such objects is small ($\sim 28$\,\%) in deriving the average relationships. Furthermore, we have found reasonable consistency in stellar mass estimates between the two methods by using objects with both NIRCam and HST photometry. Details are described in Appendix \ref{sec_app:results_metallicity_sedfit}.
Note that we cannot correct for the slit-loss of the NIRSpec spectra for the objects without NIRCam photometry. This will not affect the metallicity measurements as we only use the line ratios. However, that does prevent us from deriving the equivalent widths of the optical emission lines such as EW(\Hb), and the SFR based on the total \Hb\ luminosity. 
\\

In summary, we construct a JWST sample of 182 objects at $z=3.8-8.9$. 
The redshift distribution of the sample is shown in red in Figure \ref{fig:hist_redshift}.

\section{Results} \label{sec:results}

\subsection{Emission line ratios} \label{ssec:results_line_ratios}

To examine the nebular properties of JWST objects at $z=4-9$ and compare with those of lower-redshift galaxies, 
we plot them in the line ratio diagnostic diagram of 
(\OIII$\lambda\lambda5007,4959+$\OII$\lambda 3727$)$/$\Hb\ (R23) and 
\OIII\ $\lambda5007$$/$\OII$\lambda3727$ (O32) in Figure \ref{fig:o3o2r23} (see also \citealt{mascia2023_glass,tang2023_ceers,sanders2023_jwst}).
This diagram has been widely used to investigate the metallicity and ionization state in the local universe and up to $z=2-3$ (e.g., \citealt{KD2002,maiolino2008,NO2014,troncoso2014,sanders2016,onodera2016,strom2017,nakajima2020}).
When compared to the local Sloan Digital Sky Survey (SDSS) galaxies, which are free from AGN contamination as determined by the BPT diagram (e.g., \citealt{BPT1981, kauffmann2003}; see below for more details), we observe that the $z=2-3$ galaxies exhibit a higher O32 line ratio, indicating an overall increase in ionization parameter at higher redshifts.
Figure \ref{fig:o3o2r23} confirms the trend, and the $z=4-9$ JWST objects show a further high O32 line ratio on average than typically seen in $z=2-3$ continuum-selected galaxies. 
Remarkably, one of the ERO objects, ERO\_04590 at $z=8.5$ presents a stringent lower-limit of O32 as $>14.8$ ($3\sigma$), which corresponds to an ionization parameter of $\log U>-2.21$ according to the prescription of \citet{KD2002} (see also \citealt{fujimoto2022_jwst_alma}%
\footnote{
The authors adopts a different dust attenuation law of SMC \citep{gordon2003}, which results in a slightly lower value of the limit for the ionization parameter: $\log U>-2.27$.
}). 
The lower-limit is a factor of $\sim10$ higher value than typically seen in the SDSS galaxies.

\begin{figure}[t]
    \begin{center}
    \includegraphics[bb=29 153 549 549, width=0.99\columnwidth]{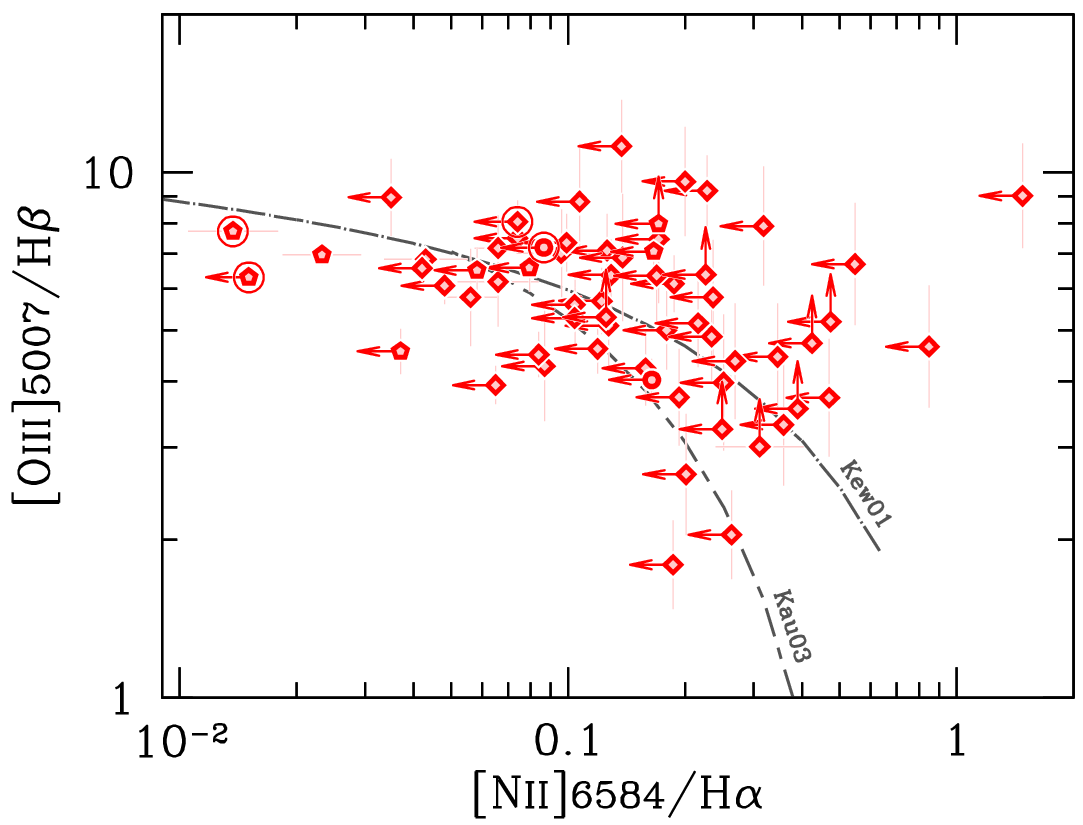}
    \caption{%
        The \NII\ BPT diagram.
        Red symbols represent the JWST objects, as shown in Figure \ref{fig:o3o2r23}, whose \Ha$+$\NII\ lines are covered in the NIRSpec spectra. Arrows indicate $3\sigma$ limits. Two popular demarcation curves between AGNs and star-forming galaxies, from \citet{kewley2001} (dot-long dashed; `Kew01') and \citet{kauffmann2003} (short dash-long dashed; `Kau03'), are also shown. None of the objects fall above the demarcation curves significantly beyond the measurement uncertainties.
    }
    \label{fig:o3hb_n2ha}
    \end{center} 
\end{figure}

At $z=2-3$, low-mass \Lya\ emitting galaxies (LAEs) are suggested to have a remarkably high O32 line ratio, which would be achieved by a hard ionizing radiation field and/or a low metallicity \citep{nakajima2016,trainor2016,erb2016}. Its implication regarding the escape of ionizing photons (and that of \Lya\ photons) are also discussed (e.g., \citealt{NO2014,izotov2016_nature,izotov2018_5more,verhamme2017,steidel2018,fletcher2019,erb2019,nakajima2020,katz2020_lyc,flury2022_II}).
The extreme O32 line ratios seen in the $z=4-9$ JWST objects are comparable to those in $z=2-3$ low-mass LAEs and $z\sim 0$ Lyman-continuum (LyC) leaking objects, suggesting that high-redshift sources have a nebular and ionization condition of gas that is similar to those typically found in lower-redshift low-mass galaxies having a strong \Lya\ and/or LyC escape with a hard ionizing spectrum. A similar finding is also discussed by \citet{mascia2023_glass} and \citet{sanders2023_jwst}, and notably by \citet{tang2023_ceers}, where the strong \Lya\ emitting galaxies at $z>7$ are directly examined with JWST and suggested to present the largest O32 line ratios.

One uncertainty in interpreting O32 arises as it also depends on the metallicity. 
In the next section, we present our method to derive the metallicities for the JWST objects using several indicators including R23, particularly by referring to the objects whose \OIII$\lambda 4363$ is detected and metallicity is precisely determined with electron temperature.
\\

Before moving to the metallicity results, we also examine our objects using another popular emission line diagram, the \NII\ BPT diagram, which plots the ratios of \OIII$\lambda 5007/$\Hb\ and \NII$\lambda 6584/$\Ha\ (\citealt{BPT1981}). This diagram is commonly used to diagnose the presence of an active galactic nucleus (AGN) and is applicable to our JWST objects up to a redshift of $z<6.9$, where the \Ha$+$\NII\ emission is covered by NIRSpec spectra. Figure \ref{fig:o3hb_n2ha} shows the \NII\ BPT diagram for our JWST objects at $z<6.9$, along with two curves that discriminate between sources dominated by AGNs and stars \citep{kewley2001,kauffmann2003}. Sources located above the curves are classified as AGN-dominated.

Figure \ref{fig:o3hb_n2ha} shows that most JWST sources have an upper limit of \NII. The \NII-detected objects fall either below or on the demarcation curves within the measurement uncertainties. None of the JWST objects are thus classified as obvious AGNs. However, it is worth noting that low-metallicity AGNs with $Z\lesssim 0.5$\,\Zsun\ may contaminate the star-forming galaxy region on the diagram (e.g., \citealt{kewley2013_theory,NM2022}). We cannot fully rule out the presence of such low-metallicity AGNs in our sample based on the \NII\ BPT result, but we suggest an absence of evolved AGNs as far as we explore ($z<6.9$).
Accordingly, we do not remove any objects from the sample for the following results based on the \NII\ BPT diagram. 
Although there are some objects whose \NII\ upper-limits are too weak to conclude their BPT diagnostics, we mention in the following sections (Sections \ref{ssec:results_MZ} and \ref{ssec:results_FMR}) that our main results are not changed by excluding these unclear objects.

As a complementary approach to our method using the \NII\ BPT diagram, a companion study by \citet{harikane2023_jwst_blagn} conducts a search for faint AGNs in our full JWST spectroscopic sample by examining the broad component (FWHM $>1000$\,\kms) around the \Ha\ emission line. The authors identify 10 objects with signatures of AGNs at redshifts $z=4.0-6.9$ (see also \citealt{kocevski2023_blagn,ubler2023_blagn}).
To avoid potential biases in our mass-metallicity relations, we remove these 10 objects from our sample when we examine the mass-metallicity relations. These objects are flagged in Table \ref{tbl:objects_all}.

\subsection{Gas-phase Metallicity} \label{ssec:results_metallicity}

\subsubsection{Direct $T_e$ method} \label{sssec:results_metallicity_directTe}

\begin{splitdeluxetable*}{lccccccccBcccc}
\tablecaption{Summary of JWST Objects with Direct $T_e$ Method
\label{tbl:objects_directTe}}
\renewcommand{\arraystretch}{1.1}
\tabletypesize{\scriptsize}
\tablehead{
\colhead{ID}
& \colhead{[O\,{\sc ii}]$^{(\star)}$}
& \colhead{[Ne\,{\sc iii}]$^{(\star)}$}
& \colhead{H$\delta$$^{(\star)}$}
& \colhead{H$\gamma$$^{(\star)}$}
& \colhead{[O\,{\sc iii}]$^{(\star)}$}
& \colhead{H$\beta$$^{(\star)}$}
& \colhead{[O\,{\sc iii}]$^{(\star)}$}
& \colhead{[O\,{\sc iii}]$^{(\star)}$}
& \colhead{redshift}
& \colhead{EW(H$\beta$)}
& \colhead{$T_e$(O\,{\sc iii})}
& \colhead{12+log(O/H)}
\\ 
& $\lambda3727$
& $\lambda3869$
& 
& 
& $\lambda4363$
& 
& $\lambda4959$
& $\lambda5007$
& & (\AA)
& ($10^4$\,K)
& 
} 
\startdata
ERO\_04590 & $<5.7$ & $<24.8$ & $21.3 \pm 2.3$ & $42.6 \pm 2.8$ & $13.8 \pm 2.7$ & $100.0 \pm 6.4$ & $110.5 \pm 6.0$ & $337.8 \pm 8.3$ & $8.496$ & $217.8 \pm 150.5$ & $2.08 \pm 0.26$ & $7.26^{+0.15}_{-0.13}$ \\ 
ERO\_05144 & $49.6 \pm 6.2$ & $63.7 \pm 8.9$ & $28.5 \pm 3.2$ & $46.8 \pm 3.1$ & $14.9 \pm 2.9$ & $100.0 \pm 3.0$ & $238.3 \pm 3.9$ & $719.0 \pm 5.5$ & $6.378$ & $150.9 \pm 51.2$ & $1.64 \pm 0.16$ & $7.82^{+0.12}_{-0.09}$ \\ 
ERO\_06355$^{(\dag)}$ & $99.6 \pm 3.7$ & $48.2 \pm 2.1$ & $24.6 \pm 1.7$ & $47.4 \pm 2.1$ & $8.7 \pm 1.8$ & $100.0 \pm 2.5$ & $267.8 \pm 7.9$ & $817.1 \pm 5.8$ & $7.665$ & $150.3 \pm 4.4$ & $1.13 \pm 0.08$ & $8.35^{+0.11}_{-0.08}$ \\ 
ERO\_10612 & $43.8 \pm 5.1$ & $54.7 \pm 4.0$ & $26.7 \pm 3.0$ & $45.7 \pm 3.1$ & $22.1 \pm 2.8$ & $100.0 \pm 3.5$ & $238.6 \pm 4.9$ & $712.3 \pm 6.7$ & $7.660$ & $210.3 \pm 16.4$ & $2.04 \pm 0.16$ & $7.59^{+0.08}_{-0.08}$ \\ 
GLASS\_100003 & $37.0 \pm 11.9$ & $29.2 \pm 8.2$ & $<7.2$ & $51.3 \pm 8.0$ & $25.7 \pm 7.7$ & $100.0 \pm 10.1$ & $259.2 \pm 12.1$ & $777.8 \pm 16.5$ & $7.877$ & $138.7 \pm 19.2$ & $1.98 \pm 0.37$ & $7.64^{+0.23}_{-0.19}$ \\ 
GLASS\_10021 & $68.3 \pm 5.1$ & $48.4 \pm 5.1$ & $24.3 \pm 3.5$ & $49.4 \pm 4.3$ & $19.8 \pm 4.1$ & $100.0 \pm 4.6$ & $269.6 \pm 5.2$ & $828.4 \pm 6.8$ & $7.286$ & $67.9 \pm 27.0$ & $1.66 \pm 0.17$ & $7.87^{+0.12}_{-0.10}$ \\ 
GLASS\_150029$^{(\ddag)}$ & $42.0 \pm 3.3$ & \nodata & $22.1 \pm 3.4$ & $47.1 \pm 4.0$ & $15.3 \pm 2.3$ & $100.0 \pm 4.5$ & $222.1 \pm 5.8$ & $631.6 \pm 7.9$ & $4.584$ & $137.9 \pm 10.3$ & $1.76 \pm 0.14$ & $7.70^{+0.09}_{-0.08}$ \\ 
GLASS\_160133$^{(\ddag)}$ & $47.5 \pm 2.1$ & $45.0 \pm 2.0$ & $23.8 \pm 1.2$ & $46.5 \pm 1.5$ & $13.4 \pm 1.4$ & $100.0 \pm 1.6$ & $255.9 \pm 2.0$ & $772.5 \pm 3.1$ & $4.015$ & $227.2 \pm 89.3$ & $1.48 \pm 0.07$ & $7.95^{+0.06}_{-0.05}$ \\ 
CEERS\_01027 & $45.4 \pm 9.2$ & $59.5 \pm 6.1$ & $25.5 \pm 5.0$ & $38.8 \pm 4.4$ & $15.6 \pm 5.0$ & $100.0 \pm 6.4$ & $220.9 \pm 7.5$ & $738.1 \pm 10.9$ & $7.820$ & $174.7 \pm 14.4$ & $1.56 \pm 0.24$ & $7.83^{+0.20}_{-0.15}$ \\ 
CEERS\_01536 & $59.5 \pm 11.6$ & $43.1 \pm 9.0$ & $<12.9$ & $63.4 \pm 10.0$ & $30.7 \pm 8.2$ & $100.0 \pm 9.6$ & $296.1 \pm 15.6$ & $805.4 \pm 16.5$ & $5.034$ & \nodata & $2.25 \pm 0.43$ & $7.61^{+0.18}_{-0.15}$ \\ 
\enddata
\tablecomments{%
($\star$) Observed flux ratios relative to \Hb\ ($\times 100$).
Upper-limit values at the $1\sigma$ level.
($\dag$) The presence of high ionization line of \NeIV\ is seen \citep{brinchmann2022_ero}.
($\ddag$) The presence of broad \Ha\ is indicated \citep{harikane2023_jwst_blagn}
}
\end{splitdeluxetable*}


\begin{figure}[t]
    \begin{center}
    \includegraphics[bb=21 169 531 696, width=0.9\columnwidth]{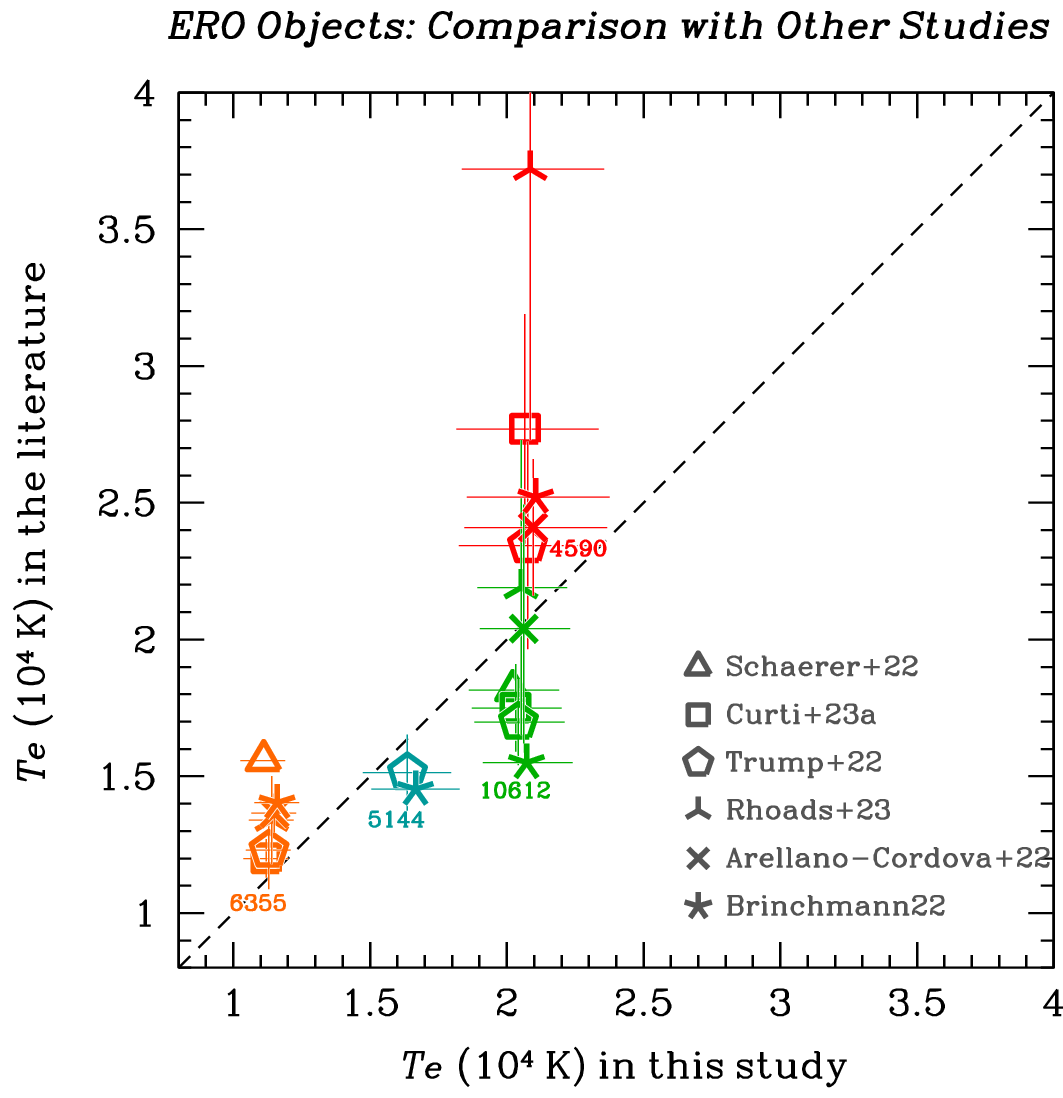}
    \caption{%
        Comparison of literature determinations of electron temperature ($T_e$) for four ERO galaxies (04590 in red, 06355 in orange, 10612 in green, 05144 in emerald green) where \OIII$\lambda 4363$ is identified. Different symbols indicate different publications, as shown in the legend. For the measurements of \citet{arellano-cordova2022_ero}, we show the average values based on the two spectra (i.e., two observing blocks: o007 and o008) for 06355 and 10612, but refer only to the o008 value for 04590.
        For each object, the symbols are slightly offset in the horizontal direction for display purposes. Some of the measurements in the literature do not present uncertainties and are plotted without error bars in the vertical direction.
    }
    \label{fig:compare_Te}
    \end{center} 
\end{figure}

\begin{figure*}[t]
    \begin{center}
        \vspace{-6pt}
        
        \subfloat{
            \includegraphics[bb=23 163 524 698, height=0.245\textheight]{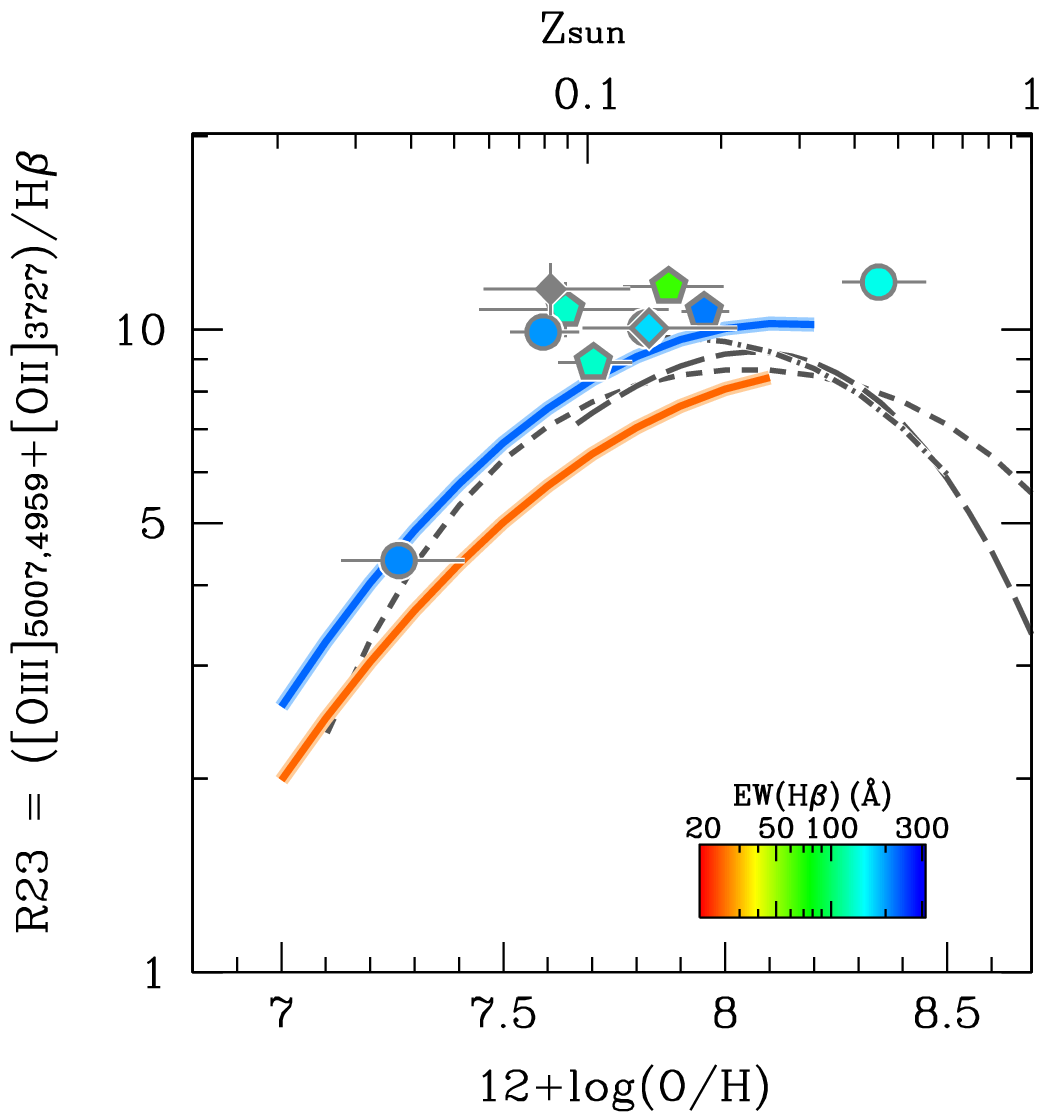}
        }
        \subfloat{
            \includegraphics[bb=23 163 524 698, height=0.245\textheight]{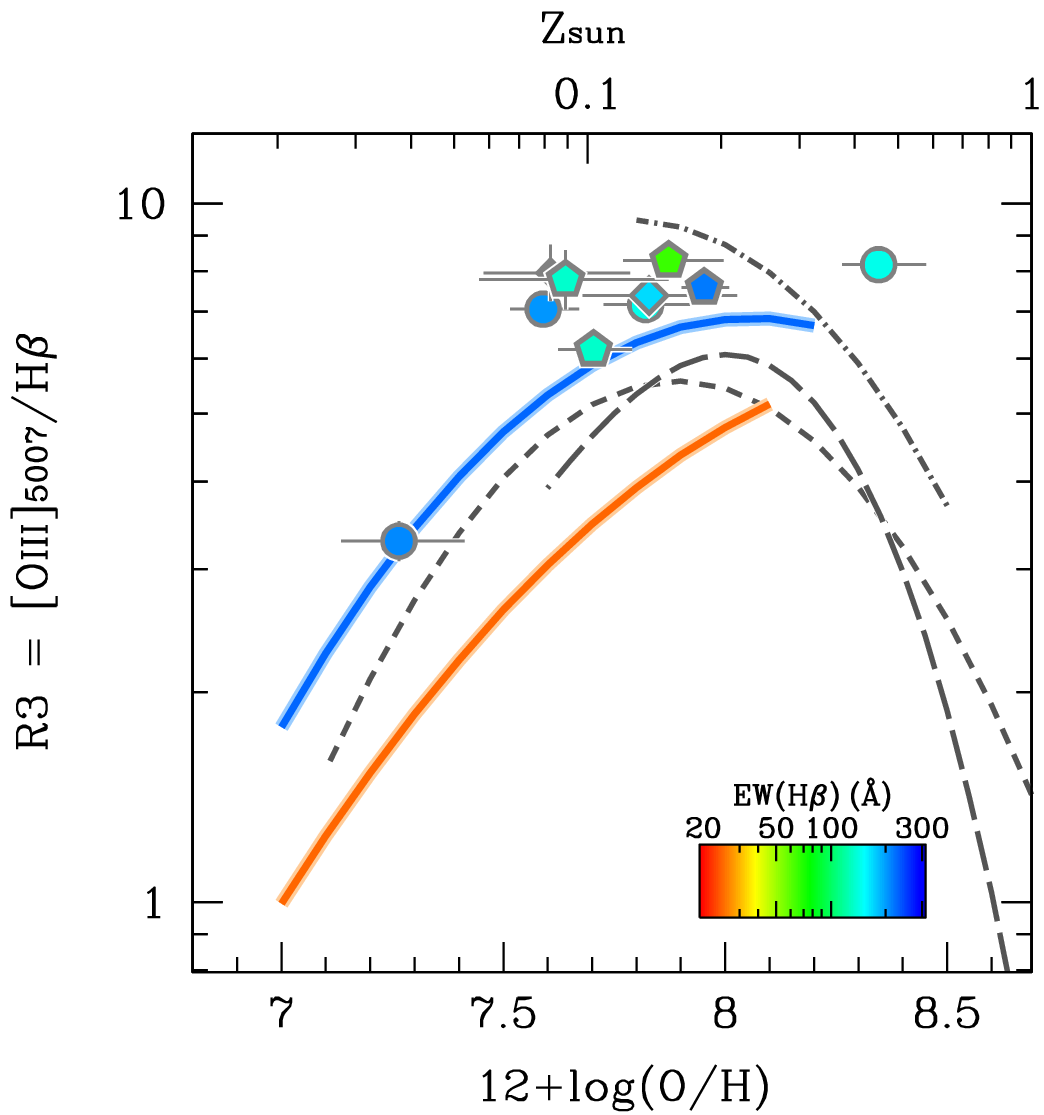}
        }      
        \subfloat{
            \includegraphics[bb=23 163 524 698, height=0.245\textheight]{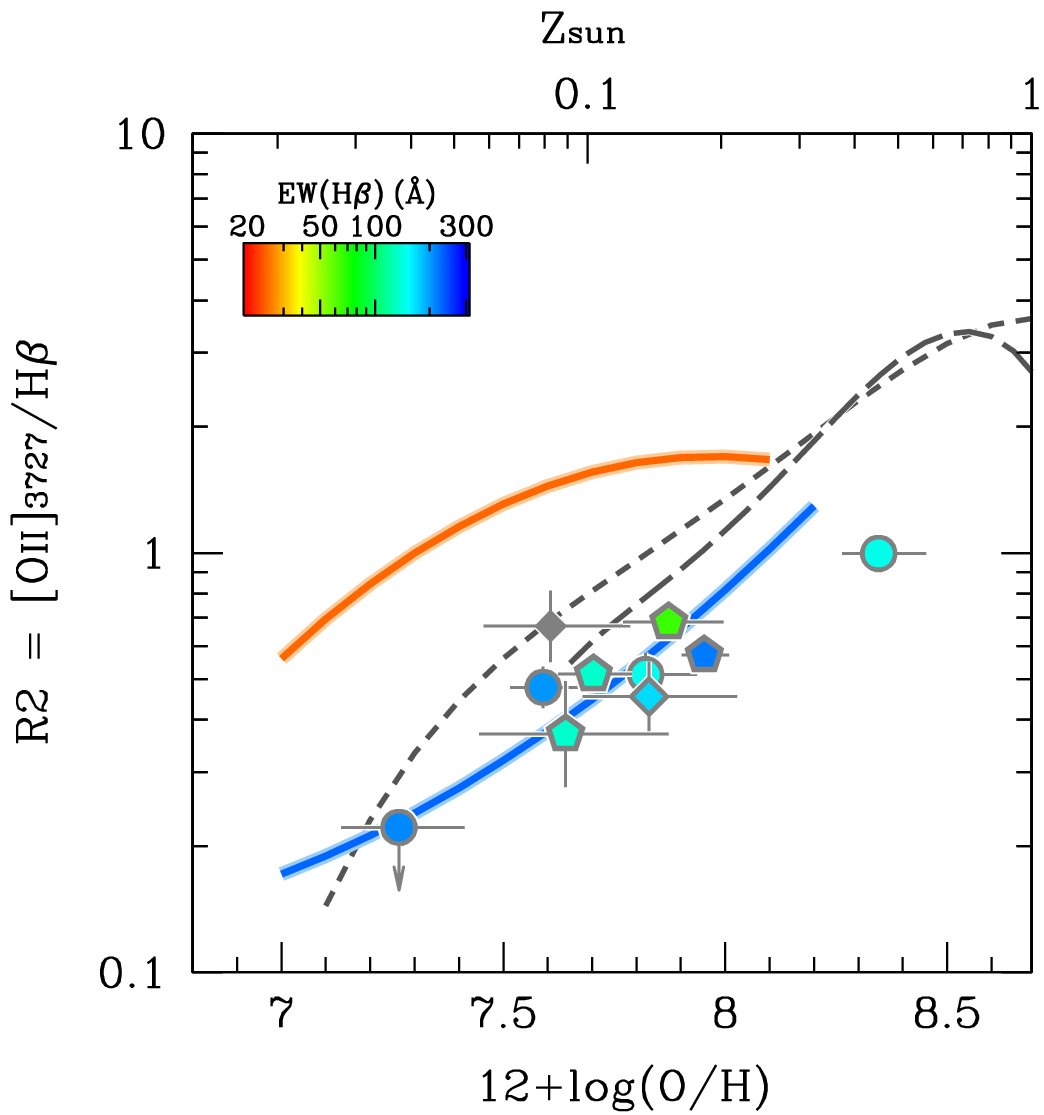}
        }
        
        \vspace{-6pt}
        
        \subfloat{
            \includegraphics[bb=23 163 524 698, height=0.245\textheight]{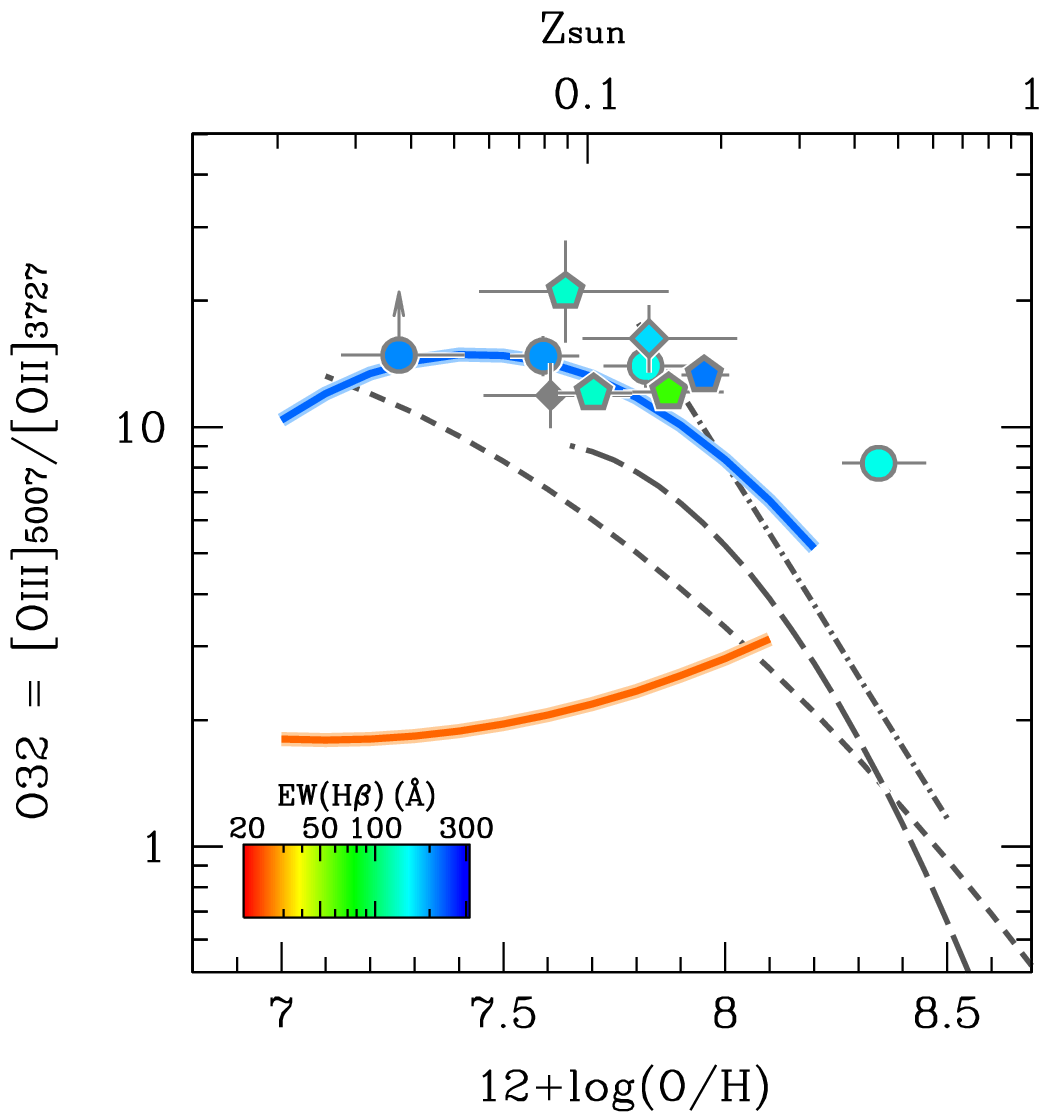}
        }
        \subfloat{
            \includegraphics[bb=23 163 524 698, height=0.245\textheight]{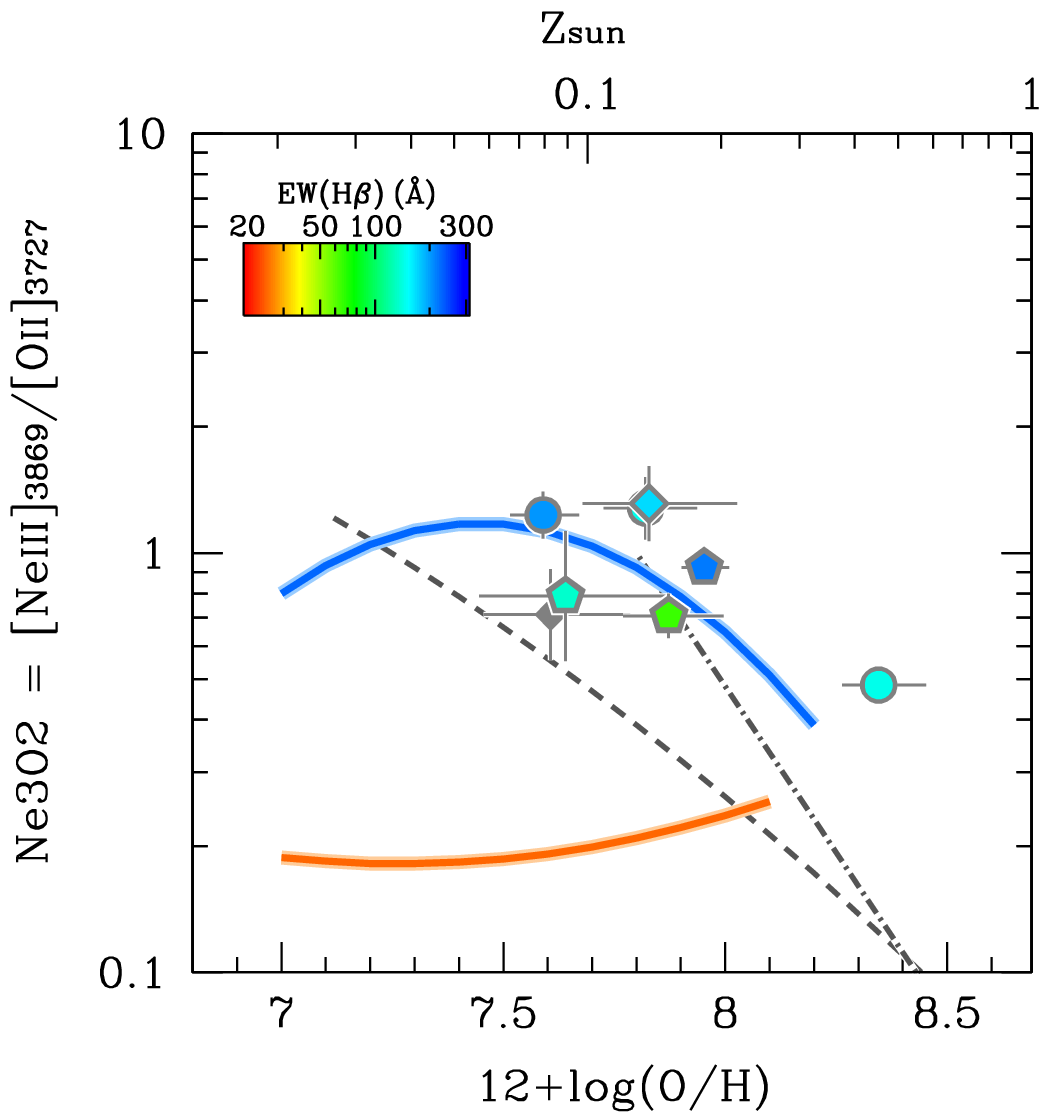}
        }
        \subfloat{
            \includegraphics[bb=23 163 524 698, height=0.245\textheight]{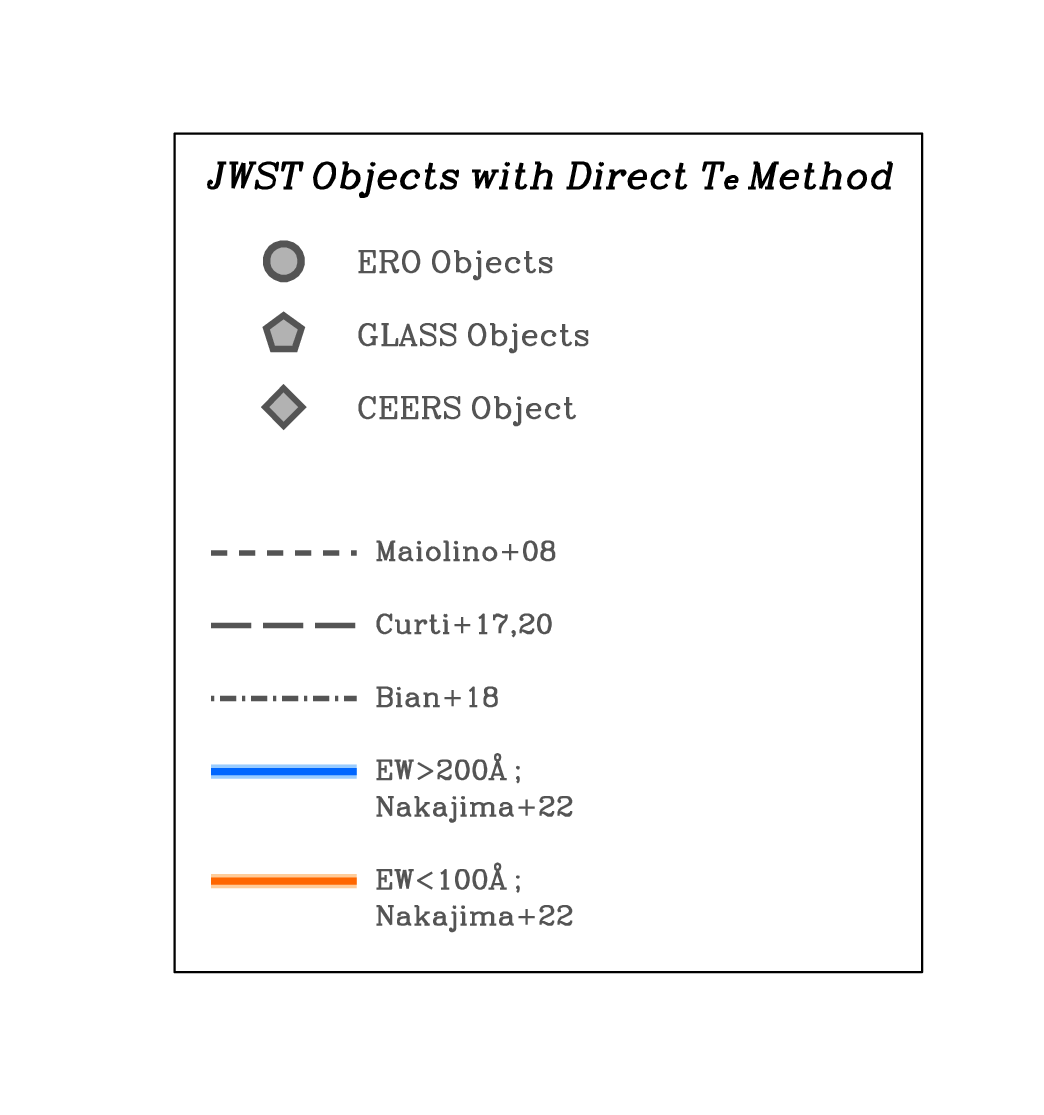}
        }
        \caption{%
            Comparison of metallicity relationships with strong line ratios (R23, R3, R2, O32, and Ne3O2 from top left to bottom right) for ERO (circles), GLASS (pentagons), and CEERS (diamonds) galaxies with metallicity measurements based on the direct $T_e$ method. Each data point is color-coded by its EW(\Hb) value, if available, as shown in the legend. The black dashed, long-dashed, and dot-dashed curves represent the metallicity relationships from \citet{maiolino2008}, \citet{curti2017,curti2020}, and \citet{bian2018}, respectively. The blue and orange curves show the functions derived in \citet{nakajima2022_empressV} for low-metallicity galaxies, with blue representing high-ionization galaxies (EW(\Hb) $>200$\,\AA) and orange representing low-ionization galaxies (EW(\Hb) $<100$\,\AA). Each curve is shown only within the metallicity range explored in the original paper.
        }
        \label{fig:Z_empirical_highz}
    \end{center} 
\end{figure*}

\begin{figure}
    \begin{center}
    \includegraphics[bb=23 163 524 698, height=0.245\textheight]{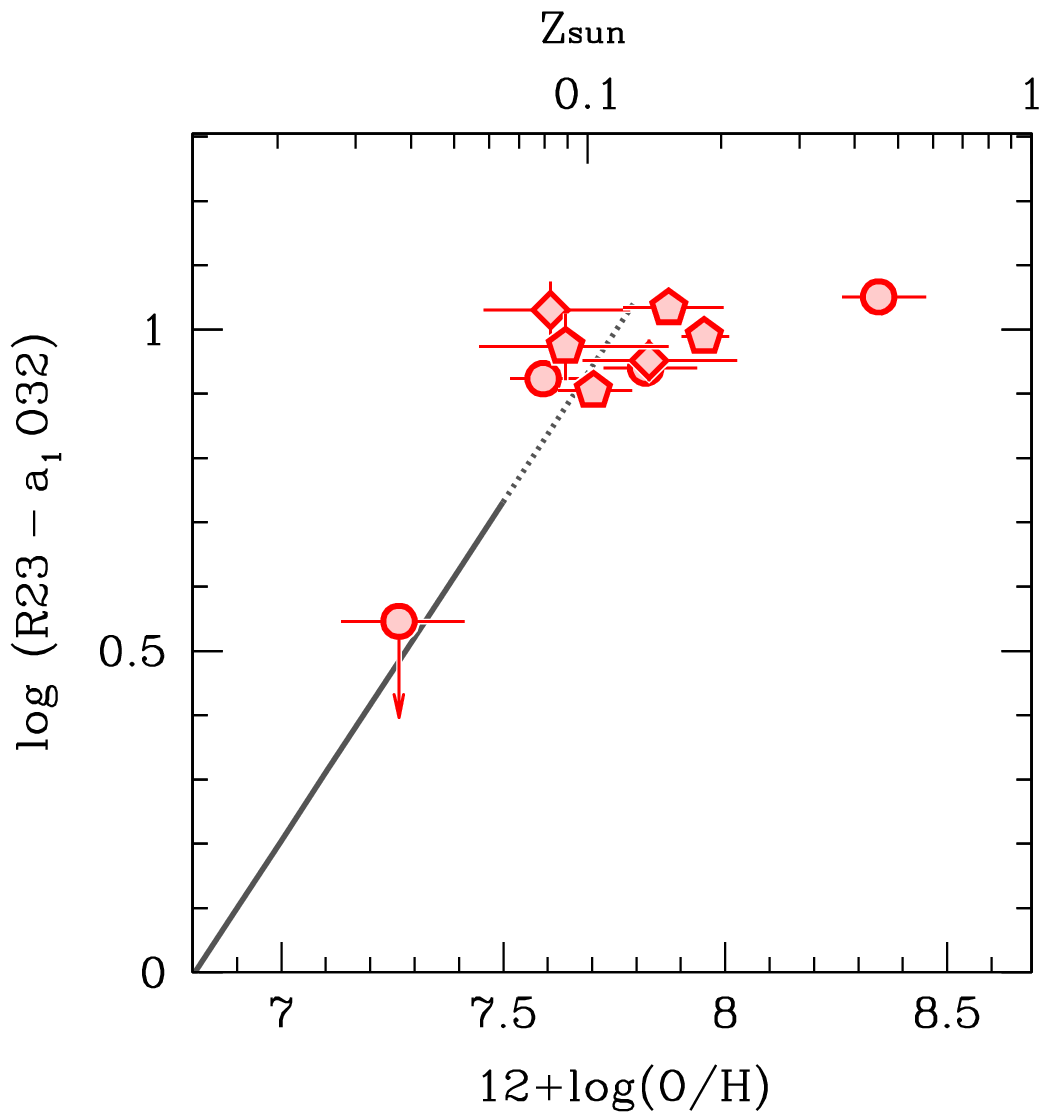}
    \caption{%
        Empirical metallicity indicator using both R23 and O32 as proposed by 
        \citet{izotov2019_lowZcandidates,izotov2021}
        in the low-metallicity regime (solid) and its extrapolation (dotted).
        The red symbols show the JWST objects as in Figure \ref{fig:Z_empirical_highz}.
    }
    \label{fig:Z_empirical_highz_ver2}
    \end{center} 
\end{figure}

Ten objects (4 from ERO, 4 from GLASS, and 2 from CEERS) present a significant detection of \OIII$\lambda 4363$ allowing us to reliably determine the oxygen abundance with the direct $T_e$ method at $z=4.0-8.5$ and use it as a proxy for gas-phase metallicity. 
We follow the procedure as summarized in \citet{nakajima2022_empressV} to derive $T_e$-based metallicities.
Briefly, we first estimate the electron temperature of O$^{2+}$ zone ($T_e$(\OIII)) using the reddening-corrected \OIII$\lambda\lambda 4363/5007$ ratio and an assumed electron number density of $100$\,cm$^{-3}$ with the task \verb+getTemDen+ from PyNeb \citep{luridiana2015}. 
Different densities such as $10-1000$\,cm$^{-3}$ do not significantly change the results (see \citealt{isobe2023_jwst} for the typical electron density in high-redshift galaxies; see also \citealt{fujimoto2022_jwst_alma}). The temperature of O$^{+}$ zone, $T_e$(\OII), is extrapolated from $T_e$(\OIII) employing the prescription of \citet{izotov2006} (cf., \citealt{brinchmann2022_ero} for a different assumption).
We derive the abundance of O$^{2+}$$/$H$^{+}$ with the fluxes of \OIII$\lambda\lambda 4959,5007$ to \Hb\ and $T_e$(\OIII), and 
O$^{+}$$/$H$^{+}$ with the \OII\ to \Hb\ reddening-corrected ratio and $T_e$(\OII)
using the PyNeb package \verb+getIonAbundance+.
We do not take into account a higher ionization abundance of O$^{3+}$$/$H$^{+}$, as we follow the approximation given by \citet{izotov2006} and confirm no clear presence of the \HeII$\lambda 4686$ emission line in the spectra.
In the spectrum of ERO\_04590, the \OII\ doublet is not detected at the $3\sigma$ level. We therefore count the O$^{2+}$ component alone for its oxygen abundance. We have checked that the $3\sigma$ upper-limit of \OII\ would increase the O/H value by $0.05$\,dex, which is small enough with respect to the measurement error ($\simeq 0.15$\,dex).
The measured oxygen abundances as well as $T_e$(\OIII) are summarized in Table \ref{tbl:objects_directTe}.

For the ERO objects, there exist several earlier studies that report the metallicities with the electron temperatures \citep{schaerer2022_ero,curti2023_ero,trump2022_ero,rhoads2023_ero,arellano-cordova2022_ero,brinchmann2022_ero}.
It is practically useful to compare our measurements with them. 
Figure \ref{fig:compare_Te} compares the $T_e$(\OIII) values based on our measurements with the earlier studies for the four \OIII$\lambda 4363$-detected ERO objects. 
Different colors correspond to different objects. 
In some previous studies, the $z=8.5$ object (ERO\_04590 ; red-marks in the plot) is claimed to present an exceptionally high electron temperature at $T_e>2.5\times 10^4$\,K, which is not usually observed in the local universe and may require some additional explanations (e.g., \citealt{katz2023_jwst,rhoads2023_ero}). 
Our re-measurement confirms a high, but not extremely high, electron temperature for ERO\_04590, $T_e=(2.08\pm 0.26)\times 10^4$\,K, which is explainable by heating of young massive stars without any special mechanisms.
It is thus important to carefully reduce and combine the spectra, as well as to extract the 1D to reliably discuss the nebular properties with the measurements of faint emission lines.
For the other objects, their electron temperatures are modest, $T_e=(1.1-2.3)\times 10^4$\,K, fairly consistent with the values in the earlier literature and similar to the electron temperatures seen in lower-redshift star-forming galaxies.
We will also discuss comparisons of metallicity determinations later on the mass-metallicity relationship.

The four ERO objects as well as the 6 newly identified GLASS and CEERS objects with \OIII$\lambda 4363$ provide the opportunity to test the empirical metallicity indicators at such high-redshift of $z>4$ \citep{curti2023_ero}.
In Figure \ref{fig:Z_empirical_highz}, we examine the following five popular metallicity indicators:
\OIII$\lambda5007$$/$\Hb\ (R3),
\OII$\lambda3727$$/$\Hb\ (R2), 
\NeIII$\lambda3869$$/$\OII$\lambda3727$ (Ne3O2), as well as 
R23 and O32
by comparing with the empirical indicators proposed by 
\citet{maiolino2008} (see also \citealt{nagao2006_metallicity}), \citet{curti2017,curti2020}, \citet{bian2018}, and \citet{nakajima2022_empressV}.
We draw each empirical relationship only in the calibrated metallicity range without showing any extrapolation.
For the relationships of \citet{nakajima2022_empressV},
two curves are depicted in each panel, showing the dependence of 
EW(\Hb) on the indicators.
The authors use EW(\Hb) as a proxy to correct for the degree of ionization state
of the gas in each galaxy, because EW(\Hb) is sensitive to the current massive-star formation efficiency 
and known to be well-correlated with the ionization state as probed by 
e.g., O32 (e.g., \citealt{NO2014,mingozzi2020,nakajima2022_empressV}).
To examine the dependencies of EW(\Hb) on the indicators, the JWST objects are color-coded by EW(\Hb) except for CEERS\_01536, whose EW(\Hb) is not derived due to the lack of NIRCam images (Section \ref{ssec:data_ceers}).

The JWST objects tend to present high ionization emission lines such as \OIII\ and \NeIII\ stronger (and low ionization lines such as \OII\ weaker) than expected from the local empirical relationships as defined by \citet{maiolino2008} and \citet{curti2017,curti2020}. Rather, they present a fairly good agreement with the relationships proposed by \citet{bian2018} and \citet{nakajima2022_empressV} for the large EW(\Hb) objects.
Because the relationships of \citet{bian2018} are based on the highly-ionized objects as typically found at $z\sim 2-3$ on the \NII\ BPT diagram 
(e.g., \citealt{BPT1981, steidel2014, shapley2015}),
those consistencies suggest the JWST objects at $z=4-8.5$ are highly ionized systems.
This trend is consistent with the analysis by \citet{sanders2023_jwst}. 
One caveat is that there is one JWST objects whose EW(\Hb) is smaller than $100$\,\AA, GLASS\_10021, falling close to the large EW(\Hb) relationships of \citet{nakajima2022_empressV} rather than the small EW ones like the other JWST objects despite the relatively small EW(\Hb). This suggests we need to test whether the prescriptions work for low ionization sources as well with a larger sample at high-redshift.

In summary, 
although we cannot fully test the indicators for low ionization sources, the current results suggest that the degree of ionization significantly influences the resulting metallicity value if one relies on an empirical metallicity indicator, and that no strong redshift evolution is seen in the strong line ratios as a function of metallicity at least for highly ionized galaxies. 
Ionization-corrections, such as those proposed in \citet{nakajima2022_empressV}, are thus crucial for metallicity estimations with the strong line methods.
In particular, we find R23, R3, and R2 show a good agreement with the observations and the empirical relationships, and confirm that they are sensitive to a small change in metallicity. 
The indicators of O32 and Ne3O2 look highly dependent on the ionization state and their plateau-like behaviors against metallicity prevent us from deriving a stable metallicity solution at the metal-poor regime.

Figure \ref{fig:Z_empirical_highz_ver2} shows another metallicity indicator proposed by \citet{izotov2019_lowZcandidates,izotov2021} which combines R23 and O32 to correct for the ionization state 
to improve the accuracy of metallicity in the low-metallicity regime. Unfortunately there is only one object, ERO\_04590, whose metallicity is low enough to be fairly compared with the method, and whose non-detection of \OII\ prevents us from confirming the reliability and accuracy of the method.
This is exactly the situation \citet{nakajima2022_empressV} have anticipated, i.e., not all of the lines are spectroscopically available at high redshift, and it is practically useful to use EW(\Hb) as a probe of ionization state. 
Still, many of the JWST objects are located along the simple extrapolation of the relationship, indicating the method can be useful even up to \Oabundance\ $\sim 7.8$.

One caveat should be noted for ERO\_06355, which has an oxygen abundance of \Oabundance\ $=8.3$ and shows an unusually strong \OIII\ line. \citet{curti2023_ero} have used the \OIII$/$\Hb\ vs. \OII$/$\Hb\ diagram and found that this object is slightly above the limit that can be explained by star-forming galaxies alone. This indicates the possibility of a hidden high-energy ionizing source. In addition, \citet{brinchmann2022_ero} tentatively suggest the presence of the high ionization line \NeIV$\lambda\lambda 2422,2424$, indicating a hard ionizing spectrum up to $\sim$60 eV, which cannot be fully explained by a conventional stellar population. These pieces of evidence suggest that the anomalous nature of this object may be explained by the presence of a high-energy ionizing source in the system, which may account for its deviation from the empirical metallicity relationships.
Another note is for GLASS\_160133 and GLASS\_150029 that have a broad component associated with their \Ha\ emission line, indicating the presence of faint AGNs in the systems. However, they still present metallicity diagnostic line ratios and EW(\Hb) that agree reasonably well with the relationships for star-forming galaxies, implying a minor contribution of AGN to the total emission strengths as well as to the optical continuum for these two objects.
In any case, one would need higher ionization lines (e.g., \NeV; \citealt{cleri2023_jwst}) to further discuss the presence of hard radiation components in these systems.

\subsubsection{Empirical method} \label{sssec:results_metallicity_empirical}

For the other JWST objects without a $T_e$-based metallicity ($N^o=182-10=172$), we adopt the empirical metallicity indicators which have been tested and confirmed to work at high-redshift (Section \ref{sssec:results_metallicity_directTe}) to estimate metallicities.
Prior to estimating metallicities empirically, we require that the objects have coverage of the \OIII$+$\Hb\ emission lines in their spectra, and that the \OIII$\lambda\lambda 5007/4959$ doublet line ratio is consistent with the theoretical value ($2.98$; \citealt{SZ2000}) at a $2.5\sigma$ significance level. We only work with objects that meet these criteria and have appropriate flux measurements. As a result, 11 objects were removed from the subsequent analysis.

For the remaining 161 objects, we use the R23-index as a primary indicator, following the overall consistency of the metallicity indicator at high-redshift as shown in Section \ref{sssec:results_metallicity_directTe}. One caveat in using R23 is that two solutions are derived for a given R23 value. We use the prescription of \citet{nakajima2022_empressV} with EW(\Hb) for the low-metallicity solutions (\Oabundance\ $\lesssim 8.0$), and the average relationship (without any ionization-correction) for the high-metallicity ones ($\gtrsim 8.0$). The latter is almost consistent with the indicators presented in \citet{curti2017,curti2020}. 
If the two metallicity solutions are not significantly separated, i.e., the observed R23 spans the peak value that appears around \Oabundance\ $=8.0$ within the uncertainty, we adopt the $1\sigma$ lower-limit of the low-metallicity solution as the lower-limit, and the upper-limit of the high-metallicity solution as the upper-limit. 
If there are two distinct metallicity solutions, we rely on the O32 line ratios to distinguish between them. It should be noted that O32 is used only for the purpose of distinguishing between the two branches, and not for calculating the actual metallicity value. Specifically, we choose the R23-based metallicity solution whose expected O32 line ratio according to the O32-metallicity indicator of \citet{nakajima2022_empressV} is more close to the observed O32 than the other. If the \OII\ is not detected and the resulting lower-limit on O32 is higher than the expected O32 at the high-metallicity branch at the $>3\sigma$ level, we choose the low-metallicity solution. 
If the lower-limit on O32 is not high enough to distinguish, we translate EW(\Hb) to O32 using the average relationship found in \citet{nakajima2022_empressV}: $\log {\rm EW}({\rm H}\beta) = 0.64\times \log {\rm O32}  +1.68$, and choose the solution.
Finally, the systematic uncertainty of the empirical method is added in quadrature to the metallicity errors.
For the other cases, we cannot reliably distinguish between the two solutions and thus leave its metallicity unconstrained based on R23.

We are unable to derive a R23-based metallicity for objects that lack \OII\ coverage and/or appropriate dust correction due to the absence of multiple Balmer emission lines. For such objects, we estimate metallicity using the R3 indicator, as it does not require a reddening correction. We follow a similar procedure as for R23 to account for the two-branch nature of the R3-index and estimate metallicities. As O32 is not applicable for these objects, we rely on EW(\Hb) to correct for the ionization state and attempt to differentiate between the two solutions.
\\

In summary, our sample consists of 86 objects with an R23-based metallicity, 49 objects with an R3-based metallicity, and 26 objects without a metallicity constraint. 
We note that for the objects in the CEERS sample that have multiple spectra taken with different gratings and/or pointings, we estimate the metallicity for each spectrum and calculate the average value and its uncertainty by combining the individual metallicity measurements.
While we understand that some of the spectra (P11 and P12) were obtained with different position angles compared to the others, we assume that the objects are compact and the different observations targeting the same objects at the center of the slit probe the same gas properties.
After excluding 10 objects with spectroscopic signatures of AGNs (Section \ref{ssec:results_line_ratios}), comprising of 8 objects with R23-based metallicity and 2 objects with $T_e$-based metallicity, 
we use $127$ objects with an R23- or R3-based metallicity, along with $8$ objects with direct $T_e$ method, for the subsequent analysis and discussion of mass-metallicity relationships.

\subsubsection{Other high-redshift galaxies from the literature} \label{sssec:results_metallicity_literature}
In addition to our JWST sample of 135 objects constructed with the ERO, GLASS, and CEERS programs, we have compiled reports of metallicity at high-redshift from the literature to discuss with a larger sample. 
Our compilation includes three objects at $z=8.1-9.5$ identified with the two DDT programs using the prism grating of NIRSpec (Proposal IDs: DD-2756 and DD-2767; \citep{williams2022_jwst, heintz2022_jwst, langeroodi2022_ddt, wang2022_jwst}), one stacked point of 117 EIGER objects at $z=5.3-6.9$ whose spectra are taken with the NIRCam slitless spectroscopy (Proposal ID: 1243; \citealt{matthee2022_eiger, kashino2022_eiger}), four objects at $z=6.1-6.4$ with commissioning data of NIRCam slitless spectroscopy \citep{sun2022_jwst1, sun2022_jwst2}, and four ALMA objects at $z=7.2-9.1$ whose metallicity is measured with \OIII$88\,\mu$m \citep{jones2020}. Their redshift distributions are illustrated in blue in Figure \ref{fig:hist_redshift}. Note that the EIGER stacked object is treated as a single object in the histogram as well as in the following figures.

For a fair comparison of metallicity in the following analysis, it is important to have metallicities that are estimated in a consistent manner, as done for our sample. All of the measurements compiled above are based on empirical indicators using strong emission lines that are calibrated with the direct $T_e$ method.
For the NIRSpec DDT objects, five objects at $z=7.9-9.5$ have reported metallicity measurements in several studies \citep{williams2022_jwst, heintz2022_jwst, langeroodi2022_ddt, wang2022_jwst}. Despite initially adopting the O32-index as the metallicity indicator in the original manuscript, \citet{heintz2022_jwst} have updated their results, now primarily using the R3-index of \citet{nakajima2022_empressV} with correction for the ionization state for large EW(\Hb) ($\geq 200$\,\AA) objects. This approach is supported by Figure \ref{fig:Z_empirical_highz} in this paper and has been adopted for some of our JWST objects. It should be noted, however, that it would be more appropriate to adopt the EW(\Hb)-dependence of the indicator following the variation of EW(\Hb) as observed in the DDT sample, instead of fixing the relation for large EW(\Hb) \citep{heintz2022_jwst}. Large EW(\Hb) of $\sim 180-250$\AA\ are confirmed for two of the five objects \citep{williams2022_jwst, wang2022_jwst}.
Among the five objects reported in \citet{heintz2022_jwst}, we use three objects, RXJ-z9500, RXJ-z8152, and RXJ-z8149, in this paper. Their metallicities are estimated to be \Oabundance\ $=$ 7.56 (+0.16/-0.17), 7.68 (+0.18/-0.19), and 7.29 (+0.22/-0.28), respectively.
One of the objects not included in our compilation, Abell-z7878, was originally suggested to have an \OIII$\lambda\lambda 5007/4959$ doublet ratio that is significantly smaller than the theoretical value. We remove this object from our compilation following the same approach used for our JWST sample.
The remaining object, Abell-z7885, still relies on the O32-index for estimating its metallicity. We have decided not to include it in our compilation due to several uncertainties associated with using the O32-index for metallicity estimations, such as the strong dependence of O32 on ionization state (Figure \ref{fig:Z_empirical_highz}) and the accuracy of dust reddening correction, as we were careful regarding the results presented in the original manuscript of \citet{heintz2022_jwst}.

\begin{figure}[t]
    \begin{center}
    \includegraphics[bb=18 169 580 564, width=0.95\columnwidth]{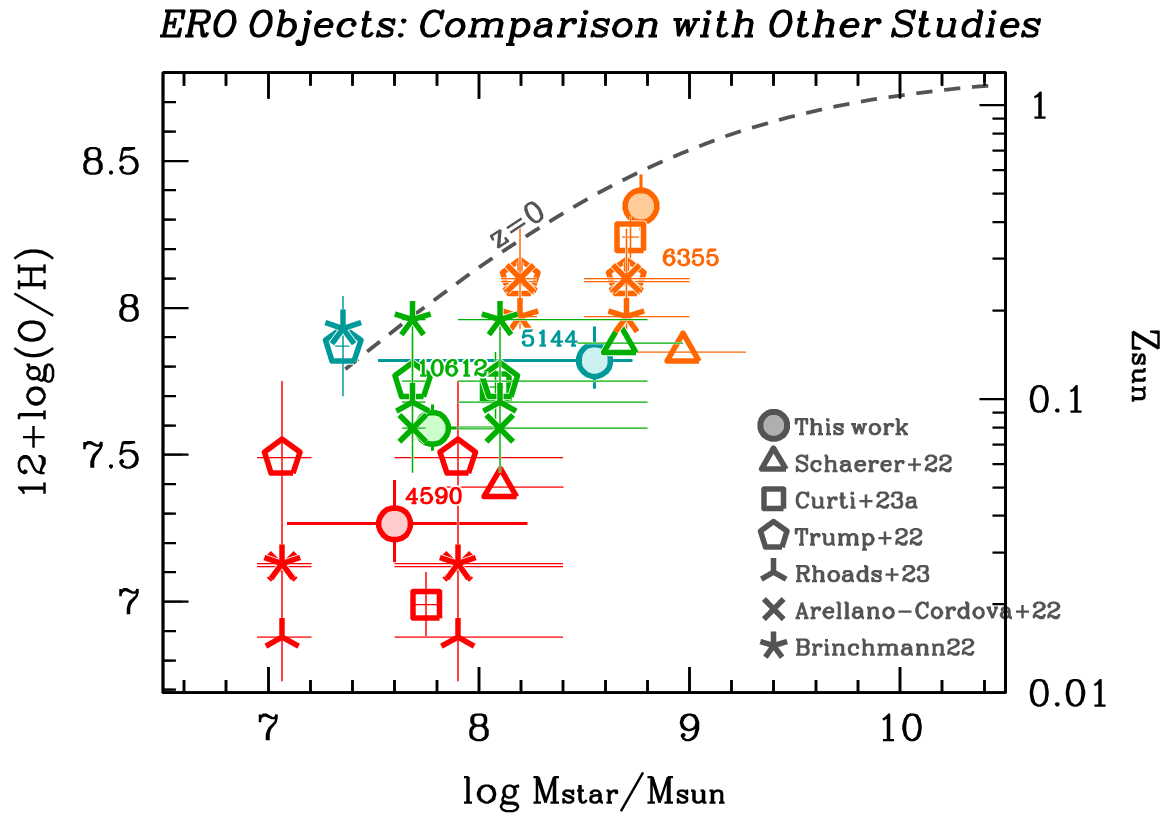}
    \caption{%
        Comparison between the mass-metallicity relation obtained in this study for ERO objects at $z=6.3-8.5$ using $T_e$-based metallicities (colored circles) and those reported in the literature. Different symbol colors indicate different objects, as shown in Figure \ref{fig:compare_Te}, and different symbols represent different publications, as indicated in the legend. For the metallicities from \citet{arellano-cordova2022_ero}, we show the average values based on the two spectra (i.e., two observing blocks: o007 and o008) for 06355 and 10612, but refer only to the o008 value for 04590. For the relationships by \citet{schaerer2022_ero} and \citet{curti2023_ero}, the stellar masses are adopted as originally estimated in each paper. For the others where stellar masses are referenced from other studies, we show the two stellar masses from \citet{carnall2023_ero} and \citet{tacchella2022_ero} for each object (only \citet{carnall2023_ero} for 05144).
        The comparison confirms a good agreement between our measurements and earlier results on the MZ relation.
    }
    \label{fig:MZ_comparison}
    \end{center} 
\end{figure}

\begin{figure*}[t]
    \begin{center}
        \subfloat{
            \includegraphics[bb=21 169 571 537, height=0.38\textheight]{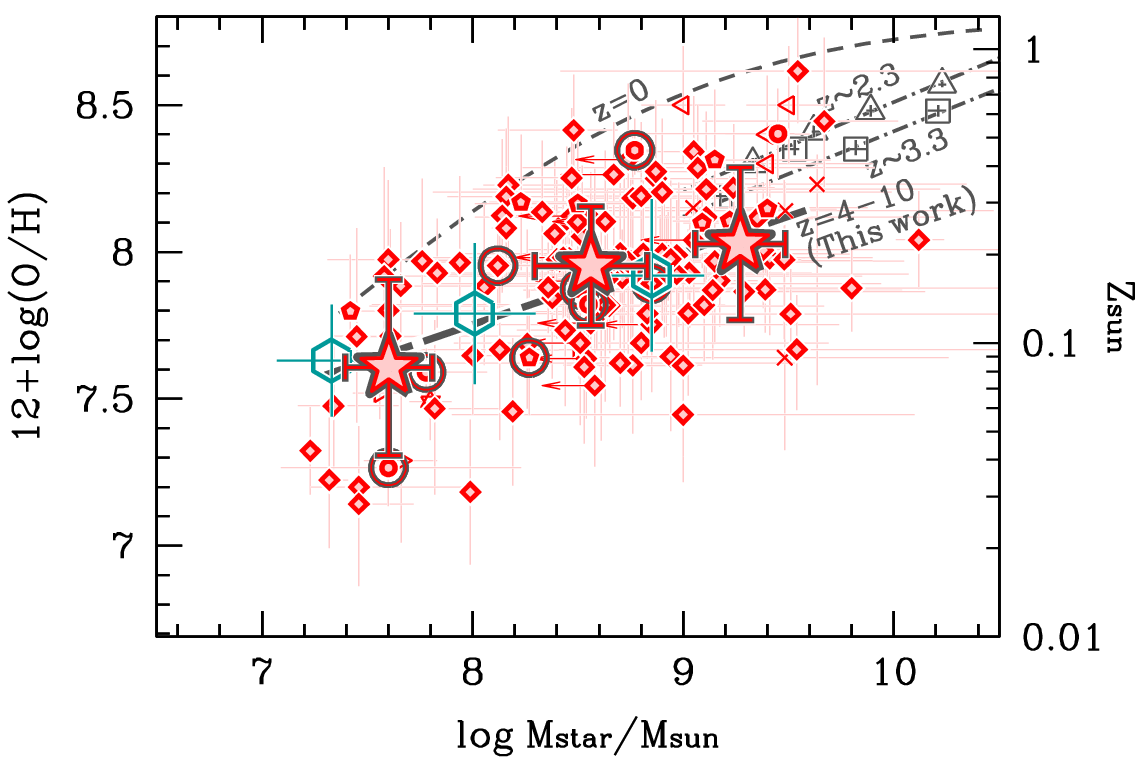}
        }
        \subfloat{
            \includegraphics[bb=89 136 366 504, height=0.255\textheight]{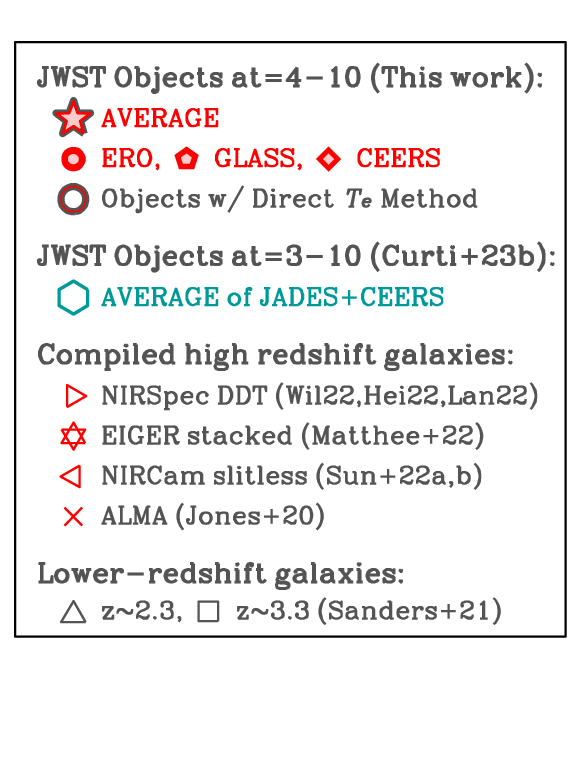}
        }
        \caption{%
            The relationship between stellar mass and metallicity. The red points represent galaxies at $z=4$--$10$, including the filled circles, pentagons, and diamonds, which represent the ERO, GLASS, and CEERS objects analyzed in this paper. Red circles denote galaxies whose metallicity is determined using the direct $T_e$ method. The large red stars represent the average relationships for the ERO, GLASS, and CEERS objects 
            in three equally separated mass ranges
            (\Mstar=$10^7-10^8$, $10^8-10^9$, and $10^9-10^{10}$\,\Msun), 
            along with the best-fit function shown as a thick long-dashed line (Equation \ref{eq:MZ}).
            Emerald green open pentagons show the average relations at $z=3-10$ based on the JADES+CEERS sample \citep{curti2023_jades}.
            Other red symbols represent high-redshift objects compiled from the literature, including open right-pointing triangles for NIRSpec DDT objects \citep{williams2022_jwst,heintz2022_jwst,langeroodi2022_ddt}, left-pointing triangles for four NIRCam objects \citep{sun2022_jwst1,sun2022_jwst2}, open hexagram for the stacked EIGER object \citep{matthee2022_eiger}, and crosses for ALMA objects \citep{jones2020}. 
            Additionally, relationships at lower redshifts are displayed, including SDSS stacked galaxies in the local universe shown as a gray dashed curve \citep{AM2013}, and those at $z\sim 2.3$ with gray open triangles and $\sim 3.3$ with gray open squares (best-fits shown as gray dot-dashed curve; \citealt{sanders2021}). The curves are displayed in the mass ranges explored in the original papers. These low-redshift metallicities are based on the direct $T_e$ method.
        }
        \label{fig:MZ}
    \end{center} 
\end{figure*}

\begin{deluxetable}{lccc}[h]
\tablecaption{Average values of stellar mass, metallicity and SFR for our JWST sample
\label{tbl:MZ_SFR_average}}
\renewcommand{\arraystretch}{1.25}
\tabletypesize{\scriptsize}
\tablehead{
\colhead{Sample} &
\colhead{$\log M_{\star}$}&
\colhead{\Oabundance} &
\colhead{$\log {\rm SFR}$} \\
 &
(\Msun) &
 &
(\Msunyr)
} 
\startdata
\multicolumn{4}{l}{== For MZ relations (incl.\,objects without SFR measurement)} \\
$z=4-10$ & $7.60 \pm 0.21$ & $7.61 \pm 0.30$ & \nodata \\
 & $8.56 \pm 0.27$ & $7.95 \pm 0.20$ & \nodata \\
 & $9.27 \pm 0.21$ & $8.03 \pm 0.26$ & \nodata \\
$z=4-6$ & $8.30 \pm 0.47$ & $7.87 \pm 0.28$ & \nodata \\
 & $9.15 \pm 0.26$ & $8.04 \pm 0.23$ & \nodata \\ 
$z=6-8$ & $7.85 \pm 0.29$ & $7.77 \pm 0.27$ & \nodata \\ 
 & $8.76 \pm 0.38$ & $7.90 \pm 0.22$ & \nodata \\ 
$z=8-10$ & $8.29 \pm 0.49$ & $7.63 \pm 0.26$ & \nodata \\
 & $9.29 \pm 0.59$ & $7.96 \pm 0.23$ & \nodata \\
\hline
\multicolumn{4}{l}{== For SFR-MZ relations} \\
$z=4-10$ & $7.56 \pm 0.21$ & $7.50 \pm 0.28$ & $0.55 \pm 0.38$ \\
 & $8.55 \pm 0.27$ & $7.91 \pm 0.21$ & $1.02 \pm 0.31$ \\
 & $9.27 \pm 0.22$ & $8.05 \pm 0.26$ & $1.45 \pm 0.44$ \\
$z=4-6$ & $8.32 \pm 0.52$ & $7.81 \pm 0.31$ & $0.84 \pm 0.40$ \\
 & $9.21 \pm 0.25$ & $8.06 \pm 0.25$ & $1.45 \pm 0.42$ \\
$z=6-8$ & $7.94 \pm 0.29$ & $7.72 \pm 0.26$ & $0.77 \pm 0.29$ \\
 & $8.79 \pm 0.34$ & $7.90 \pm 0.21$ & $1.19 \pm 0.32$ \\
$z=8-10$ & $8.13 \pm 0.53$ & $7.54 \pm 0.28$ & $1.03 \pm 0.37$ \\
 & $9.53 \pm 0.59$ & $7.84 \pm 0.20$ & $1.37 \pm 0.60$ \\
\hline
\enddata
\tablecomments{%
The first half of this table displays the average masses and metallicities of the samples used for the MZ relations, while the second half shows the samples used for the SFR-MZ relations. The second half samples exclude objects whose SFRs are not measured or constrained (Section \ref{ssec:results_FMR}), resulting in a smaller sample size compared to the first half.
}
\end{deluxetable}

The metallicity values for RXJ-z9500, RXJ-z8152, and RXJ-z8149 in \citet{heintz2022_jwst} are consistent with those reported in other studies. The metallicity of RXJ-z9500 is first reported by \citet{williams2022_jwst}, where it was estimated to be \Oabundance\ $=7.48\pm 0.08$ using the R23+O32 method of \citet{izotov2019_lowZcandidates, izotov2021} (Figure \ref{fig:Z_empirical_highz_ver2}). An ionization correction is thus taken into account for the metallicity measurement there. Similarly, \citet{langeroodi2022_ddt} estimate metallicities for RXJ-z8152 and RXJ-z8149 (RX2129-ID11002 and RX2129-ID11022 in \citeauthor{langeroodi2022_ddt}) using the R23+O32 method \citep{izotov2019_lowZcandidates, izotov2021}, and report \Oabundance\ $=7.65\pm 0.07$ and $<7.51$ ($1\sigma$), respectively. \citet{wang2022_jwst} independently suggest a consistent metallicity for RXJ-z8152, although the ionization correction is not fully taken into account. 
We use the properties, including metallicities, that are calculated by averaging the values from \citet{williams2022_jwst}, \citet{heintz2022_jwst}, and \citet{langeroodi2022_ddt} for the three DDT objects in our compilation.

For the other compilation, we adopt the metallicity values as derived in the original papers. The stacked EIGER object has already derived a metallicity fairly consistent with the value based on our method using R23. The four NIRCam objects are all located in the high-metallicity branch based on the \NII$\lambda 6564$ detection, and the indicator of \citet{bian2018} is used. 
We also note that the stellar masses and SFRs are all corrected for different IMFs to have the same \citet{chabrier2003} IMF using the conversion factors shown in \citet{MD2014}.
\\

In summary, our total sample consists of 147 galaxies (135 from our reduction of ERO, GLASS, and CEERS, and 12 from the compilation) with metallicity measurements at $z=4-10$. This sample is used in the following analysis of the metallicity relationships.

\subsection{Mass-Metallicity Relation} \label{ssec:results_MZ}

The main focus of this paper is to discuss the evolution of metallicity in star-forming galaxies. In this section, we specifically present the stellar mass-metallicity relation at high redshift using the JWST observations, as well as data compiled from relevant literature.

Before presenting the full results we firstly focus on the four ERO objects and compare in Figure \ref{fig:MZ_comparison} our mass-metallicity relation with those presented in the earlier studies \citep{schaerer2022_ero,curti2023_ero,trump2022_ero, rhoads2023_ero, arellano-cordova2022_ero,brinchmann2022_ero}.
The metallicities are all derived based on the direct $T_e$ method as summarized in Figure \ref{fig:compare_Te}.
The stellar masses are corrected for different IMFs to have the same \cite{chabrier2003} IMF. \citet{schaerer2022_ero} and \citet{curti2023_ero} derive the masses in their own papers, while the others refer to either \citet{carnall2023_ero} or \citet{tacchella2022_ero} and hence both values are shown in the plot.
We confirm our mass-metallicity relations for the four ERO objects overall show a good agreement with the earlier results. 
One note is that we confirm the metallicity of ERO\_04590 is not extremely low, \Oabundance\ $=7.26$ (+0.15/-0.13) for its stellar mass.

Figure \ref{fig:MZ} illustrates the relation between stellar mass and metallicity (MZ) with the full sample at $z=4-10$. The MZ relations determined at $z=0$, $\simeq 2.3$, and $\simeq 3.3$ with the direct $T_e$ method are also plotted (\citealt{AM2013, sanders2021}; see also \citealt{nishigaki2023} that the $T_e$-based MZ relation at $z=0$ continues decreasing down to \Mstar\ $\sim 10^5$\,\Msun). 
The red symbols denote the galaxies at $z=4-10$, including our 135 ERO, GLASS, and CEERS galaxies, 3 NIRSpec DDT objects, as well as the massive galaxies provided by NIRCam slitless and ALMA spectroscopy, and the low-mass stacked object as presented in Section \ref{sssec:results_metallicity_literature}. 
We note again that the $10$ objects with spectroscopic signatures of AGNs \citep{harikane2023_jwst_blagn} have been excluded here to conservatively discuss the results free from any AGN biases (Section \ref{ssec:results_line_ratios}).
The objects analyzed using the direct $T_e$ method are marked with a red open circle, suggesting a positive correlation between mass and metallicity in place in the redshift range $z=5-8.5$. Furthermore, we divide our full sample of ERO, GLASS, and CEERS galaxies into three according to the stellar mass: \Mstar\ $=10^{7}-10^{8}$, $10^{8}-10^{9}$, and $10^{9}-10^{10}$\,\Msun, and obtain the average MZ relations as shown with the large open stars in Figure \ref{fig:MZ}
and as given in Table \ref{tbl:MZ_SFR_average}. 
Note that the compiled objects, as well as the CEERS objects with only an upper-limit on \Mstar, are not used in deriving the average relations.
Following the single power law form of \citet{sanders2021}, the average MZ relation of the $z=4-10$ galaxies can be approximated as:
\begin{equation}
12+\log ({\rm O/H}) = {\rm Z}_{10} + \gamma \log (M_{\star}/10^{10}\,M_{\odot})
\label{eq:MZ}
\end{equation} 
with the best-fit parameters of ${\rm Z}_{10}=8.24\pm 0.05$ and $\gamma=0.25\pm 0.03$
in the mass range of \Mstar\ $\sim 10^{7.5}-10^{9.5}$\,\Msun, as shown with a gray long-dashed line. 
The parameter $\gamma$ corresponds to the slope of the MZ relation, and its best-fit value and uncertainty confirm an increasing trend of metallicity with stellar mass, as tentatively seen with the three ERO objects (e.g., \citealt{schaerer2022_ero,curti2023_ero,trump2022_ero}), and as widely known at low-redshift.

Compared to the $z=0$ MZ relation, these high-redshift galaxies clearly present a metallicity lower than typical galaxies at $z=0$ for a given stellar mass. The decrease is typically $\sim 0.5$\,dex around \Mstar\ $\sim 10^{9}$\,\Msun, but it becomes smaller at the low-mass end ($\sim 0.3$\,dex).
Interestingly, a similar offset of $\sim 0.3$\,dex is observed between $z=0$ and $z\sim 2-3$, suggesting that the evolution of MZ relation is small from $z\sim 2-3$ to $z=4-10$. Although there may be a decrease of $\sim 0.2$\,dex in the typical metallicity at the high-mass end of \Mstar\ $\sim 10^{9.5}$\,\Msun\ from $z\sim 2-3$ to $z=4-10$, no strong evolution is found beyond the error.
The same conclusion can be drawn from comparisons with the MZ relations at $z\lesssim 4$ whose metallicities are empirically estimated with the strong line indicators, 
as presented in Appendix \ref{sec_app:results_MZ_w_empirical}.

\begin{figure}[t]
    \begin{center}
        \subfloat{
            \includegraphics[bb=21 169 571 537, width=0.895\columnwidth]{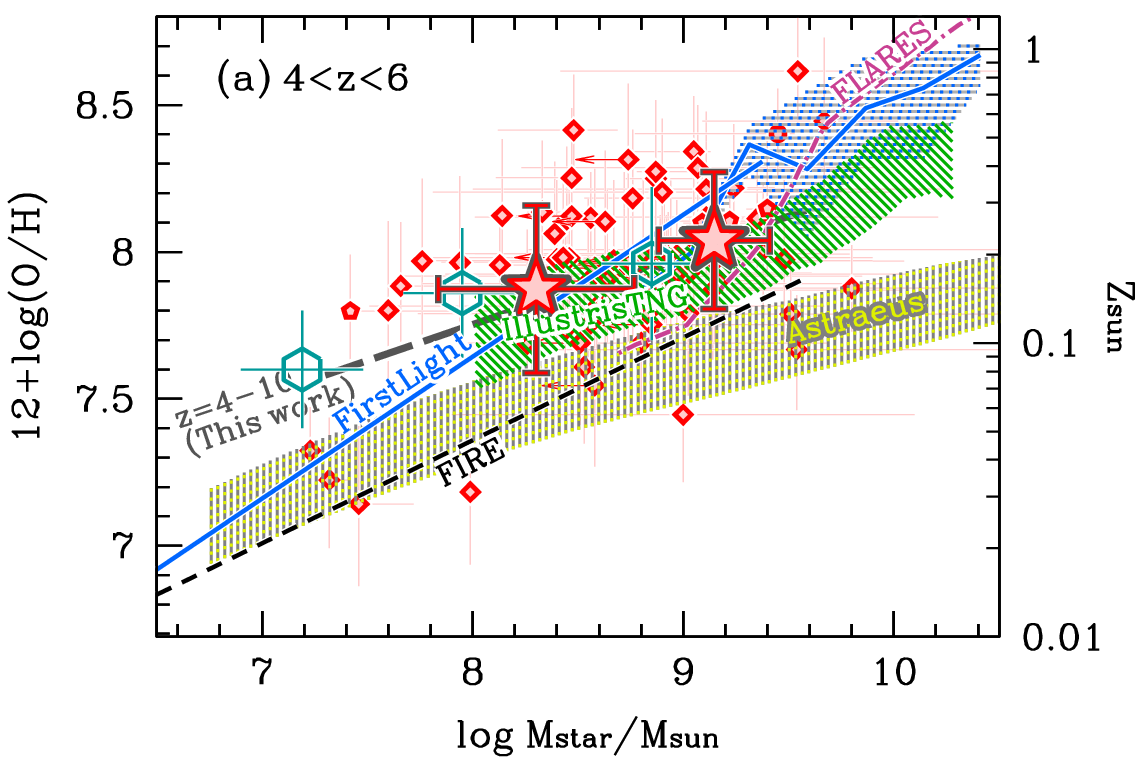}
        }
        
        \vspace{-11pt} 
        \subfloat{
            \includegraphics[bb=21 169 571 537, width=0.895\columnwidth]{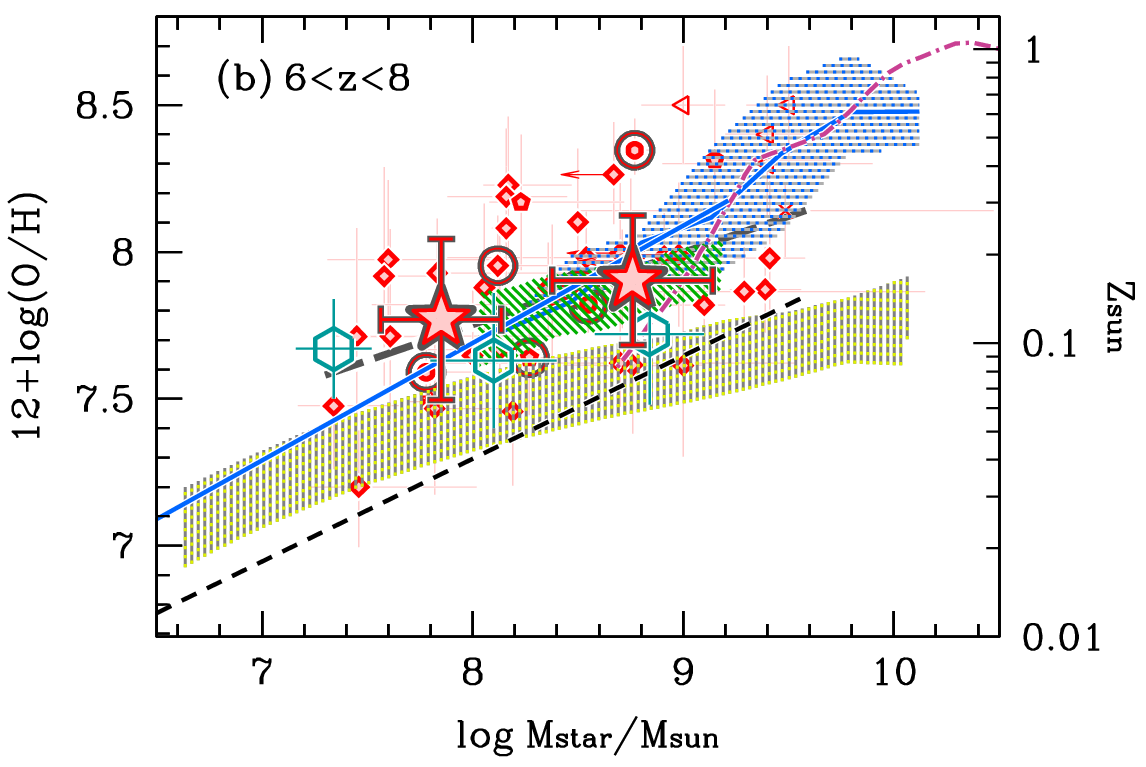}
        }      
        
        \vspace{-11pt} 
        \subfloat{
            \includegraphics[bb=21 169 571 537, width=0.895\columnwidth]{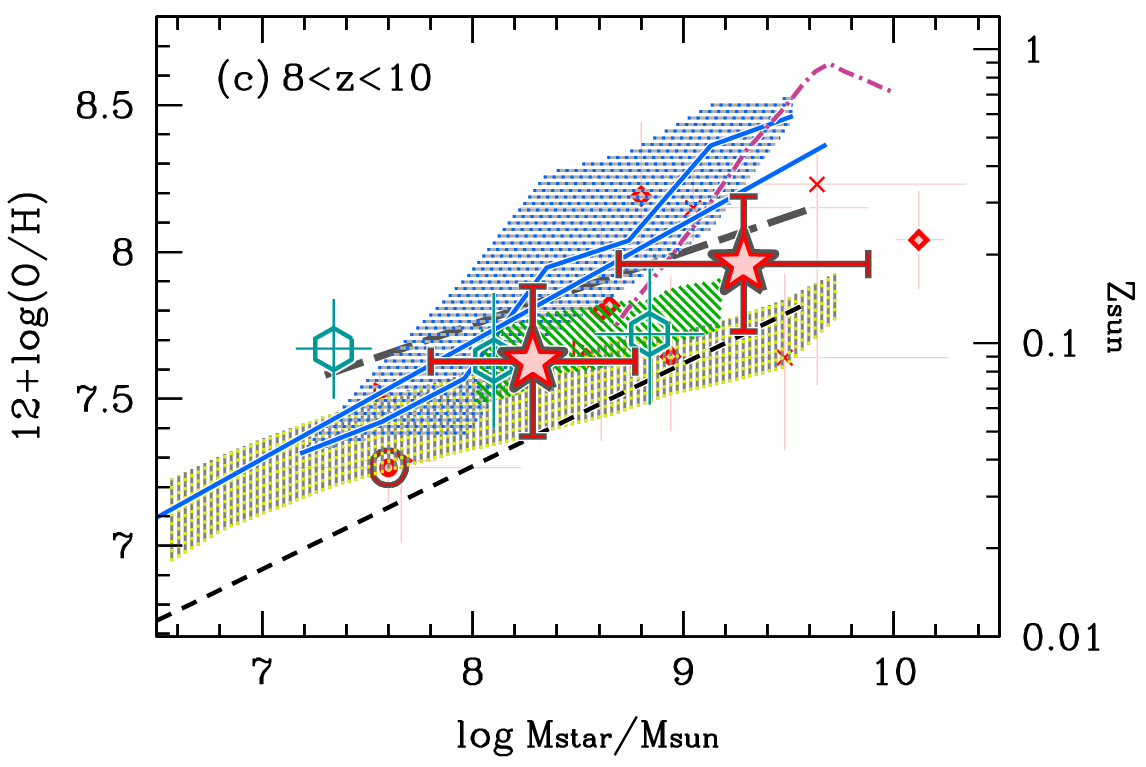}
        }
        
        \vspace{-7pt} 
        \caption{%
            The MZ relation in three different redshift bins: (a) $z=4$--$6$, (b) $z=6$--$8$, and (c) $z=8$--$10$. The red symbols and the best-fit function (Eq.\,\ref{eq:MZ}) are as shown in Figure \ref{fig:MZ}. The large stars represent the average MZ relations re-derived in each redshift bin, by splitting the sample into two groups based on stellar mass to have the equal numbers of galaxies. 
            For the JADES+CEERS relations (\citealt{curti2023_jades} in emerald green), we plot their $z=3-6$ sub-sample's relations in Panel (a), while adopt the $z=6-10$ relations in Panels (b) and (c).
            In addition, the cosmological simulation results at $z=5$, $7$, and $9$ are displayed in Panel (a), (b), and (c), respectively; 
            \textsc{FIRE} in black (\citealt{ma2016_MZR}), 
            \textsc{IllustrisTNG} in green (\citealt{torrey2019}), 
            \textsc{FirstLight} in blue (\citealt{langan2020} in the low-mass regime with solid curves, and \citealt{nakazato2023} in the high-mass regime with shades),
            \textsc{Astraeus} in yellow (\citealt{ucci2021}), and 
            \textsc{FLARES} in magenta (\citealt{wilkins2022_MZR}).
            Some extrapolations are applied, as detailed in the text. 
        }
        \label{fig:MZ_redshift_sim}
    \end{center} 
\end{figure}

Figure \ref{fig:MZ} also includes a comparison with the latest results from the JADES observations. \citet{curti2023_jades} very recently report on the metallicity measurements of galaxies at $z=3-10$ based on deep JADES spectroscopic observations (see also \citealt{cameron2023_jades}). In the figure, their MZ relation is shown using emerald green open pentagons, which represent the combination of the low-mass JADES sample with the high-mass CEERS sample, latter of which is provided in this paper.
We find that their MZ relation, despite covering slightly different redshift ranges, agrees well with ours (Eq.\,\ref{eq:MZ}). It is not surprising to see a good agreement between our results and those reported by \citet{curti2023_jades} in the high-mass end (\Mstar\ $\gtrsim 10^8$\,\Msun), since the regime is dominated by the CEERS objects presented in this paper. The JADES objects confirm that the MZ relation we obtained in Eq.\,\ref{eq:MZ} continues down to \Mstar\ $\simeq 10^{7}$\,\Msun\ on average.

In Appendix \ref{sec_app:results_metallicity_sedfit}, we investigate our MZ relation using only galaxies for which the stellar mass is well-determined through SED fitting to the JWST/NIRCam photometry, after excluding CEERS objects whose masses are estimated empirically from \Muv\ (as discussed in Section \ref{ssec:data_ceers}). 
Because we have already excluded the CEERS objects with only an upper-limit on \Mstar, all of which are \Muv-based, in deriving the average relations, the fraction of objects without NIRCam photometry is small ($\sim 28$\,\%). Indeed, our conclusions remain unchanged when considering only the objects with reliable measures of \Mstar, as demonstrated in the figures in Appendix \ref{sec_app:results_metallicity_sedfit}.
Moreover, we mention in the \NII\ BPT diagram in Section \ref{ssec:results_line_ratios} that some of the JWST objects have \NII\ upper-limits that are too weak to conclude the ionization nature (stars vs. AGNs). By removing these unclear objects and using only galaxies that are surely diagnosed as star-forming galaxies in the \NII\ BPT diagram (i.e., those with an (upper-limit on) \NII/\Ha\ falling below the demarcation curve), we obtain a fully consistent MZ relation as found in the full sample (Figure \ref{fig:MZ}). This implies a negligible contribution of AGNs in our sample.
\\

To further examine any redshift evolution among our sample from $z=4$ to $10$, we plot in Figure \ref{fig:MZ_redshift_sim} the MZ relations for the three different redshift bins, $z=4-6$, $6-8$, and $8-10$. 
In each panel, the sub-sample is further divided into two groups based on their masses, and their average values are shown with large stars
as summarized in Table \ref{tbl:MZ_SFR_average}. 
Although there are large error bars, these average points suggest that the slope and normalization of the MZ relation in different redshift bins are consistent with those based on the full sample at $z=4-10$ (Eq.\,\ref{eq:MZ}). Figure \ref{fig:MZ_redshift_sim} indicates that there is no significant evolution in the MZ relation among our $z=4-10$ sample.
We note that there may be a weak trend towards lower metallicity in the highest-redshift bin, albeit with a small sample size. 
\citet{curti2023_jades} also tentatively suggest a similar evolution.
That will be further explored in the discussion.

The no/weak evolution is consistent with the predictions of some cosmological simulations, as presented and compared in Figure \ref{fig:MZ_redshift_sim}. We compile the hydrodynamic, $N$-body, and/or semi-numerical simulations showing the MZ relation at high-redshift; \textsc{FIRE} by \citet{ma2016_MZR} in black, \textsc{IllustrisTNG} by \citet{torrey2019} in green, \textsc{FirstLight} by \citet{langan2020} and \citet{nakazato2023} in blue (see also \citealt{ceverino2017}), \textsc{Astraeus} by \citet{ucci2021} in yellow, and \textsc{FLARES} by \citet{wilkins2022_MZR} in magenta (see also \citealt{lovell2021}). We plot the theoretical predictions at $z=5$, $7$, and $9$ for the redshift bin of $z=4-6$, $6-8$, and $8-10$, respectively.
For the \textsc{FIRE} curves, we extrapolate the result at $z=6$ to $z=7$ and $9$ using their redshift evolution function, although the evolution is tiny ($\Delta\log{\rm (O/H)}=-0.025$\,dex from $z=6$ to $7$, and $-0.05$\,dex from $z=6$ to $9$). 
Similarly, the \textsc{IllustrisTNG} predictions at $z=5$, $z=7$, and $z=9$ are extrapolated from the results at $z=4$, $z=6$, and $z=6$, respectively, using their redshift evolution function ($\Delta\log{\rm (O/H)}=-0.08$\,dex from $z=4$ to $5$, $-0.075$\,dex from $z=6$ to $7$, and $-0.1$\,dex from $z=6$ to $9$). 
For the \textsc{FirstLight} results, we refer to \citet{langan2020} and \citet{nakazato2023} to prove the low-mass and the massive regime, respectively. We assume the $z=8$ relation of \citet{langan2020} in panel (c), and the $z=6$ relation of \citet{nakazato2023} in panel(a), assuming no strong evolution at $z=8-9$ and $z=5-6$, respectively.
For the \textsc{FLARES} results, we adopt the stellar metallicities of only young ($<10$\,Myr) star particles and assume the stellar and gas-phase metallicities in the region of massive-star formation are comparable. 
Likewise, the metallicities for the \textsc{IllustrisTNG} and \textsc{FirstLight} model are SFR-weighted and mass-weighted of young ($<100$\,Myr) star particles, respectively, allowing a fair comparison with our metallicity measurements based on the nebular emission lines.
On the other hand, we note that the \textsc{FIRE} model adopts the mass-weighted metallicity of all gas particle that belong to the ISM, and the \textsc{Astraeus} model counts the oxygen mass in the halo without any weighting which could result in an inequitable comparison with the observations.

Comparing the observations with the simulations in Figure \ref{fig:MZ_redshift_sim}, we find the observed MZ relation and its weak evolution over $z=4-10$ are in good agreement particularly with \textsc{IllustrisTNG}, \textsc{FirstLight}, and \textsc{FLARES} in the mass range of \Mstar\ $=10^{8}-10^{9.5}$\,\Msun. Notably, the slope of the MZ-relation found in the \textsc{IllustrisTNG} results is likely coincide with the observations. 
On the other hand, the simulations of \textsc{FIRE} and \textsc{Astraeus} are generally suggested to under-predict metallicities except for some metal-poor galaxies such as ERO\_04590 found in the highest-redshift bin at $z>8$. That is probably due to their implementation of feedbacks, i.e., galaxies eject too many metals from galaxies and/or accreting gas is too efficient in lowering the metal contents in galaxies
\citep{ma2016_MZR, ucci2021}. 
Finally, we note that the average metallicity measured for low-mass galaxies below \Mstar\ $<10^{8}$\,\Msun\ can be slightly higher than the existing predictions by $\sim 0.2$\,dex. It is thus necessary to increase the sample size to determine the MZ relation in the low-mass end as well as to extend the theoretical predictions such as \textsc{IllustrisTNG} towards lower-mass to conclude the consistency and to better-understand the early chemical enrichment in the low-mass systems at high-redshift.

\begin{figure}[t]
    \begin{center}
    \includegraphics[bb=19 169 506 537, width=0.98\columnwidth]{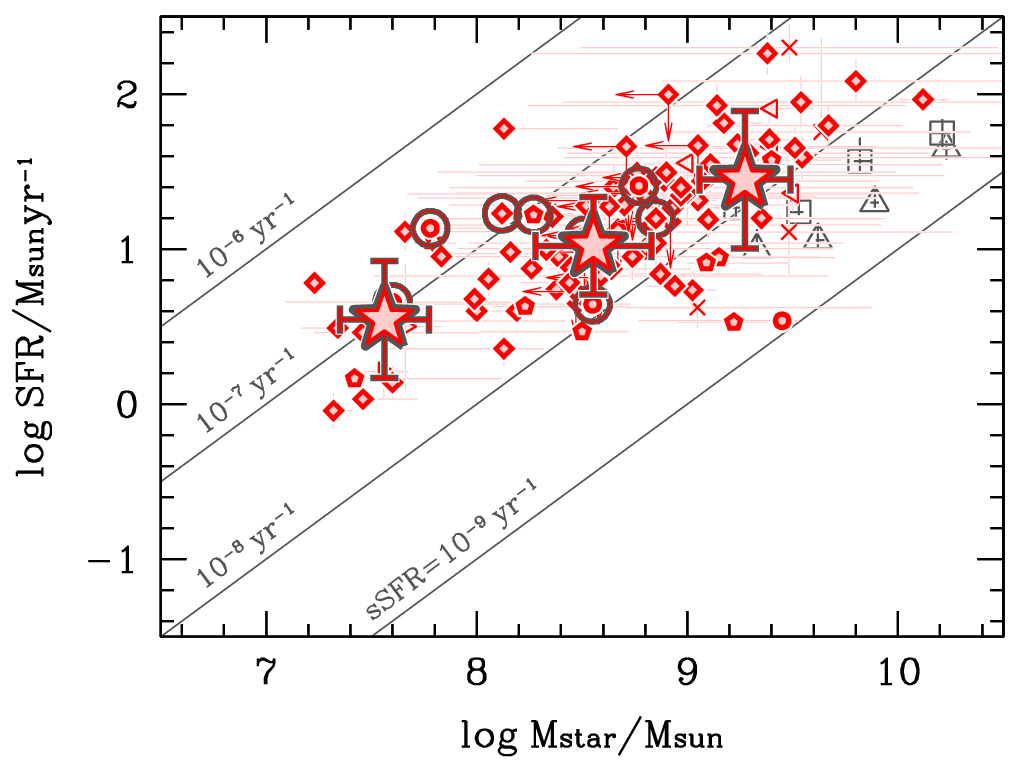}
    \caption{%
        The relationship between stellar mass and SFR, with SFR derived based on the total \Hb\ luminosity (as described in the text) for use in the SFR-MZ relation. The symbols used are consistent with those in Figure \ref{fig:MZ}. The background lines represent sSFR values ranging from $10^{-9}$ to $10^{-6}$\,yr$^{-1}$, from bottom to top. The JWST objects are distributed along the sequence of sSFR values around $10^{-8}-10^{-7}$\,yr$^{-1}$.
    }
    \label{fig:SFMS}
    \end{center} 
\end{figure}

\begin{figure*}[t]
    \begin{center}
        \subfloat{
           \includegraphics[bb=21 179 571 547, width=0.515\textwidth]{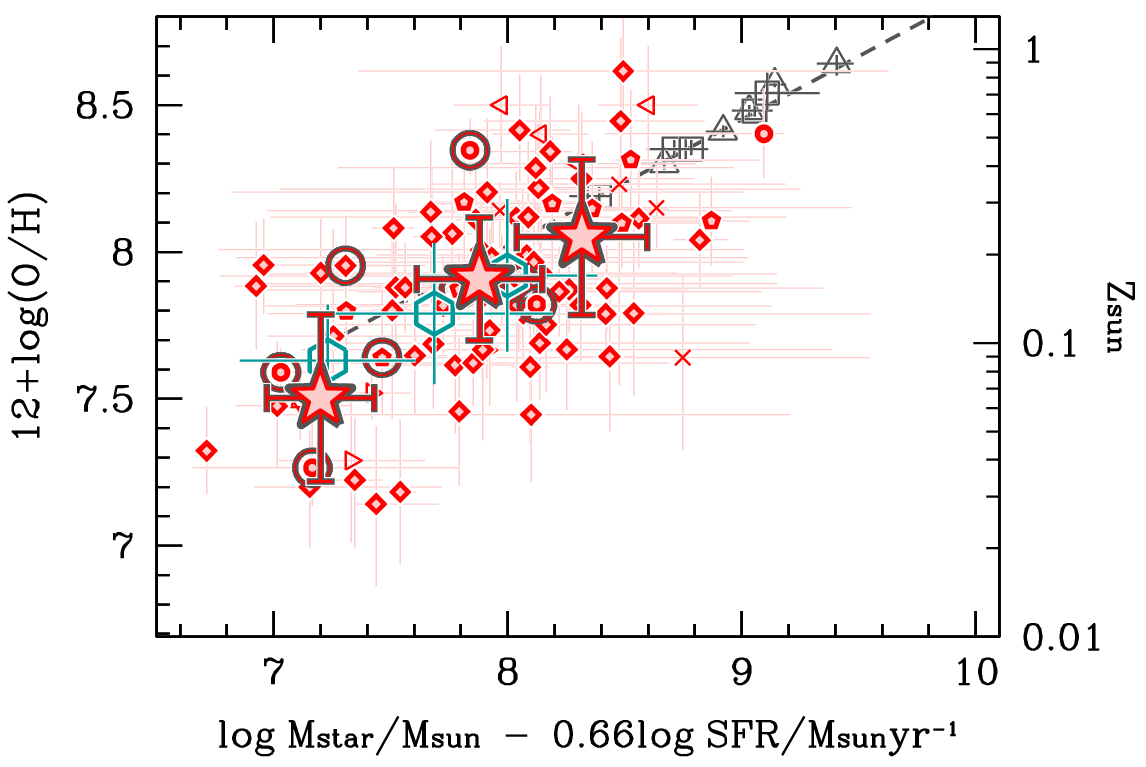}
        }
        \subfloat{
           \includegraphics[bb=21 179 571 547, width=0.515\textwidth]{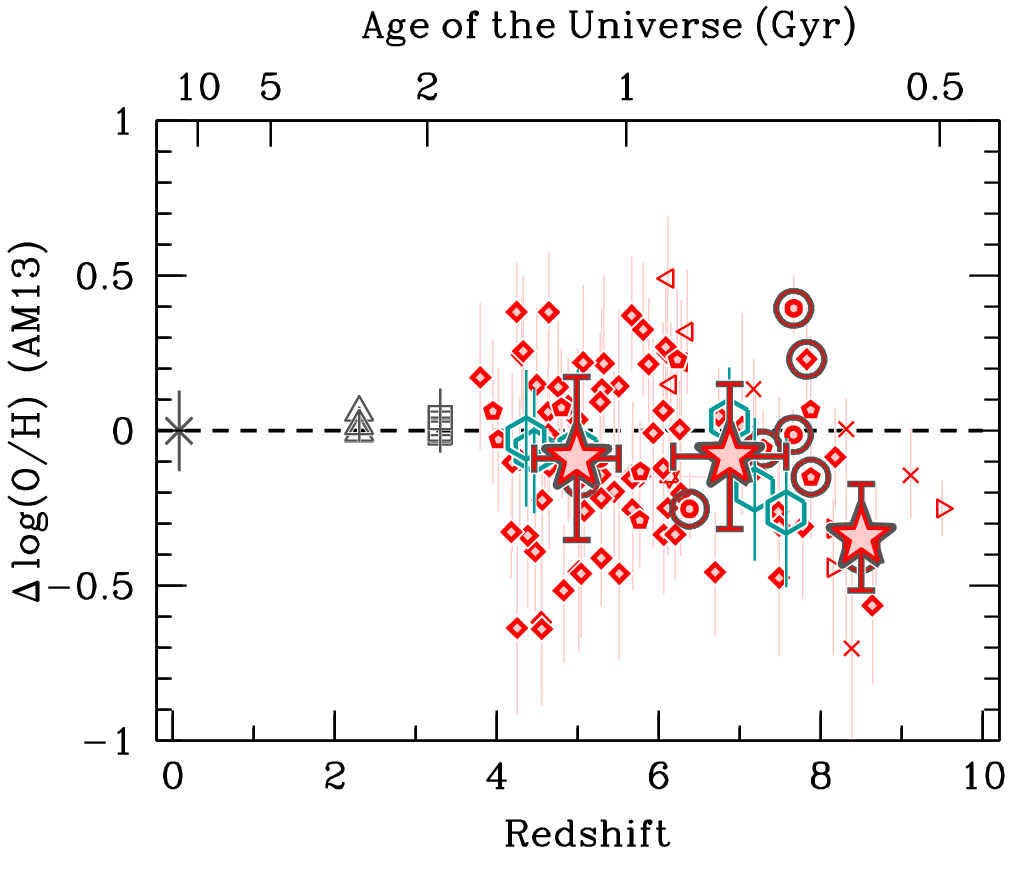}
        } 
        \caption{%
            \textbf{(a: Left:)}
            Star-formation rate dependence of the Mass-Metallicity (SFR-MZ) relation. The symbols and curves are consistent with those in Figure \ref{fig:MZ}, including the average points which are based on the three mass ranges that are equally separated.
            The coefficient of $0.66$ for the combination of stellar mass and SFR is as indicated by \citet{AM2013} in the local universe, down to \Mstar\ $\sim 10^{7.4}$\,\Msun. 
            Stacked galaxies at $z=2-3$ with $T_e$-based metallicities are also suggested to fall on the $z\sim 0$ relation in this parameter space \citep{sanders2021}.
            Notably, the JWST objects analyzed in this study at $z=4-10$ (red) also fall on the same SFR-MZ relation on average. 
            A consistent view is obtained with the JADES+CEERS average points at $z=3-10$ (emerald green; \citealt{curti2023_jades}).
		\textbf{(b: Right:)}
            Metallicity difference from the SFR-MZ relation of \citet{AM2013} ($\Delta\log$(O/H) $=$ observed $-$ predicted metallicity) for different redshift sources. The symbols are consistent with Panel (a), except for the average points (large red stars), which are recalculated on this panel to represent the average $\Delta\log$(O/H) values for the objects found in 
            three different redshift bins, $z=4-6$, $6-8$, and $8-10$.
            For the JADES+CEERS sample, average points of three mass bins are plotted for the two sub-samples at $z=3-6$ and $6-10$, as found in Figure \ref{fig:MZ_redshift_sim}.
            This plot illustrates that no evolution is seen up to $z\sim 8$, but a significant decrease in metallicity is observed beyond $z>8$, albeit with a small sample size in this highest-redshift bin.
        }
        \label{fig:FMR}
    \end{center} 
\end{figure*}

\begin{figure*}[t]
    \begin{center}
        \subfloat{
           \includegraphics[bb=21 169 571 566, width=0.49\textwidth]{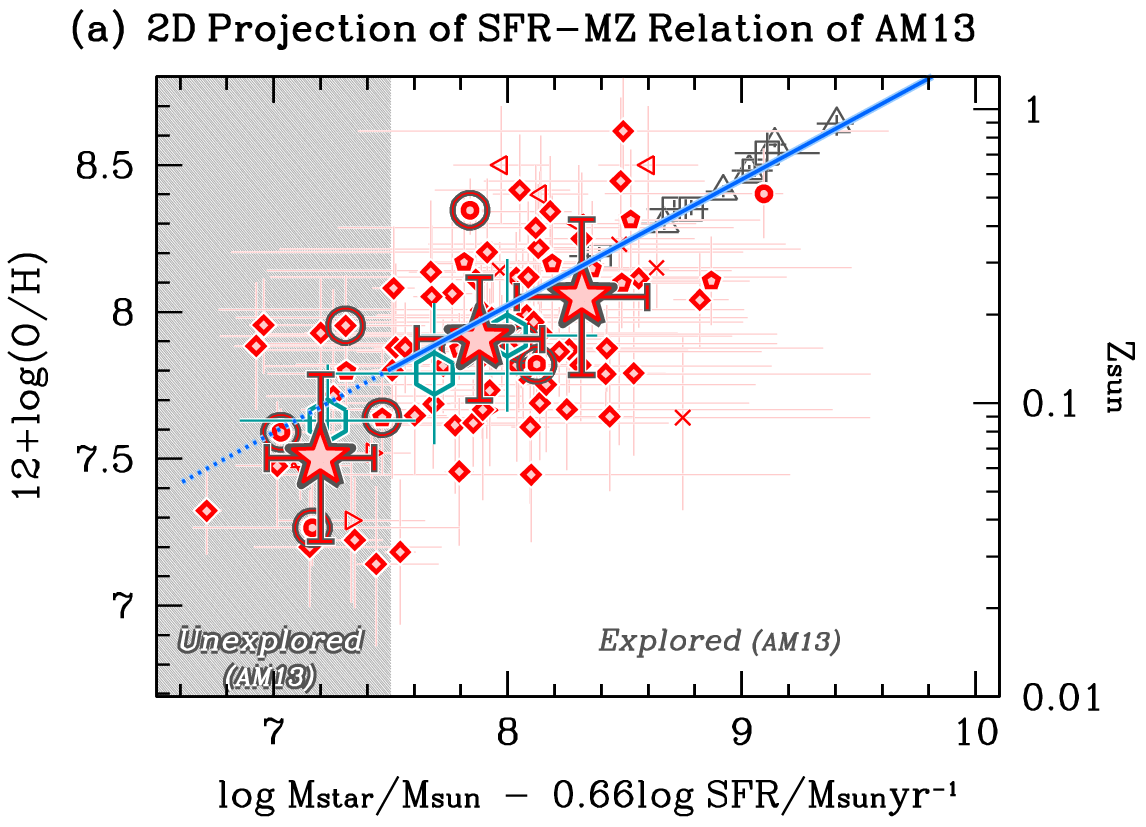}
        }
        \subfloat{
           \includegraphics[bb=21 169 571 566, width=0.49\textwidth]{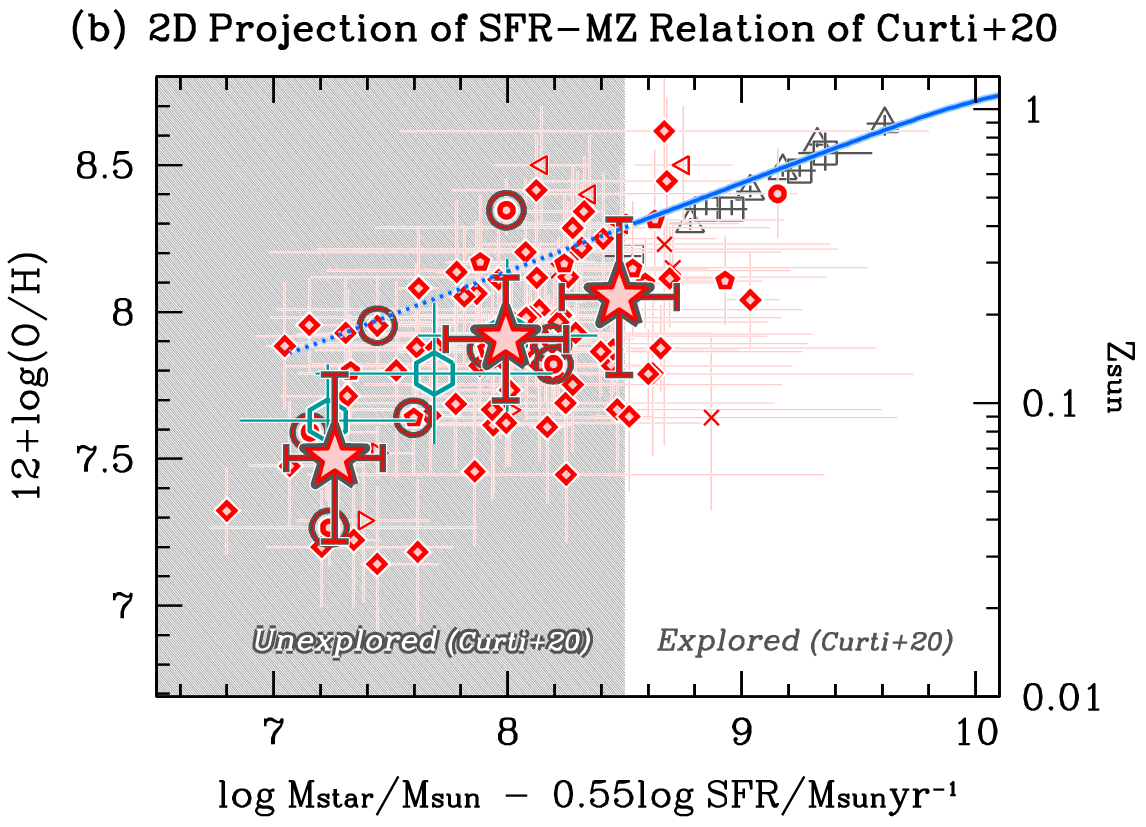}
        } 
        \caption{%
            \textbf{(a: Left:)}
            Similar to the left panel of Figure \ref{fig:FMR}, but with the regions of $\mu_{0.66}$ explored and unexplored by the original paper of \citet{AM2013} highlighted. The blue solid curve represents the best-fit 2D projection of the SFR-MZ relation by \citet{AM2013}, determined down to $\mu_{0.66} \sim 7.5$, with extrapolation towards lower mass shown by the dotted curve. Among the JWST objects, 83\,\%\ have $\mu_{0.66}>7.5$ and can be directly compared with \citeauthor{AM2013}'s relation.
		\textbf{(b: Right:)}
            Similar to the left panel, but with a different coefficient of $0.55$ adopted for the combination of mass and SFR on the abscissa axis, as presented by \citet{curti2020}. The blue solid curve represents the best-fit 2D projection of \citet{curti2020}'s SFR-MZ relation, determined down to $\mu_{0.55} \sim 8.5$, and extrapolated towards lower mass with the dotted curve. Only approximately 15\,\%\ of the JWST objects have $\mu_{0.55}>8.5$, and as such, the vast majority of the sample occupies a parameter space that is not explored by \citet{curti2020}.
        }
        \label{fig:FMR_comparison}
    \end{center} 
\end{figure*}

\begin{figure}[t]
    \begin{center}
        \subfloat{
            \includegraphics[bb=21 169 508 543, width=0.96\columnwidth]{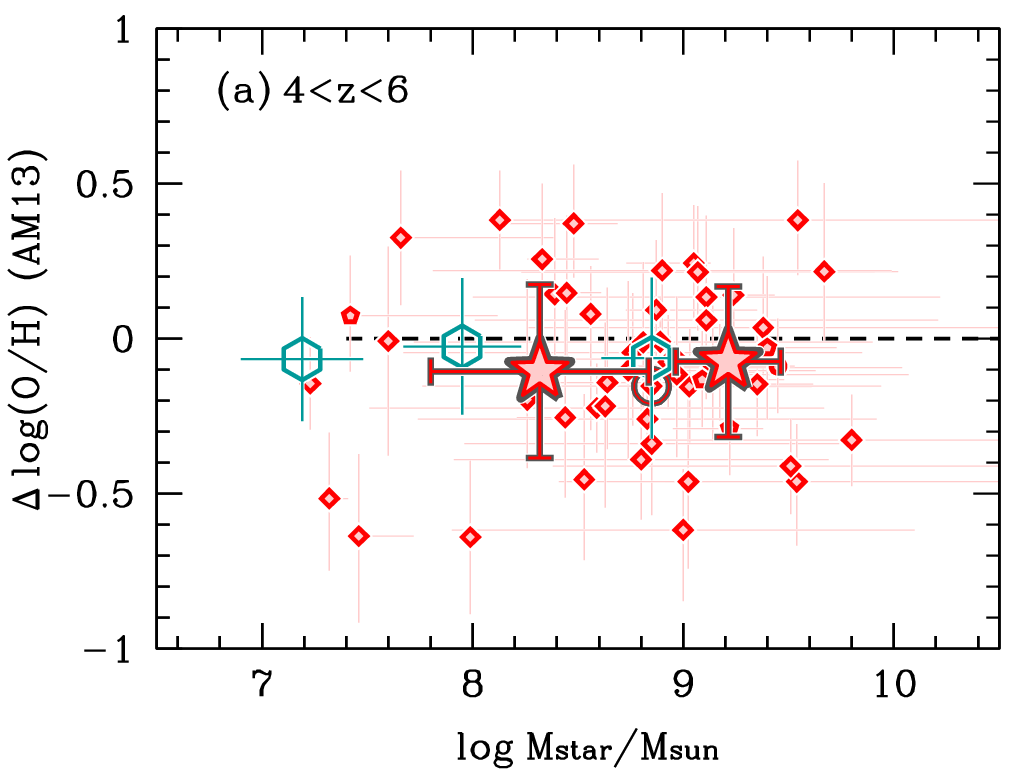}
        }
        
        \vspace{-6pt} 
        \subfloat{
            \includegraphics[bb=21 169 508 543, width=0.96\columnwidth]{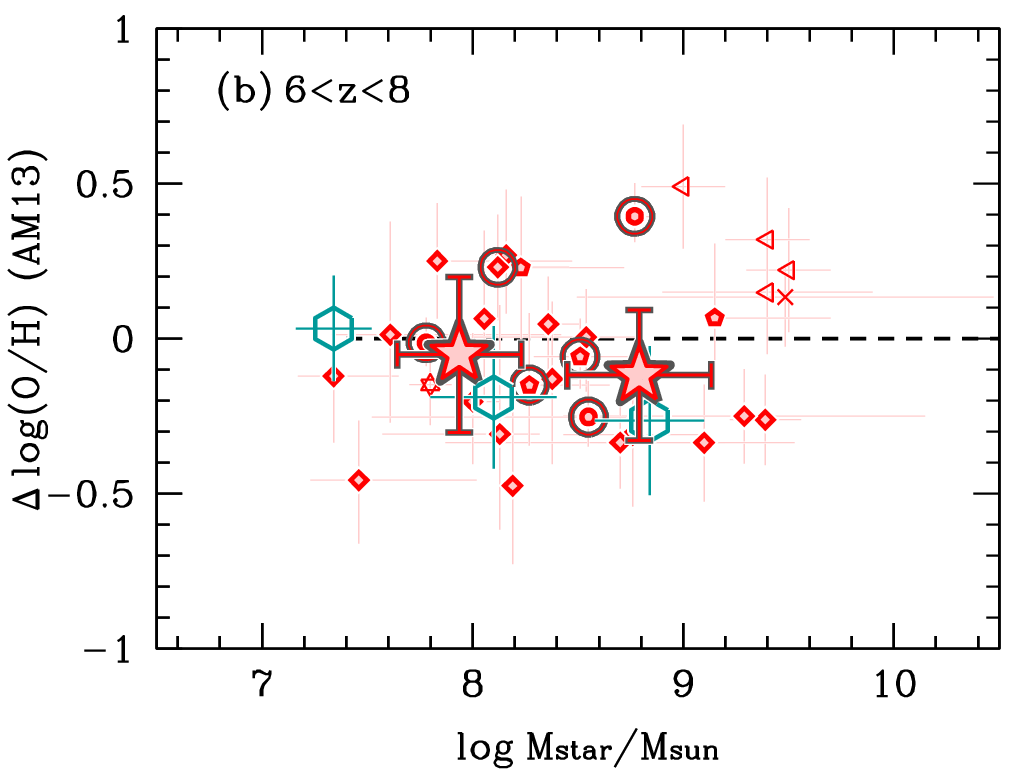}
        }      
        
        \vspace{-6pt} 
        \subfloat{
            \includegraphics[bb=21 169 508 543, width=0.96\columnwidth]{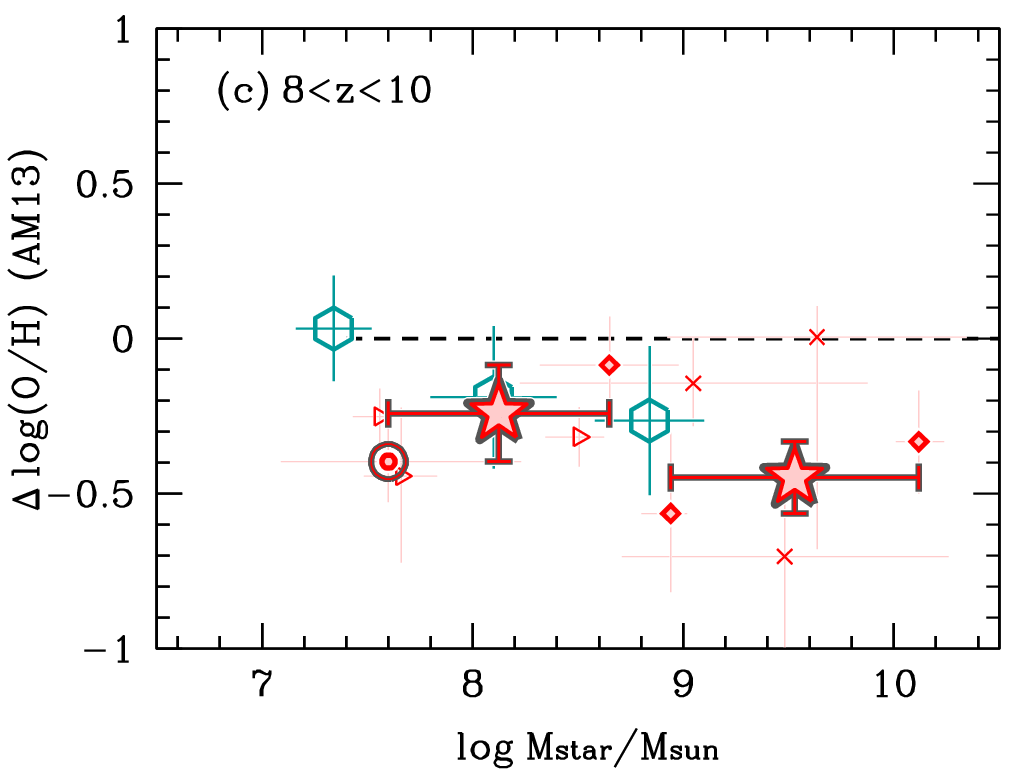}
        }
        
        \vspace{-6pt} 
        \caption{%
            Metallicity difference from the SFR-MZ relation of \citet{AM2013} as a function of stellar mass in three different redshift bins:
            (a) $z=4-6$, (b) $z=6-8$, and (c) $z=8-10$.
            The symbols are consistent with those in Figure \ref{fig:MZ_redshift_sim}. No clear dependence of $\Delta \log$(O/H) on stellar mass is evident beyond the statistical errors.
        }
        \label{fig:Offset_FMR_Mass}
    \end{center} 
\end{figure}

\subsection{Mass-Metallicity-SFR Relation} \label{ssec:results_FMR}

The next important aspect is the SFR-dependence of the MZ relation to discuss the chemical evolution, given 
the claim of a redshift-invariant \textit{fundamental} relation between mass, metallicity, and SFR (SFR-MZ relation)
out to $z\sim 2-3$ \citep{mannucci2010,sanders2021}.
There are several expressions to describe the mutual dependencies between the three quantities \citep{mannucci2010,lara-lopez2010,AM2013,sanders2017,curti2020,sanders2021}.
In this paper, we primarily use the variable $\mu_{\alpha}$ originally proposed by \citet{mannucci2010}:
$\mu_{\alpha} = \log({\rm M}_{\star}) - \alpha \log({\rm SFR})$, with $\alpha=0.66$ being adopted here that minimizes the scatter of the local low-metallicity galaxies with a direct $T_e$-based metallicity in the $\mu_{\alpha}$--metallicity plane \citep{AM2013}\,%
\footnote{
We have changed the IMFs from \citet{kroupa2001} to \citet{chabrier2003} for the masses and SFR of \citet{AM2013} following the conversion factors of \citet{MD2014}, although the shift in $\mu_{0.66}$ is tiny ($0.01$\,dex).
}:
\begin{equation}
12+\log ({\rm O/H}) = 0.43 \times \mu_{0.66} + 4.58.
\label{eq:FMR_AM2013}
\end{equation} 
This is advantageous to be directly compared with a majority of the JWST objects down to \Mstar\ $\sim 10^{7.4}$\,\Msun. Interestingly, the $z\simeq 2.3$ and $\simeq 3.3$ stacked objects of \citet{sanders2021}, whose metallicities are reliably determined with the direct $T_e$ method, fall directly on the same relationship. 
Another popular form of the SFR-MZ relation is provided by \citet{curti2020}. Our choice of the relation of \citet{AM2013} against that of \citet{curti2020} will be revisited later.

To estimate SFRs for the JWST objects, we adopt the total, reddening-corrected \Hb\ luminosity for each object, as it is the best indicator for the on-going ($\lesssim 10$\,Myr) star-formation activity. The SFR relation of \citet{kennicutt1998} is adopted for the Balmer line with a correction of IMF to \citet{chabrier2003} using the conversion factor of \citep{MD2014}.
Accordingly, the SFR-MZ relation is examined only for the JWST objects whose reddening correction is successfully applied and whose spectrum is slit-loss corrected. 
The latter constraint depends on whether the object has the NIRCam coverage (only in the CEERS field; Section \ref{ssec:data_ceers}). Among the CEERS objects lacking a slit-loss correction, we rescue 42 objects whose UV stellar emission is constrained with HST. For these objects, we translate \Muv\ into SFR(\Hb) assuming a typical ionizing photon production efficiency as indicated at high-redshift ($\log \xi_{\rm ion} = 25.6\pm 0.2$; e.g., \citealt{endsley2022_jwst,matthee2022_eiger}) and the factor of $0.44$ difference between the reddening \ebv\ for the nebular and stellar emission \citep{calzetti2000}.
The assumption of \xiion\ will be revisited elsewhere (K. Nakajima et al. in prep.) using the JWST sample presented in this paper.
The objects lacking a proper dust reddening correction or 
having just upper-limits on both \Mstar\ and SFR are excluded.
In short, 96 out of the 147 objects are used for the SFR-MZ relation at high-redshift.
The second half of Table \ref{tbl:MZ_SFR_average} presents the average SFR values for different masses and redshifts.

In Figure \ref{fig:SFMS}, we show the individual and average distributions of the JWST objects on the stellar mass-SFR plane using the SFR values as derived above (i.e., mainly from \Hb). The high-redshift objects are distributed along the sequence of sSFR $=10^{-8}-10^{-7}$\,yr$^{-1}$. This is overall consistent with the star-formation main sequence of galaxies at $z=4-7$ where sSFR of $10^{-8.5}-10^{-7.5}$\,yr$^{-1}$ is typically suggested (e.g., \citealt{stark2013,debarros2014,santini2017,popesso2023}). We note that some of our galaxies are above sSFR $\gtrsim 10^{-7.5}$\,yr$^{-1}$, in particular in the low-mass regime. This can be because the current spectroscopically-confirmed sample is partly biased toward actively star-forming systems, and/or because the sample contains higher-redshift objects at $z>7$.

Figure \ref{fig:FMR}(a) show the SFR-MZ relation of the $z=4-10$ galaxies and their average points as shown in Figure \ref{fig:MZ} on the $\mu_{0.66}$-metallicity plane, together with the $z=0$ average relation (Eq.\,\ref{eq:FMR_AM2013}) and the data-points of $z\simeq 2.3$ and $3.3$ stacked galaxies \citep{sanders2021}.
Adopting the best-fit parameter $\alpha=0.66$ found in the low-redshift metal-poor star-forming galaxies \citep{AM2013}, the $z=4-10$ objects interestingly fall on the same SFR-MZ relationship. 
The JADES+CEERS sample, notably including lower-mass galaxies than ours, is also illustrated to show a consistent SFR-MZ relation on average.

We adopt the SFR-MZ relation of \citet{AM2013} at $z=0$ that is compared with the relation at $z=4-10$ because the sample of \citet{AM2013} contains comparably low-mass, actively star-forming galaxies. Figure \ref{fig:FMR_comparison}(a) clarifies the parameter space that is explored by \citet{AM2013}, demonstrating that the majority ($83$\,\%) of the JWST objects have $\mu_{0.66}>7.5$ and can be directly compared with the relation of \citet{AM2013}. Note that local extremely metal-poor galaxies with \Mstar\ $\sim 10^{5}-10^{7}$\,\Msun\ and \Oabundance\ $=7.0-7.7$ are suggested to be reproduced by the chemical evolution models \citep{lilly2013} with the same parameters as found for the \citeauthor{AM2013}'s galaxies \citep{nishigaki2023}. This evidence supports the simple extrapolation of the relation towards lower masses.
On the other hand in Figure \ref{fig:FMR_comparison}(b), we show another form of the SFR-MZ relation derived by \citet{curti2020}. The authors find another best-fit $\alpha=0.55$ as the best 2D projection of the SFR-MZ relation on the $\mu_{\alpha}$-metallicity plane by using the global sample of stacked SDSS galaxies. The relation is determined over $\mu_{0.55}\sim 8.5-11.0$. Because $\sim 85$\,\%\ of the JWST sample have $\mu_{0.55}$ below $8.5$, most of the $z=4-10$ galaxies occupies the parameter space that is not explored by \citet{curti2020}. 
Moreover, \citet{curti2020} clarify a different degree of SFR-dependence of the MZ relation for the low- and high-sSFR subsamples in the local universe. For the low-sSFR subsample (sSFR $<10^{-9.5}$\,yr$^{-1}$), a weaker SFR dependence is indicated ($\alpha=0.22$). On the other hand, the high-sSFR subsample with sSFR $>10^{-9.5}$\,yr$^{-1}$ presents a stronger SFR dependence with $\alpha=0.65$, in close agreement with the best-fit value found by \citet{AM2013}. 
Given the high sSFRs for the JWST objects (sSFR $\sim 10^{-7}-10^{-8}$\,yr$^{-1}$; see Figure \ref{fig:SFMS}) and also for the $z=2-3$ galaxies (sSFR $\sim 10^{-8.5}$\,yr$^{-1}$, we conclude the strong SFR dependence of $\alpha=0.66$ as found by \citet{AM2013} is preferred and adopted in this paper to be compared with the high-redshift galaxies. 
We discuss the evolution of the \citeauthor{curti2020}'s SFR-MZ relation later in Section \ref{sec:discussion}.

Figure \ref{fig:FMR}(b) clarifies the redshift evolution on the SFR-MZ relation of \citet{AM2013} by showing the residual metallicity for a given $\mu_{0.66}$ with respect to the $z=0$ relation: $\Delta\log({\rm O/H}) =$ \Oabundance$_{\rm obs} -$ \Oabundance$(\mu_{0.66}$; Eq.\,\ref{eq:FMR_AM2013}) for each galaxy's redshift. In this panel, the average points are derived for the objects found in the different redshift bins, at $z=4-6$, $6-8$, and $8-10$.

Two interesting results arise in Figure \ref{fig:FMR}(b).
One is that the SFR-MZ relation shows no evolution up to $z\sim 8$ within $\Delta\log{\rm O/H}\simeq 0.3$\,dex. 
Another is that a significant decrease of metallicity is found beyond the error at $z>8$. 
The decrease at the highest-redshift bin is not visible in Panel (a) due to the small sample size, but hinted by the MZ relation at $z=8-10$ (Figure \ref{fig:MZ_redshift_sim}c).
To further examine the evolution of the SFR-MZ relation, we plot in Figure \ref{fig:Offset_FMR_Mass} the residual metallicity as a function of stellar mass for each of the three redshift bins. 
According to the two average points probing different mass ranges,
no significant dependency of the residual metallicity on stellar mass is indicated in each redshift bin with the current sample. For the highest redshift bin, the decrease of metallicity is suggested for individual galaxies regardless of their stellar masses. 
Although the JADES+CEERS sub-sample at $z=6-10$ appears to exhibit a mass dependence in the residual metallicity, we suggest this would be caused by a bias introduced by the dominance of lower-redshift galaxies (i.e., $z=6-8$) in the lower-mass regime.
This is consistent with the MZ plane in Figure \ref{fig:MZ_redshift_sim}, where the lowest-mass point of the JADES+CEERS sub-sample at $z=6-10$ is more consistent with our $z=6-8$ relation rather than the $z=8-10$ one.
These results suggest that chemical properties of star-forming galaxies up to $z\sim 8$ are very similar to those of local and $z=2-3$ counterparts, while a break of the metallicity equilibrium state may be in place beyond $z\sim 8$. More discussions follow in Section \ref{sec:discussion}.

Finally, we have conducted several tests to demonstrate the robustness of the results presented in this section.
In Appendix \ref{sec_app:results_metallicity_sedfit}, we examine the SFR-MZ relation using only the JWST galaxies whose stellar mass is well-determined with SED fit to the NIRCam photometry and whose SFR is obtained with the slit-loss corrected \Hb. Our conclusions remain unchanged even with the smaller sample that has good measurements of \Mstar\ and SFR, although the scatter of individual data points, as seen in Figure \ref{fig:Offset_FMR_Mass} for example, becomes less prominent. 
In Appendix \ref{sec_app:results_FMR_mu066_covered}, we further investigate the SFR-MZ relation by using only the JWST galaxies with $\mu_{0.66}>7.5$, where the relation is explored by \citet{AM2013}, allowing for a more robust comparison with high-redshift galaxies to discuss their evolution. We confirm that almost the same SFR-MZ relation and its redshift evolution are obtained based on this subsample, suggesting that the extrapolation of the relation at $\mu_{0.66}\sim 7-7.5$ for the low-mass end galaxies in our sample does not introduce biases. 
Furthermore, we note that a consistent view of the SFR-MZ relations is also supported by using only galaxies that are surely diagnosed as star-formation dominated in the \NII\ BPT diagram (see also Section \ref{ssec:results_MZ}).

\section{Discussion and summary} \label{sec:discussion}

We have conducted a (re-)analysis of the JWST/NIRSpec ERO data, as well as the ERS data of GLASS and CEERS, to investigate the chemical properties of galaxies at redshifts ranging from $z=4$ to $10$. Our analysis includes the use of 135 JWST objects based on our improved reduction and calibration of the NIRSpec data, as well as the compilation of 12 objects from the other recent JWST observations and previous literature.
We confirm that our new emission-line flux measurements and errors successfully address the issues reported in the literature related to Balmer decrements and electron temperatures. 
The estimated electron temperatures for the four ERO objects, along with the 6 GLASS + CEERS objects with \OIII$\lambda 4363$ at $z=4-8.5$ range from $T_e=1.1\times10^4$ to $2.3\times 10^4$\,K. These temperatures are similar to those found in lower-redshift star-forming galaxies and can be fully explained by heating of young massive stars, without the need for additional ionizing mechanisms. This is particularly evident for the $z=8.5$ object, whose $T_e$ has been updated from $\gtrsim 25000$\,K to $(2.08\pm 0.26)\times 10^4$\,K.

We have determined the mass-metallicity (MZ) relation for galaxies at $z=4-10$ and find no significant evolution compared to $z\sim 2-3$ when extrapolating to the low-mass regime. This result is consistent with the early JWST study results reported shortly after the ERO data release, except for the $z=8.5$ object whose deviation from the others on the MZ relation becomes small after re-measurement of $T_e$. 
Furthermore, our compilation of results does not show any significant evolution among the $z=4-10$ sample. 
Theoretical simulations also predict a similar trend of small redshift evolution in the MZ relation at $z\sim 5-9$ ($\Delta \log ({\rm O/H})\lesssim 0.1$\,dex; \citealt{ma2016_MZR, langan2020, ucci2021, nakazato2023}), with some simulations suggesting a potentially observable decrease in metallicity with redshift for a given mass ($\sim 0.26$\,dex from $z=5$ to $9$; \citealt{torrey2019}).
Although there may be a weak evolution towards lower metallicity at $z>8$, the current sample has relatively large uncertainties in metallicity ($\Delta\log$(O/H) $\sim 0.3$\,dex), which makes it difficult to fully distinguish between different predictions of redshift evolution at high redshifts $z>4$, especially for low-mass galaxies with \Mstar\ $<10^8\,$\Msun. Further detailed observational and theoretical studies are needed to address the apparent discrepancy and understand the early chemical enrichment in low-mass systems at high redshifts.

We have also investigated the SFR dependence of the MZ relation (SFR-MZ relation) at high redshifts, finding (i) no evolution from $z=0$ to $z=4-8$ within $\Delta \log \mathrm{(O/H)}=0.3$\,dex, and (ii) a significant decrease of metallicity at $z>8$.
The former finding supports the idea that the SFR-MZ relation, also known as the Fundamental Metallicity Relation (FMR), can indeed describe the properties of galaxies with no redshift evolution. This suggests the existence of a metallicity equilibrium state via mechanisms such as star-formation, metal-poor gas inflow, and outflow (e.g., \citealt{lilly2013}) that persists up to $z\sim 8$.
This sounds inconsistent with the potential deviation from the FMR at $z\gtrsim 2.5$ as originally indicated by \citet{mannucci2010}.
The authors have suggested, based on metallicities estimated using local empirical relations, that the same SFR-MZ relation shows no redshift evolution up to $z\sim 2.5$. They claimed that galaxies at $z\sim 3$ fall below the relation by approximately $\sim 0.6$\,dex, with a combination of \Mstar\ and SFR of $\alpha=0.32$ in $\mu_{\alpha}$. However, when using $T_e$-based metallicities from \citet{sanders2021} and adopting a different parameterization of $\alpha=0.66$ in $\mu_{\alpha}$ from \citet{AM2013}, no clear evolution from $z=0$ up to $z\simeq 3.3$ is found. 
Our findings, which utilize the parameterization of $\alpha=0.66$, support the claim of no average evolution of the SFR-MZ relation up to $z\sim 8$.
This apparent inconsistency is likely attributed to the differences in metallicity estimations. High-redshift galaxies are known to have a higher ionization state of gas, parameterized by the ionization parameter, compared to local galaxies (Figure \ref{fig:o3o2r23}; see also e.g., \citealt{NO2014, sanders2023_jwst}). Consequently, local empirical relations that assume a relatively low ionization parameter may result in biased metallicity estimates for high-redshift galaxies, particularly at $z\gtrsim 2.5$ where the \Ha+\NII$\lambda 6584$ lines are not available in the pre-JWST era. In fact, Figure \ref{fig:Z_empirical_highz} demonstrates that JWST objects with a large EW(\Hb) exhibit a systematic offset from the empirical relations of \citet{maiolino2008}, which is used for metallicity estimations in \citet{mannucci2010}. Therefore, accurate metallicity estimations using the reliable direct $T_e$ method, or accounting for the ionization state evolution as prescribed in \citet{nakajima2022_empressV} and \citet{izotov2019_lowZcandidates,izotov2021}, are essential for future metallicity studies of the high-redshift universe with JWST.

\begin{figure}[t]
    \begin{center}
        \subfloat{
           \includegraphics[bb=21 169 508 587, width=0.45\textwidth]{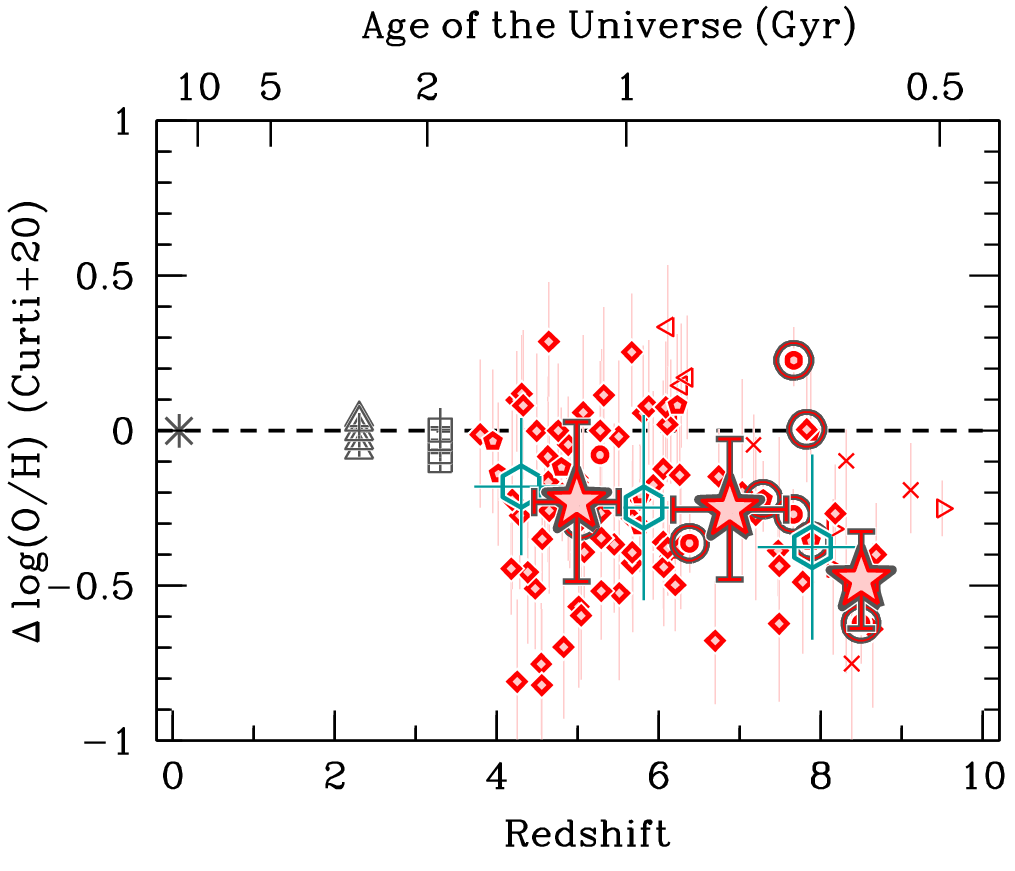}
        }
        \caption{%
            Similar to the right panel of Figure \ref{fig:FMR}, but employing the SFR-MZ relation of \citet{curti2020}. 
            As similarly found in Figure \ref{fig:FMR}, our analysis (red) and the JADES+CEERS study by \citet{curti2023_jades} (emerald green) suggest a consistent evolution of the SFR-MZ relation at $z=4-10$ on average. If we adopt the formalism of \citet{curti2020} which is defined at $z=0$, our JWST objects begin to deviate from the SFR-MZ relation at $z\sim 6$. This finding contrasts with the use of another $z=0$ SFR-MZ relation of \citet{AM2013}, which suggests deviations starting at $z\sim 8$ (Figure \ref{fig:FMR}b).
            It is worth noting, however, that a simple extrapolation of the relation towards lower mass and higher SFR is assumed for the majority of the JWST objects (see Figure \ref{fig:FMR_comparison}(b)).
        }
        \label{fig:Offset_FMR_curti20}
    \end{center} 
\end{figure}

The latter finding of the potential evolution of the SFR-MZ relation beyond $z\sim 8$ is intriguing. 
A similar result is supported by \citet{heintz2022_jwst} as independently illustrated with the three DDT objects in our figures (red open right-pointing triangles),
and also by the recent study of \citet{curti2023_jades} based on the deep JADES observations. However, in contrast to our results using the SFR-MZ relation of \citet{AM2013}, \citet{curti2023_jades} argue that galaxies begin to deviate at $z\sim 6$, if they adopt the SFR-MZ relation of \citet{curti2020}.
Because we confirm that our results and the JADES study yield the similar MZ and SFR-MZ relations, as evidenced by the consistency between the red and emerald green symbols in Figures \ref{fig:MZ}, \ref{fig:MZ_redshift_sim}, and \ref{fig:FMR}, we attribute the redshift differences where deviations occur to the different functional forms of the SFR-MZ relation defined at $z=0$. This is also supported by the evolution of the SFR-MZ relation based on the \citet{curti2020}'s formalism for our JWST objects, as shown in Figure \ref{fig:Offset_FMR_curti20}, which indicates a consistent evolution with deviations starting at $z\sim 6$ as reported by \citet{curti2023_jades}. 
While the use of a 3D SFR-MZ relation can account for the SFR-dependence of the MZ relation's slope and capture the metallicity variations more accurately than a simple 2D projection using $\mu_{\alpha}$ as assumed in the \citeauthor{AM2013}'s relation (e.g., \citealt{curti2020}), this is probably not an essential issue in this study. 
This is because the JWST objects analyzed so far present a narrow distribution of sSFR (Figure \ref{fig:SFMS}; see also Figure 4 of \citealt{curti2023_jades}).
We acknowledge that extended local-baseline studies that cover low-mass and high sSFR galaxies at $z=0$ are necessary to discuss the chemical evolution, including the fundamental SFR-MZ relation, and determine when and if high-redshift galaxies begin to deviate from the relation. In this work, however, we regard that the best estimates are the results based on the SFR-MZ relation by \citet{AM2013} as it provides a more appropriate comparison by covering the low-mass regime where most of the JWST objects are found (Section \ref{ssec:results_FMR} and Figure \ref{fig:FMR_comparison}).

If we assume that the first star formation began to occur at $z=16-27$ (\citealt{harikane2022}; see also e.g., \citealt{abel2002,bromm2002,DF2018}), galaxies at $z=8-10$ would be at most $\sim 200-500$\,Myr old. During such a short timescale after the Big Bang, there could be a higher probability for galaxies at higher redshifts to be in the early stages of formation, where they have not yet reached the metallicity equilibrium state via processes such as star formation, inflow, and outflow.
An alternative explanation could be varying degrees of feedback processes, including inflow and outflow, in the early universe. As evident from the comparison of cosmological simulation results in Figure \ref{fig:MZ_redshift_sim}, galaxies could exhibit lower metallicities if metals are efficiently ejected from galaxies or diluted by gas accretion from inflow of pristine gas, as observed in simulations such as \textsc{FIRE} and \textsc{Astraeus}. Furthermore, other events such as galaxy mergers and AGN activities may also play a role in modulating star-formation, causing metal redistribution, and potentially influencing the SFR-MZ relationship at high-redshift (e.g., \citealt{springel2005,torrey2012,weinberger2018}).
Due to the limited sample size and associated metallicity errors, particularly at $z>8$, it is challenging to make definitive conclusions from the current study. Further statistical investigations of metallicity are crucial to confirm the suggested lack of evolution in the SFR-MZ relation up to $z\sim 8$, followed by a decrease at higher redshifts. These findings need to be rigorously compared with theoretical studies to better understand the underlying physics that govern early galaxy evolution.


\section*{Acknowledgements}

We are grateful to Hidenobu Yajima, Hajime Fukushima, Chris Lovell, Yurina Nakazato, and the anonymous referee for useful comments and discussions that improved our manuscript.
This work is based on observations made with the NASA/ESA/CSA James Webb Space Telescope. The data were obtained from the Mikulski Archive for Space Telescopes at the Space Telescope Science Institute, which is operated by the Association of Universities for Research in Astronomy, Inc., under NASA contract NAS 5-03127 for JWST. These observations are associated with program \#1324 (ERS-GLASS), \#1345 (ERS-CEERS), \#2561 (UNCOVER), and \#2736 (ERO).
The authors acknowledge the teams of JWST commissioning, ERO, GLASS, UNCOVER, and CEERS for developing their observing programs with a zero-exclusive-access period.
Moreover, this work is based in part on observations taken by the CANDELS Multi-Cycle Treasury Program with
the NASA/ESA HST, which is operated by the Association of Universities for Research in Astronomy, Inc., under NASA contract NAS5-26555.
This paper is supported by World Premier International Research Center Initiative (WPI Initiative), MEXT, Japan, as well as the joint research program of the Institute of Cosmic Ray Research (ICRR), the University of Tokyo. This work is supported by KAKENHI (JP19H00697, JP20H00180, and JP21H04467) Grant-in-Aid for Scientific Research (A) through the Japan Society for the Promotion of Science. 
In addition, KN acknowledges support from JSPS KAKENHI Grant JP20K22373.
YI is supported by JSPS KAKENHI Grant JP21J20785, and also acknowledges funding from the Hayakawa Satio Fund awarded by the Astronomical Society of Japan.
YH acknowledges support from JSPS KAKENHI Grant JP21K13953.



\bibliographystyle{aasjournal}{}
\bibliography{Refs_paper.bib}{}


\appendix
\restartappendixnumbering

\section{MZ and SFR-MZ relations based only on galaxy properties derived from SED fitting to NIRCam photometry}
\label{sec_app:results_metallicity_sedfit}

In the main text, we present the mass-metallicity (MZ) and the SFR-MZ relations for the full JWST sample with the NIRSpec spectra. Out of the 135 JWST objects with metallicity measurements, 81 objects have NIRCam imaging data, and their stellar masses \Mstar\ are derived based on SED fitting to the NIRCam photometry. 
The remaining 54 objects, all of which are in the CEERS field, are not covered by NIRCam, and their \Mstar\ is empirically estimated using UV luminosities from HST photometry (Section \ref{ssec:data_ceers}). Additionally, their SFRs are estimated from \Muv\ with some assumptions for comparison with SFR(\Hb) (Section \ref{ssec:results_FMR}). 
We note that 23 out of the 54 objects are not detected in HST and have only upper limits on \Mstar\ and SFR. Since objects with only an upper limit on \Mstar\ are not considered when deriving the average MZ relations and best-fit, there are practically 31 objects (i.e., $=54-23$), accounting for $\sim 28$\,\% of the sample, which may introduce bias due to different methods used for \Mstar\ and SFR estimation. 
In this appendix, we present the main figures with only the JWST objects having NIRCam photometry and properties derived from SED fitting, in order to examine how the main results can be affected by excluding the 31 objects with less certain estimations of \Mstar\ and SFR.

\begin{figure}[h]
    \begin{center}
        \subfloat{
           \includegraphics[bb=21 169 571 537, width=0.45\textwidth]{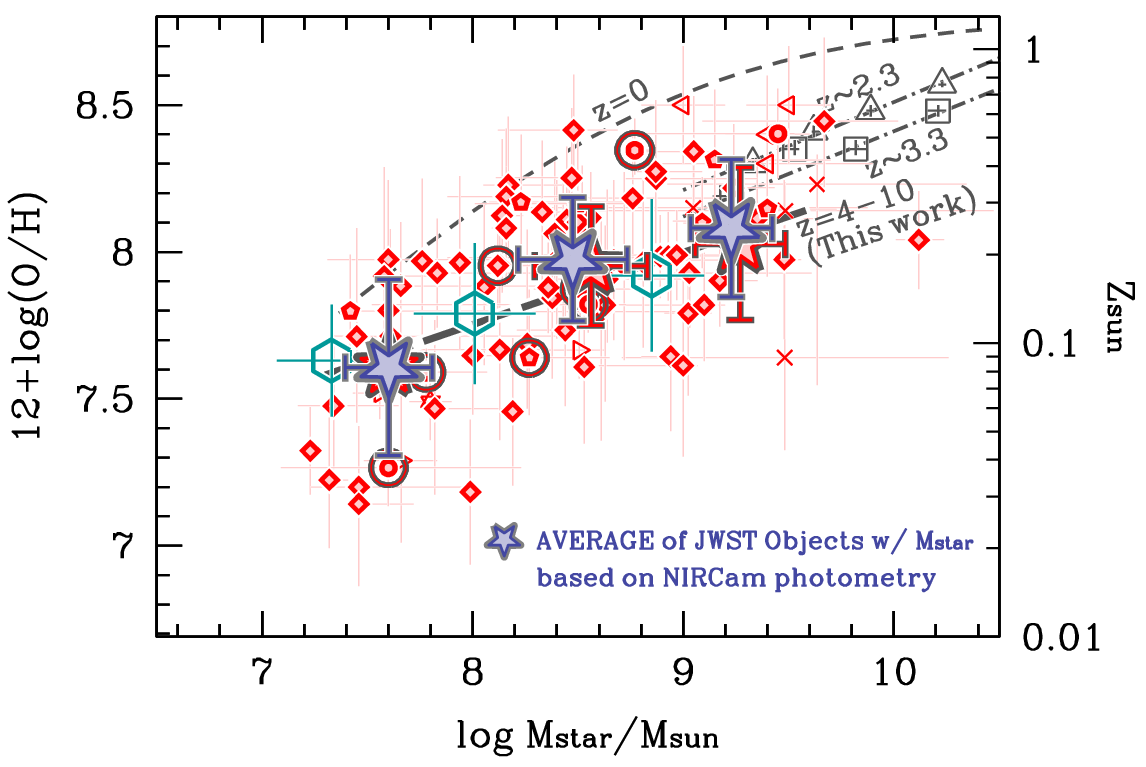}
        }
        \caption{%
            Similar to Figure \ref{fig:MZ}, but excluding the individual JWST objects whose stellar masses are empirically derived, so all the plotted data-points have stellar masses that are derived based on SED fitting to NIRCam photometry.
            The average of the JWST objects plotted on this figure are represented by purple inverted stars.
            As a comparison, the average relations based on the full sample are shown with the red stars and the gray long-dashed line, as presented in \ref{fig:MZ}, which are positioned almost behind the purple inverted stars. This confirms that our results remain unchanged when adopting the subsample with certain estimations of \Mstar. 
        }
        \label{fig:MZ_sedfit}
    \end{center} 
\end{figure}

Figure \ref{fig:MZ_sedfit} presents the MZ relation using only the JWST objects with \Mstar\ based on NIRCam photometry. Their average points are shown with the inverted purple stars. The red stars and the gray long-dashed line represent the average relations based on the full sample of the JWST objects as shown in Figure \ref{fig:MZ}.
 We confirm that the difference between the two is negligibly small, and the MZ relation based only on the JWST objects with reliable \Mstar\ estimates is fully consistent with that based on the full sample, within the uncertainty.

\begin{figure}
    \begin{center}
        \subfloat{
            \includegraphics[bb=21 169 571 537, width=0.32\textwidth]{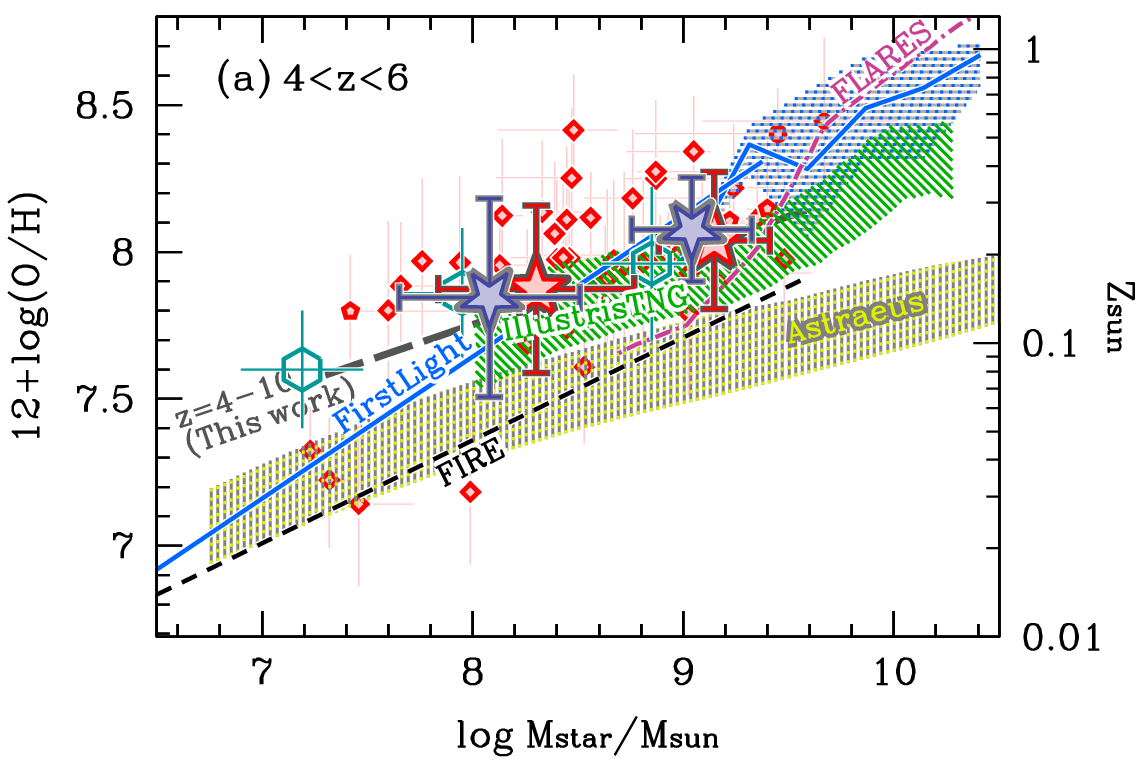}
        }
        \subfloat{
            \includegraphics[bb=21 169 571 537, width=0.32\textwidth]{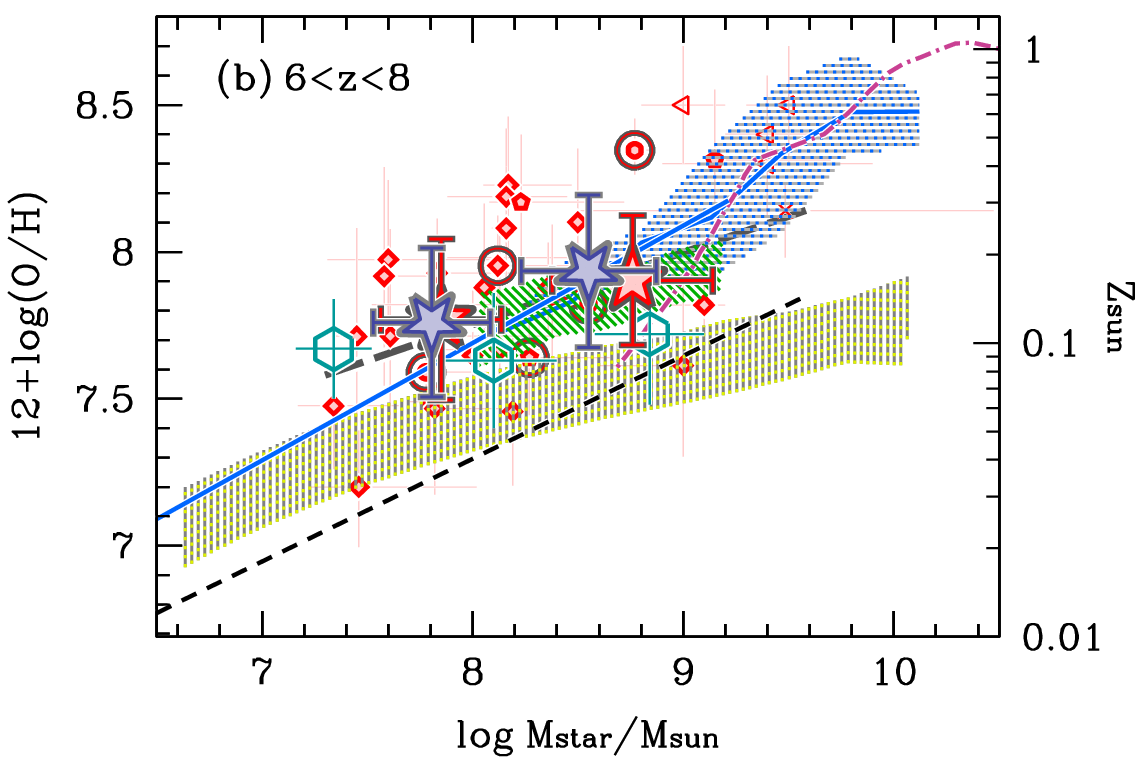}
        }       
        \subfloat{
            \includegraphics[bb=21 169 571 537, width=0.32\textwidth]{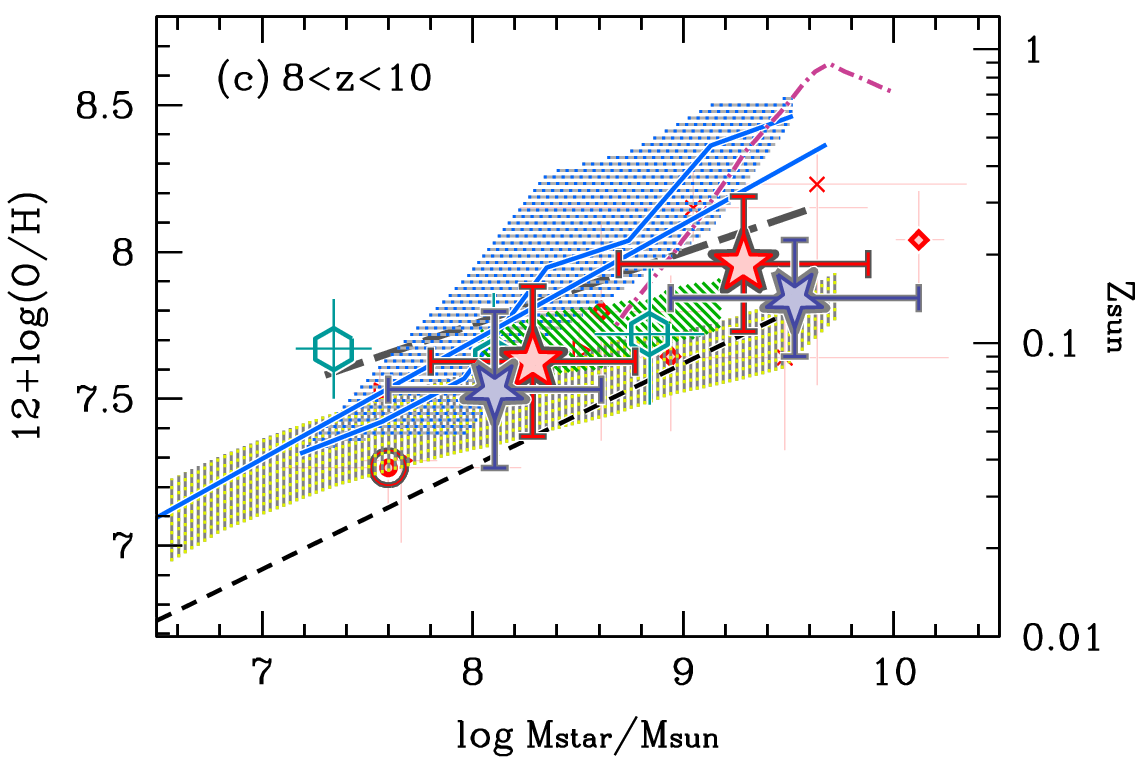}
        }
        \caption{%
            Similar to Figure \ref{fig:MZ_redshift_sim}, but excluding the individual JWST objects without JWST/NIRCam photometry.
            The average relations of the JWST objects in each panel are represented by purple inverted stars.
        }
        \label{fig:MZ_redshift_sim_sedfit}
    \end{center} 
\end{figure}

\begin{figure}
    \begin{center}
        \subfloat{
           \includegraphics[bb=21 169 571 537, width=0.45\textwidth]{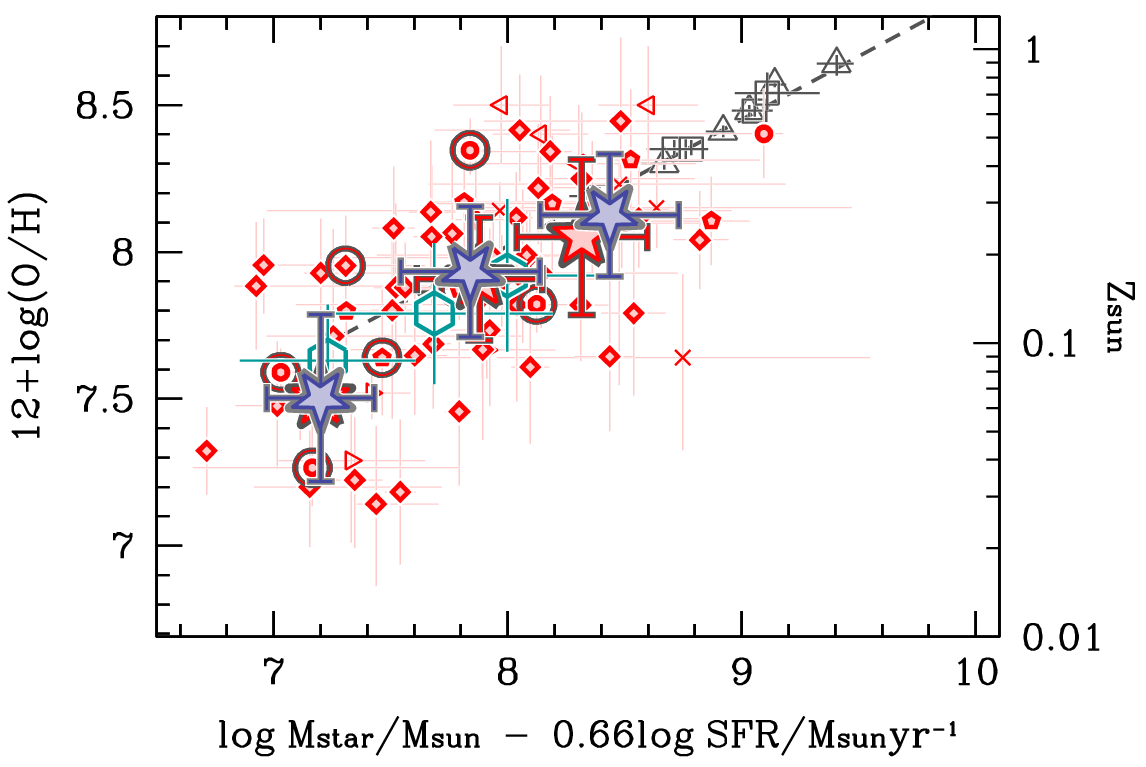}
        }
        \subfloat{
           \includegraphics[bb=21 169 508 587, width=0.45\textwidth]{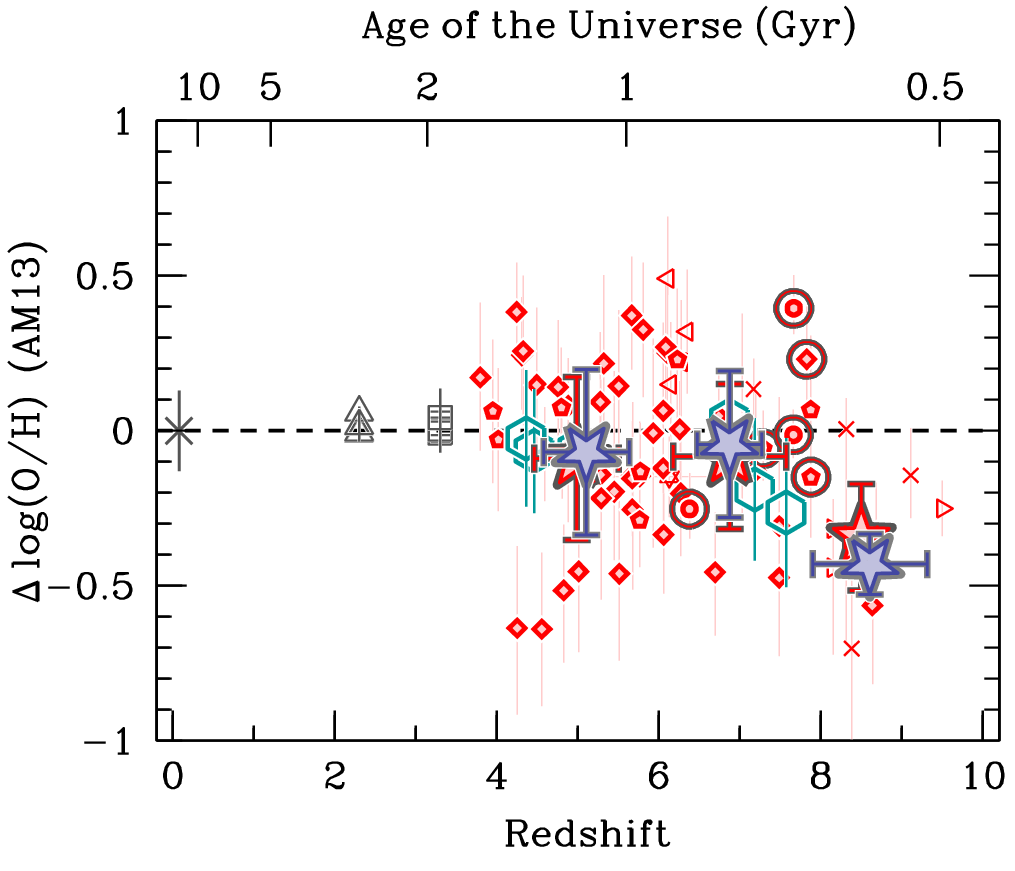}
        } 
        \caption{%
            Similar to Figure \ref{fig:FMR}, but excluding the individual JWST objects without JWST/NIRCam photometry.
            The average relations of the JWST objects in each panel are represented by purple inverted stars.
        }
        \label{fig:FMR_sedfit}
    \end{center} 
\end{figure}

\begin{figure}
    \begin{center}
        \subfloat{
            \includegraphics[bb=21 169 508 543, width=0.32\textwidth]{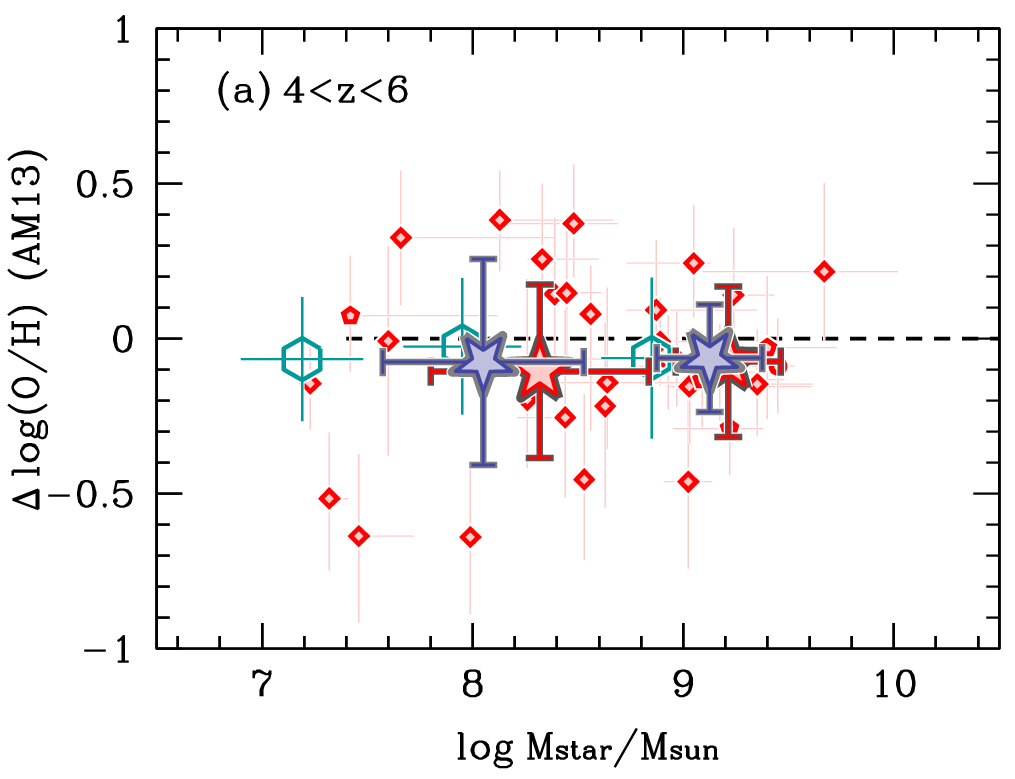}
        }
        \subfloat{
            \includegraphics[bb=21 169 508 543, width=0.32\textwidth]{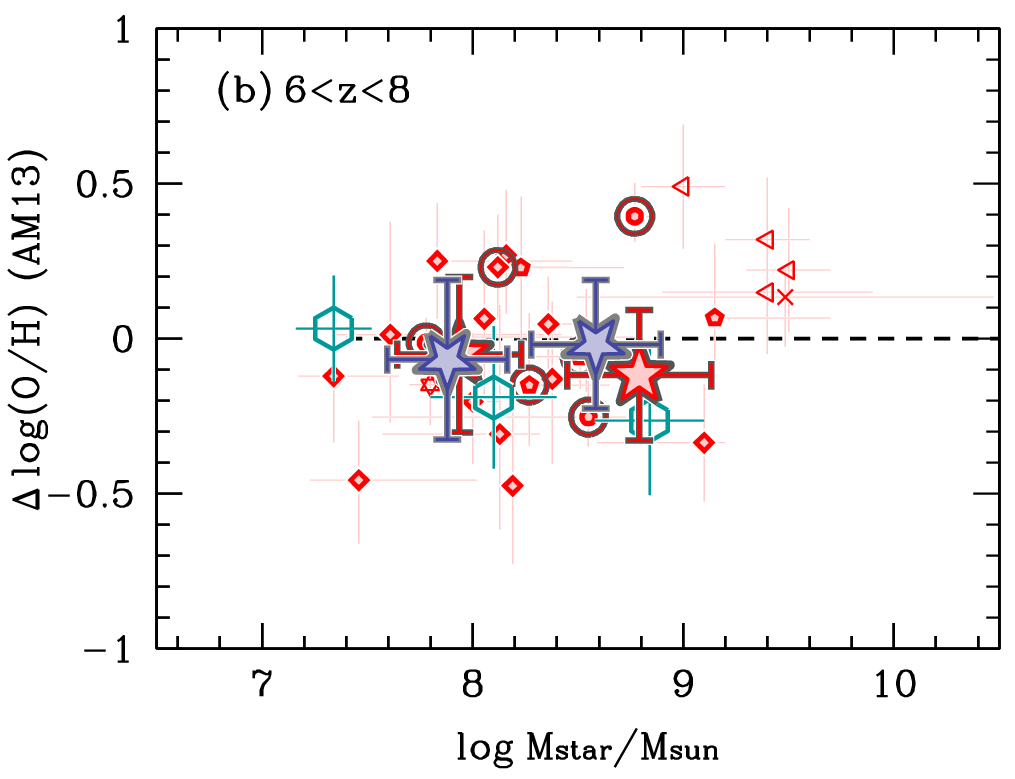}
        }       
        \subfloat{
            \includegraphics[bb=21 169 508 543, width=0.32\textwidth]{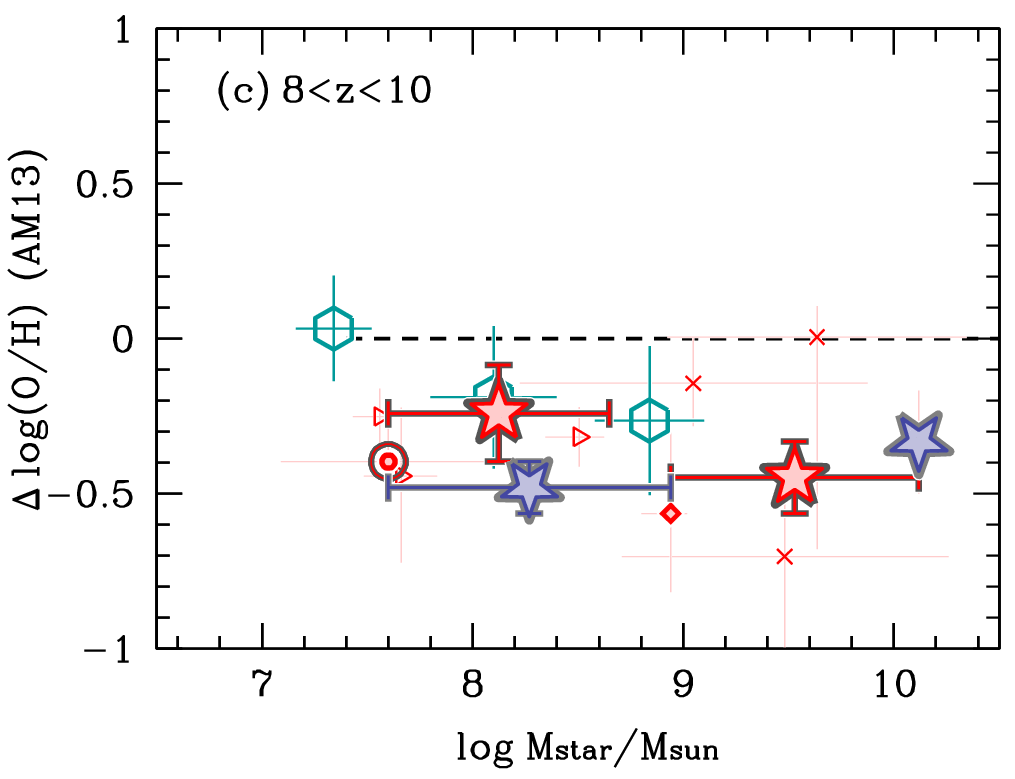}
        }
        \caption{%
            Similar to Figure \ref{fig:Offset_FMR_Mass}, but excluding the individual JWST objects without JWST/NIRCam photometry. The average relations of the JWST objects in each panel are represented by purple inverted stars. In the $z=8-10$ panel, only a single average point is displayed due to the limited sample size.
        }
        \label{fig:Offset_FMR_Mass_sedfit}
    \end{center} 
\end{figure}

Likewise, Figures \ref{fig:MZ_redshift_sim_sedfit}, \ref{fig:FMR_sedfit}, and \ref{fig:Offset_FMR_Mass_sedfit} are regenerated plots of Figures \ref{fig:MZ_redshift_sim}, \ref{fig:FMR}, and \ref{fig:Offset_FMR_Mass} from the main text, respectively, using only the JWST objects with NIRCam photometry. These figures confirm that our results remain unchanged when adopting the subsample, indicating that no clear biases are introduced by using the 31 objects with \Mstar\ and SFR estimated in a different manner.
We have also checked the consistency of \Mstar\ between the two methods by comparing the estimates from the 81 CEERS objects with both measurements, and found that $\sim 90$\,\%\ of them have consistent \Mstar\ values at the $2\sigma$ level. The remaining 8 objects show overestimated \Muv-based \Mstar\ values compared to NIRCam-based \Mstar\ at $>2\sigma$ level. Although such a small fraction of outliers may exist among the objects lacking NIRCam photometry and introduce additional scatter in the MZ and MZ-SFR relations (e.g., Fig \ref{fig:MZ_redshift_sim}(a) versus Fig \ref{fig:MZ_redshift_sim_sedfit}(a) at the high-mass region), the impact is minor as demonstrated in this appendix.

\section{Comparing with empirically estimated metallicity relations at low-redshifts using strong line indicators}
\label{sec_app:results_MZ_w_empirical}

In the main text we refer to \citet{AM2013} and \citet{sanders2021} to compare our results with the MZ relations at lower redshifts. This is because their metallicities are reliably determined with the direct $T_e$ method. Moreover, the $z=0$ relation of \citet{AM2013} is explored down to \Mstar\ $\sim 10^{7.4}$\,\Msun\ and can be directly compared with most of the JWST objects.
In this appendix, we also present comparisons with the other MZ relations at low-redshifts whose metallicities are empirically estimated with the strong line indicators.

\begin{figure}[h]
    \begin{center}
        \subfloat{
            \includegraphics[bb=-4 169 546 537, width=0.45\textwidth]{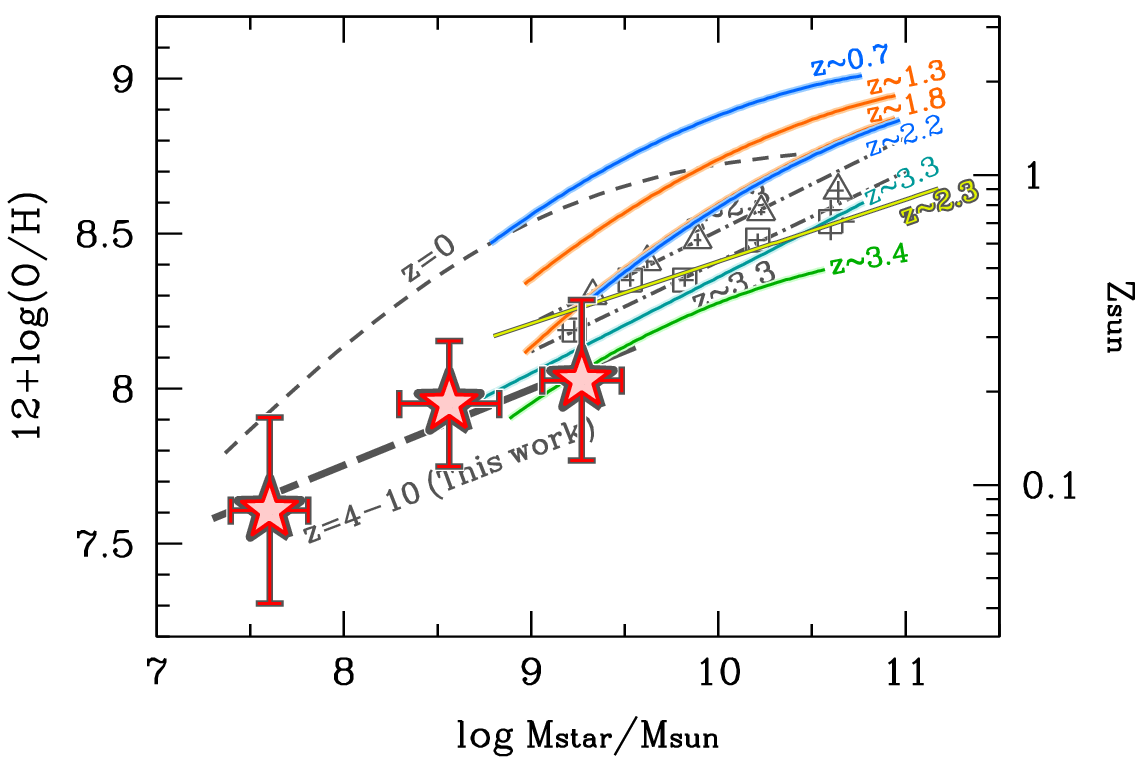}
        }
        \subfloat{
            \includegraphics[bb=-4 169 546 537, width=0.45\textwidth]{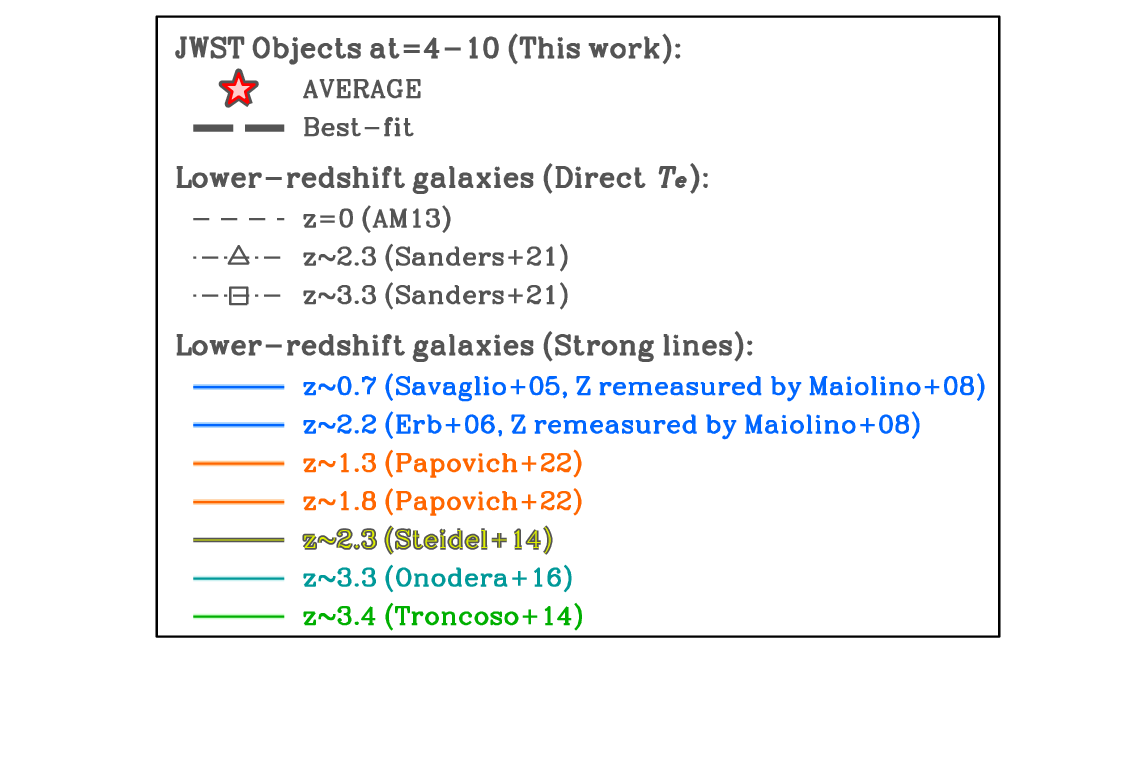}
        } 
        \caption{%
            Similar to Figure \ref{fig:MZ}, but excluding the individual JWST objects, and instead focusing on comparisons with low-redshift MZ relations based on empirically estimated metallicities using strong line indicators, as indicated in the legend. Each MZ relation is depicted in the mass range explored in the original paper.
        }
        \label{fig:MZ_w_empirical}
    \end{center} 
\end{figure}

Figure \ref{fig:MZ_w_empirical} shows the strong line-based MZ relations at $z\lesssim 3$ in addition to our results (the average points and the best-fit) and the $T_e$-based MZ relations at $z=0-3$ \citep{AM2013,sanders2021} as in Figure \ref{fig:MZ}.
The strong line-based MZ relations include those for $z\sim 0.7$ galaxies \citep{savaglio2005}, $z\sim 1-2$ \citep{papovich2022}, $z\sim 2-3$ \citep{erb2006_MZR,steidel2014}, and $z\sim 3-4$ (\citealt{troncoso2014} (see also \citealt{maiolino2008,mannucci2009}), \citealt{onodera2016}).
The results by \citet{savaglio2005} and \citet{erb2006_MZR} are revisited by \citet{maiolino2008} and their metallicities are re-measured. We also apply corrections to the mass scale to have the same \cite{chabrier2003} IMF.
For the $z=2-3$ galaxies the \NII$/$\Ha\ index is mainly used for the metallicity estimations, while for the other redshift studies combinations of \OII, \OIII\ and \Hb\ (i.e., R23-index and O32-index in practice) are utilized, in addition to \NeIII\ when available.
Although there are some differences in the MZ relations between different studies and methods for a similar redshift, several factors account for the scatter, such as different SFR activities for the different samples and variations of the metallicity indicators.
Despite the slight differences, these overall confirm a redshift evolution of MZ relation, i.e., a decreasing trend of metallicity in the mass range of $10^9-10^{11}$\,\Msun. 
Comparing with these strong line-based MZ relations, our main conclusion does not change that the JWST objects at $z=4-10$ are distributed along the simple extrapolation of the MZ relations at $z\sim 2-3$ towards lower mass.

\section{Evolution of the SFR-MZ relation without the low-mass end galaxies}
\label{sec_app:results_FMR_mu066_covered}

Figure \ref{fig:FMR_comparison}(a) highlights that the majority of the JWST sample ($83$\,\%) has $\mu_{0.66}>7.5$ and is directly compared with the lower-redshift galaxies in the SFR-MZ relation of \citet{AM2013}. Still, we rely on the simple extrapolation of the relation towards lower $\mu_{0.66}$ for the remaining galaxies and discuss the evolution of the SFR-MZ relation together with the direct comparisons of galaxies at $\mu_{0.66}>7.5$.
In this appendix, we present how the low-mass end galaxies at $\mu_{0.66}<7.5$ can impact the view of the evolution of the SFR-MZ relation.

\begin{figure}[h]
    \begin{center}
        \subfloat{
           \includegraphics[bb=21 169 508 587, width=0.45\textwidth]{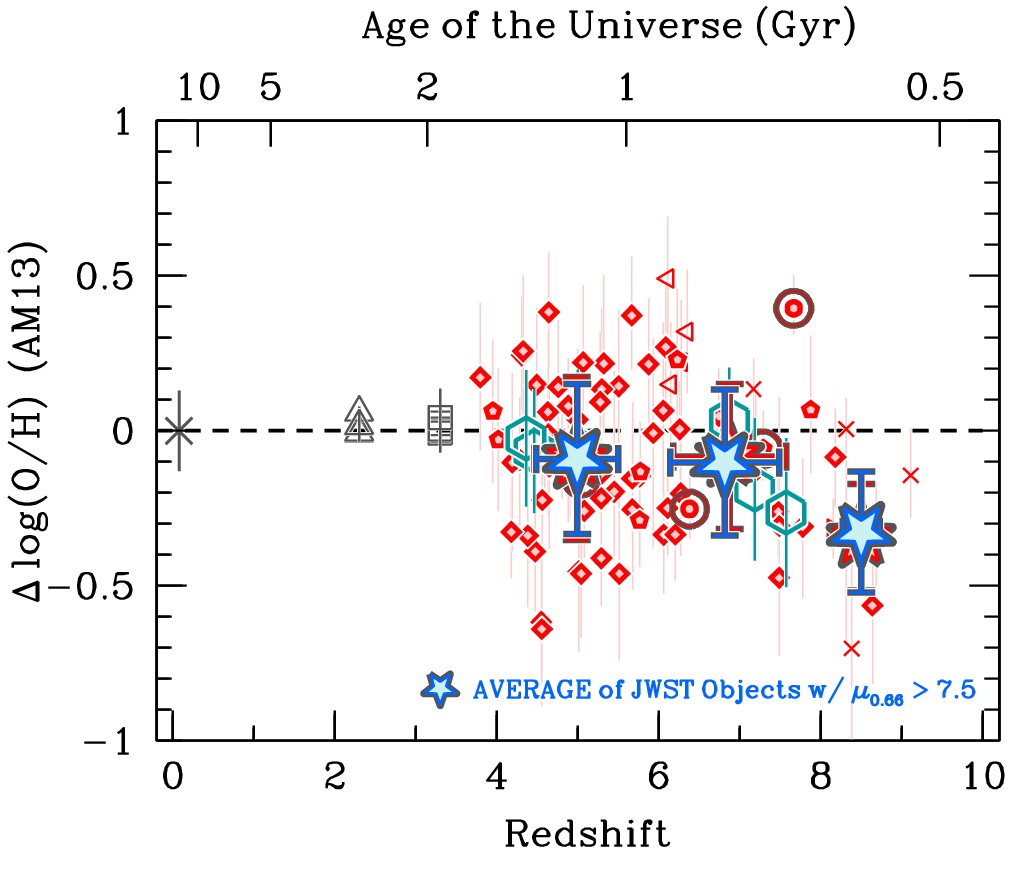}
        }
        \caption{%
            Similar to the right panel of Figure \ref{fig:FMR}, but excluding the individual JWST objects at $\mu_{0.66}<7.5$. The average relations of the subsample are denoted with the blue inverted stars.
            As a comparison, the average relations based on the full sample are shown with the red stars, as presented in Figure \ref{fig:FMR}, 
            which are positioned almost behind the blue inverted stars. This confirms that our results on the redshift evolution of the SFR-MZ relation remain unchanged when considering only the objects at $\mu_{0.66}>7.5$.
        }
        \label{fig:Offset_FMR_mu066_covered}
    \end{center} 
\end{figure}

Figure \ref{fig:Offset_FMR_mu066_covered} shows the redshift evolution of the SFR-MZ relation by only using the JWST objects having $\mu_{0.66}>7.5$. Their average are represented by the inverted blue stars. As a comparison, the average relations based on the full sample are shown with the red stars, as presented in Figure \ref{fig:FMR}, which are positioned almost behind the blue stars. This confirms that the same results arise from the subsample as discussed in the main text, i.e., (i) the same \citet{AM2013} relation between mass, metallicity, and SFR can explain the properties of galaxies up to $z\sim 8$, and (ii) a deficit in metallicity is indicated at $z>8$. 
We therefore conclude that there is no critical impact of the low-mass end galaxies and the extrapolation of the relation at least in the range $\mu_{0.66}\sim 7-7.5$ for the discussion of the evolution of the SFR-MZ relation.

\section{Summary of Properties for JWST Objects in a Tabular Format}
\label{sec_app:tbl_all}

In Table \ref{tbl:objects_all}, we have compiled the essential properties of the complete sample of JWST objects presented in this paper, based on our improved reduction and calibration of the NIRSpec data from ERO, ERS-GLASS, and ERS-CEERS.

\movetabledown=45pt
\begin{longrotatetable}
\begin{deluxetable*}{lcccccccccccccl}
\tablecaption{Summary of the full JWST Sample analyzed in this paper
\label{tbl:objects_all}}
\tabletypesize{\scriptsize}
\tablehead{
\colhead{ID}
& \colhead{R.A.}
& \colhead{Decl.}
& \colhead{Redshift}
& \colhead{\Muv}
& \colhead{log \Mstar}
& \colhead{log SFR}
& \colhead{EW(H$\beta$)}
& \colhead{R3}
& \colhead{R23}
& \colhead{O32}
& \colhead{log(O/H)}
& \multicolumn{2}{c}{Flags}
& \colhead{Note} \\ 
\cline{13-14}
 & (deg)
& (deg)
& 
& (mag)
& (\Msun)
& (\Msun\,yr$^{-1}$)
& (\AA)
& ($\star$)
& ($\star$)
& ($\star$)
& +12
& P($\dag$)
& M($\ddag$)
& \multicolumn{1}{c}{($\P$)}
} 
\startdata
ERO\_04590 & $110.8593287$ & $-73.4491656$ & $8.496$ & $-17.95^{+0.23}_{-0.23}$ & $7.60^{+0.63}_{-0.51}$ & $0.66^{+0.03}_{-0.03}$ & $218 \pm 150$ & $3.4 \pm 0.2$ & $4.5 \pm 0.3$ & $>14.8$ & $7.26^{+0.15}_{-0.13}$ & 1 & \textit{direct} & $\mu=8.70$ \\ 
ERO\_05144 & $110.8396739$ & $-73.4453570$ & $6.378$ & $-18.94^{+0.34}_{-0.34}$ & $8.55^{+0.18}_{-1.03}$ & $0.65^{+0.01}_{-0.01}$ & $151 \pm 51$ & $7.2 \pm 0.2$ & $9.9 \pm 0.3$ & $18.6 \pm 3.3$ & $7.82^{+0.12}_{-0.09}$ & 1 & \textit{direct} & $\mu=3.18$ \\ 
ERO\_06355 & $110.8445942$ & $-73.4350590$ & $7.665$ & $-20.33^{+0.00}_{-0.00}$ & $8.77^{+0.08}_{-0.01}$ & $1.41^{+0.01}_{-0.01}$ & $150 \pm 4$ & $8.2 \pm 0.2$ & $11.8 \pm 0.3$ & $8.2 \pm 0.3$ & $8.35^{+0.11}_{-0.08}$ & 1 & \textit{direct} & $\mu=1.78$ \\ 
ERO\_08140 & $110.7880022$ & $-73.4618680$ & $5.275$ & $-19.63^{+0.00}_{-0.00}$ & $9.45^{+0.08}_{-0.68}$ & $0.54^{+0.04}_{-0.04}$ & $19 \pm 3$ & $4.0 \pm 0.4$ & $6.8 \pm 0.7$ & $2.9 \pm 0.2$ & $8.40^{+0.15}_{-0.15}$ & 1 & \textit{R23} & $\mu=3.23$ \\ 
ERO\_10612 & $110.8339649$ & $-73.4345232$ & $7.660$ & $-20.07^{+0.08}_{-0.08}$ & $7.78^{+0.29}_{-0.03}$ & $1.14^{+0.01}_{-0.02}$ & $210 \pm 16$ & $7.1 \pm 0.3$ & $9.8 \pm 0.4$ & $21.7 \pm 4.0$ & $7.59^{+0.08}_{-0.08}$ & 1 & \textit{direct} & $\mu=1.86$ \\ 
GLASS\_10000 & $3.6013389$ & $-30.3792254$ & $7.881$ & $-20.36^{+0.16}_{-0.14}$ & $8.16^{+0.27}_{-0.07}$ & $1.25^{+0.03}_{-0.04}$ & $195 \pm 32$ & \nodata & \nodata & \nodata & \nodata & 1 & \nodata & $\mu=1.38$ \\ 
GLASS\_100001 & $3.6038450$ & $-30.3822348$ & $7.874$ & $-20.29^{+0.50}_{-0.29}$ & $9.15^{+0.55}_{0.00}$ & $0.95^{+0.10}_{-0.13}$ & $40 \pm 11$ & $4.7 \pm 1.3$ & $7.8 \pm 2.1$ & $2.8 \pm 0.4$ & $8.31^{+0.24}_{-0.20}$ & 1 & \textit{R23} & $\mu=1.41$ \\ 
GLASS\_100003 & $3.6045089$ & $-30.3804439$ & $7.877$ & $-20.69^{+0.10}_{-0.17}$ & $8.27^{+0.38}_{-0.04}$ & $1.22^{+0.04}_{-0.05}$ & $139 \pm 19$ & $7.8 \pm 0.8$ & $10.8 \pm 1.1$ & $20.4 \pm 5.9$ & $7.64^{+0.23}_{-0.19}$ & 1 & \textit{direct} & $\mu=1.34$ \\ 
GLASS\_100005 & $3.6064583$ & $-30.3809939$ & $7.879$ & $-20.06^{+0.45}_{-0.27}$ & $8.38^{+0.43}_{-0.13}$ & $1.25^{+0.13}_{-0.18}$ & $252 \pm 108$ & \nodata & \nodata & \nodata & \nodata & 1 & \nodata & $\mu=1.33$ \\ 
GLASS\_10021 & $3.6085111$ & $-30.4185405$ & $7.286$ & $-21.44^{+0.34}_{-0.28}$ & $8.51^{+0.30}_{-0.22}$ & $1.09^{+0.02}_{-0.02}$ & $68 \pm 27$ & $8.3 \pm 0.4$ & $11.7 \pm 0.6$ & $12.1 \pm 0.9$ & $7.87^{+0.12}_{-0.10}$ & 1 & \textit{direct} & $\mu=1.72$ \\ 
GLASS\_150008 & $3.6025299$ & $-30.4192324$ & $6.229$ & $-19.71^{+1.01}_{-0.15}$ & $8.23^{+0.49}_{-0.15}$ & $0.63^{+0.02}_{-0.02}$ & $86 \pm 26$ & $6.4 \pm 0.4$ & \nodata & \nodata & $8.17^{+0.23}_{-0.23}$ & 1 & \textit{R3} & $\mu=2.06$ \\ 
GLASS\_50038 & $3.5651997$ & $-30.3942643$ & $5.772$ & $-20.12^{+1.36}_{-0.19}$ & $9.09^{+0.39}_{-0.09}$ & $0.91^{+0.04}_{-0.05}$ & $36 \pm 7$ & $7.1 \pm 0.7$ & $11.2 \pm 1.2$ & $3.9 \pm 0.9$ & $8.10^{+0.16}_{-0.15}$ & 1 & \textit{R23} & $\mu=1.76$ \\ 
GLASS\_110000 & $3.5706428$ & $-30.4146381$ & $5.763$ & $-19.91^{+0.64}_{-0.34}$ & $9.22^{+0.16}_{-0.27}$ & $0.53^{+0.04}_{-0.05}$ & $31 \pm 10$ & $6.6 \pm 0.7$ & $9.6 \pm 1.1$ & $6.4 \pm 1.1$ & $8.11^{+0.15}_{-0.15}$ & 1 & \textit{R23} & $\mu=1.40$ \\ 
GLASS\_160122 & $3.5649015$ & $-30.4249568$ & $5.331$ & \nodata & \nodata & \nodata & \nodata & \nodata & \nodata & \nodata & \nodata & \nodata & \nodata & $\mu=1.19$ \\ 
GLASS\_50002 & $3.5770012$ & $-30.4155218$ & $5.133$ & $-19.44^{+0.65}_{-0.44}$ & $8.87^{+0.29}_{-0.08}$ & $0.47^{+0.18}_{-0.31}$ & $47 \pm 25$ & $>8.0$ & $>12.6$ & $4.4 \pm 1.2$ & \nodata & 1 & \nodata & $\mu=1.70$ \\ 
GLASS\_80070 & $3.5823211$ & $-30.3876547$ & $4.797$ & $-17.77^{+1.01}_{-0.19}$ & $7.42^{+0.70}_{-0.04}$ & $0.17^{+0.04}_{-0.05}$ & $100 \pm 14$ & $4.6 \pm 0.5$ & $7.0 \pm 0.7$ & $4.6 \pm 0.6$ & $7.80^{+0.19}_{-0.18}$ & 1 & \textit{R23} & $\mu=7.91$ \\ 
GLASS\_150029 & $3.5771664$ & $-30.4225760$ & $4.584$ & $-19.20^{+0.10}_{-1.08}$ & $9.12^{+0.03}_{-0.33}$ & $1.04^{+0.02}_{-0.02}$ & $138 \pm 10$ & $6.3 \pm 0.3$ & $8.9 \pm 0.4$ & $12.0 \pm 1.0$ & $7.70^{+0.09}_{-0.08}$ & 1 & \textit{direct} & $\mu=1.48$ , \textit{Broad} \\ 
GLASS\_40066 & $3.5997164$ & $-30.4318948$ & $4.020$ & $-20.43^{+1.86}_{-0.26}$ & $9.40^{+0.33}_{-0.10}$ & $1.57^{+0.02}_{-0.02}$ & $84 \pm 18$ & $7.0 \pm 0.3$ & \nodata & \nodata & $8.15^{+0.23}_{-0.23}$ & 1 & \textit{R3} & $\mu=1.40$ \\ 
GLASS\_160133 & $3.5802754$ & $-30.4244041$ & $4.015$ & $-18.94^{+0.08}_{-0.09}$ & $8.11^{+0.41}_{-0.22}$ & $1.16^{+0.01}_{-0.01}$ & $227 \pm 89$ & $7.7 \pm 0.1$ & $10.7 \pm 0.2$ & $13.3 \pm 0.6$ & $7.95^{+0.06}_{-0.05}$ & 1 & \textit{direct} & $\mu=1.54$ , \textit{Broad} \\ 
GLASS\_80029 & $3.6031803$ & $-30.4157094$ & $3.951$ & $-18.92^{+1.00}_{-0.50}$ & $8.50^{+0.22}_{-0.49}$ & $0.47^{+0.04}_{-0.05}$ & $62 \pm 25$ & $6.5 \pm 0.7$ & \nodata & \nodata & $8.16^{+0.23}_{-0.23}$ & 1 & \textit{R3} & $\mu=2.39$ \\ 
CEERS\_00003 & $215.0051890$ & $52.9965800$ & $8.005$ & $-18.76^{+1.02}_{-0.52}$ & $8.61^{+0.48}_{-0.43}$ & \nodata & $285 \pm 202$ & $6.3 \pm 2.1$ & \nodata & \nodata & $7.80^{+0.31}_{-0.44}$ & 1 & \textit{R3} & \\ 
CEERS\_00007 & $215.0117060$ & $52.9883030$ & $8.873$ & $-20.31^{+0.35}_{-0.32}$ & $8.01^{+0.23}_{-0.05}$ & \nodata & $102 \pm 70$ & $>2.0$ & \nodata & \nodata & \nodata & 1 & \nodata & \\ 
CEERS\_00044 & $215.0011150$ & $53.0112690$ & $7.104$ & $-19.38^{+0.13}_{-0.33}$ & $7.94^{+0.28}_{-0.42}$ & \nodata & $86 \pm 52$ & $>9.6$ & \nodata & \nodata & \nodata & 1 & \nodata & \\ 
CEERS\_00067 & $215.0155970$ & $53.0118570$ & $6.212$ & $-15.87^{+0.15}_{-0.38}$ & $6.11^{+0.35}_{-0.05}$ & \nodata & $578 \pm 225$ & $>3.3$ & \nodata & \nodata & \nodata & 1 & \nodata & \\ 
CEERS\_00314 & $214.8689530$ & $52.8772060$ & $5.270$ & $-19.04^{+0.40}_{-0.08}$ & $8.72^{+0.09}_{-0.22}$ & \nodata & $69 \pm 58$ & $>4.2$ & \nodata & \nodata & $7.95^{+0.32}_{-0.32}$ & 1 & \textit{R3} & \\ 
CEERS\_00323 & $214.8725560$ & $52.8759490$ & $5.668$ & $-18.70^{+0.04}_{-0.10}$ & $7.23^{+0.06}_{-0.05}$ & $0.78^{+0.03}_{-0.04}$ & $417 \pm 43$ & $3.9 \pm 0.3$ & $5.1 \pm 0.4$ & $>1.4$ & $7.32^{+0.15}_{-0.15}$ & 1 & \textit{R23} & $N^o=2$ \\ 
CEERS\_00355 & $214.8064820$ & $52.8788270$ & $6.105$ & $-19.79^{+0.19}_{-0.47}$ & $7.83^{+0.64}_{-0.07}$ & $0.95^{+0.05}_{-0.05}$ & $144 \pm 24$ & $6.1 \pm 0.8$ & $9.1 \pm 1.2$ & $7.1 \pm 2.0$ & $7.93^{+0.19}_{-0.19}$ & 1 & \textit{R23} & $N^o=2$ \\ 
CEERS\_00356 & $214.8052900$ & $52.8778610$ & $5.641$ & $-18.26^{+0.63}_{-0.42}$ & $8.41^{+0.17}_{-0.26}$ & \nodata & $28 \pm 30$ & $>2.7$ & \nodata & \nodata & \nodata & 1 & \nodata & \\ 
CEERS\_00362 & $214.8126890$ & $52.8815360$ & $6.051$ & $-18.55^{+0.03}_{-0.28}$ & $7.34^{+0.31}_{-0.17}$ & $0.49^{+0.08}_{-0.11}$ & $252 \pm 99$ & $4.8 \pm 1.1$ & $6.4 \pm 1.5$ & $>6.3$ & $7.48^{+0.22}_{-0.21}$ & 1 & \textit{R23} & \\ 
CEERS\_00381 & $214.8196710$ & $52.8797550$ & $5.513$ & $-18.54^{+0.34}_{-0.89}$ & $9.03^{+0.11}_{-0.12}$ & $0.74^{+0.12}_{-0.17}$ & $81 \pm 33$ & $4.0 \pm 1.4$ & $5.5 \pm 1.9$ & $>2.8$ & $7.79^{+0.29}_{-0.28}$ & 1 & \textit{R23} & $N^o=2$ \\ 
CEERS\_00386 & $214.8321840$ & $52.8850830$ & $6.618$ & $-18.48^{+0.81}_{-0.32}$ & $7.60^{+0.54}_{-0.29}$ & \nodata & $38 \pm 93$ & $>5.2$ & \nodata & \nodata & $7.97^{+0.27}_{-0.27}$ & 1 & \textit{R3} & $N^o=2$ \\ 
CEERS\_00397 & $214.8361970$ & $52.8826930$ & $6.005$ & $-21.23^{+0.17}_{-0.24}$ & $8.45^{+0.53}_{-0.06}$ & $1.76^{+0.02}_{-0.03}$ & $139 \pm 18$ & $7.5 \pm 0.5$ & $11.0 \pm 0.7$ & $6.7 \pm 0.9$ & $7.87^{+0.14}_{-0.14}$ & 1 & \textit{R23} & $N^o=2$ , \textit{Broad} \\ 
CEERS\_00403 & $214.8289680$ & $52.8757010$ & $5.766$ & $-20.30^{+0.69}_{-0.28}$ & $9.35^{+0.27}_{-0.09}$ & $1.20^{+0.06}_{-0.07}$ & $34 \pm 8$ & $5.7 \pm 0.9$ & $10.9 \pm 1.7$ & $1.6 \pm 0.2$ & $8.11^{+0.18}_{-0.17}$ & 1 & \textit{R23} & $N^o=2$ \\ 
CEERS\_00407 & $214.8393160$ & $52.8825650$ & $7.029$ & $-18.55^{+0.11}_{-0.48}$ & $7.61^{+0.81}_{-0.28}$ & $0.54^{+0.09}_{-0.12}$ & $219 \pm 117$ & $6.5 \pm 1.6$ & $8.5 \pm 2.1$ & $>6.5$ & $7.71^{+0.36}_{-0.28}$ & 1 & \textit{R23} & \\ 
CEERS\_00439 & $214.8253640$ & $52.8630650$ & $7.179$ & $-19.28^{+0.08}_{-0.22}$ & $7.45^{+0.40}_{-0.05}$ & \nodata & $233 \pm 42$ & $5.9 \pm 1.0$ & \nodata & \nodata & $7.71^{+0.37}_{-0.29}$ & 1 & \textit{R3} & \\ 
CEERS\_00498 & $214.8130450$ & $52.8342490$ & $7.179$ & $-20.20^{+0.05}_{-0.22}$ & $7.82^{+0.32}_{-0.05}$ & \nodata & $232 \pm 58$ & $4.5 \pm 1.2$ & \nodata & \nodata & $7.47^{+0.30}_{-0.29}$ & 1 & \textit{R3} & \\ 
CEERS\_00515 & $214.8785340$ & $52.8741420$ & $5.664$ & $-16.89^{+0.98}_{-2.12}$ & $8.48^{+0.21}_{-0.36}$ & $0.65^{+0.07}_{-0.09}$ & $96 \pm 41$ & $5.0 \pm 0.9$ & $6.6 \pm 1.3$ & \nodata & $8.41^{+0.19}_{-0.17}$ & 1 & \textit{R23} & \\ 
CEERS\_00534 & $214.8591170$ & $52.8536400$ & $7.115$ & $-20.10^{+0.51}_{-0.24}$ & $8.72^{+0.28}_{-0.40}$ & \nodata & \nodata & \nodata & \nodata & \nodata & \nodata & 1 & \nodata & \\ 
CEERS\_00542 & $214.8316240$ & $52.8315050$ & $7.064$ & $-19.89^{+0.38}_{-0.06}$ & $8.25^{+0.07}_{-0.10}$ & \nodata & $14 \pm 26$ & $>2.4$ & \nodata & \nodata & \nodata & 1 & \nodata & \\ 
CEERS\_00545 & $214.8644110$ & $52.8536570$ & $5.668$ & $-20.86^{+1.32}_{0.00}$ & $9.03^{+0.58}_{-0.11}$ & $1.34^{+0.07}_{-0.09}$ & $72 \pm 21$ & $5.7 \pm 1.1$ & $8.7 \pm 1.7$ & $4.8 \pm 1.2$ & $7.93^{+0.19}_{-0.19}$ & 1 & \textit{R23} & \\ 
CEERS\_00577 & $214.8928610$ & $52.8651570$ & $6.695$ & $-18.35^{+0.70}_{-0.43}$ & $7.46^{+0.56}_{-0.23}$ & $0.46^{+0.10}_{-0.12}$ & $260 \pm 124$ & $2.9 \pm 0.8$ & $4.0 \pm 1.1$ & \nodata & $7.20^{+0.19}_{-0.20}$ & 1 & \textit{R23} & \\ 
CEERS\_00603 & $214.8672470$ & $52.8367370$ & $6.059$ & $-19.98^{+0.17}_{-1.06}$ & $9.10^{+0.10}_{-0.51}$ & $1.19^{+0.12}_{-0.16}$ & $251 \pm 69$ & $6.4 \pm 1.1$ & $9.0 \pm 1.5$ & $11.5 \pm 3.2$ & $7.82^{+0.19}_{-0.19}$ & 1 & \textit{R23} & $N^o=3$ \\ 
CEERS\_00618 & $214.8764690$ & $52.8394120$ & $6.055$ & $-19.57^{+0.18}_{-0.10}$ & $8.06^{+0.06}_{-0.06}$ & $0.81^{+0.11}_{-0.14}$ & $127 \pm 45$ & $3.7 \pm 1.1$ & $5.0 \pm 1.2$ & $>2.7$ & $7.88^{+0.28}_{-0.27}$ & 1 & \textit{R23} & $N^o=4$ \\ 
CEERS\_00669 & $214.8963810$ & $52.8385140$ & $5.273$ & $-19.55^{+0.93}_{-0.38}$ & $8.76^{+0.41}_{-0.23}$ & \nodata & $34 \pm 26$ & $>3.5$ & \nodata & \nodata & $8.18^{+0.23}_{-0.23}$ & 1 & \textit{R3} & \\ 
CEERS\_00670 & $214.9036810$ & $52.8449140$ & $5.804$ & $-18.05^{+0.53}_{-0.35}$ & $7.66^{+0.73}_{-0.05}$ & $1.11^{+0.12}_{-0.17}$ & $141 \pm 46$ & $7.5 \pm 2.4$ & $10.7 \pm 3.5$ & $>3.0$ & $7.88^{+0.22}_{-0.22}$ & 1 & \textit{R23} & \\ 
CEERS\_00672 & $214.8896770$ & $52.8329770$ & $5.666$ & $-16.71^{+0.31}_{-1.16}$ & $9.03^{+0.05}_{-0.18}$ & $1.20^{+0.11}_{-0.15}$ & $37 \pm 11$ & $2.7 \pm 0.8$ & $5.2 \pm 1.8$ & $>0.8$ & $8.57^{+0.30}_{-0.23}$ & 1 & \textit{R23} & \textit{Broad} \\ 
CEERS\_00686 & $215.1508619$ & $52.9895618$ & $7.752$ & $>-20.42$ & $<8.44$ & \nodata & \nodata & $5.7 \pm 0.9$ & \nodata & \nodata & \nodata & 0 & \nodata & \\ 
CEERS\_00689 & $214.9990525$ & $52.9419767$ & $7.549$ & $>-20.93$ & $<8.70$ & \nodata & \nodata & $7.0 \pm 1.0$ & \nodata & \nodata & $8.00^{+0.24}_{-0.24}$ & 0 & \textit{R3} & $N^o=2$ \\ 
CEERS\_00698 & $215.0503166$ & $53.0074406$ & $7.471$ & $-21.60^{+0.11}_{-0.11}$ & $9.39^{+0.17}_{-0.17}$ & $1.71^{+0.14}_{-0.21}$ & \nodata & $7.9 \pm 0.5$ & $11.1 \pm 0.8$ & $14.9 \pm 3.6$ & $7.87^{+0.14}_{-0.15}$ & 0 & \textit{R23} & \\ 
CEERS\_00707 & $214.9919080$ & $52.9250450$ & $4.899$ & $-19.74^{+0.20}_{-0.20}$ & $8.76^{+1.09}_{-1.09}$ & $1.36^{+0.16}_{-0.26}$ & \nodata & $9.0 \pm 1.7$ & $12.0 \pm 3.4$ & $6.0 \pm 1.2$ & $7.92^{+0.19}_{-0.24}$ & 0 & \textit{R23} & $N^o=2$ \\ 
CEERS\_00716 & $215.0803487$ & $52.9932406$ & $6.961$ & $-21.65^{+0.17}_{-0.17}$ & $9.42^{+0.18}_{-0.18}$ & \nodata & \nodata & \nodata & \nodata & \nodata & \nodata & 0 & \nodata & \\ 
CEERS\_00717 & $215.0814058$ & $52.9721795$ & $6.934$ & $-21.49^{+0.12}_{-0.12}$ & $9.34^{+0.17}_{-0.17}$ & $1.72^{+0.15}_{-0.23}$ & \nodata & $7.4 \pm 1.3$ & $11.1 \pm 2.0$ & $7.2 \pm 2.0$ & $7.87^{+0.17}_{-0.18}$ & 0 & \textit{R23} & $N^o=4$ , \textit{Broad} \\ 
CEERS\_00746 & $214.8091416$ & $52.8684835$ & $5.626$ & $-17.82^{+0.33}_{-0.61}$ & $9.30^{+0.05}_{-0.14}$ & $0.92^{+0.11}_{-0.14}$ & $30 \pm 11$ & $5.8 \pm 1.7$ & $6.8 \pm 2.0$ & $>1.0$ & $8.29^{+0.24}_{-0.22}$ & 1 & \textit{R23} & $N^o=2$ , \textit{Broad} \\ 
CEERS\_00792 & $214.8717663$ & $52.8331673$ & $6.257$ & $-18.84^{+0.24}_{-0.33}$ & $8.54^{+0.24}_{-0.66}$ & $0.97^{+0.07}_{-0.08}$ & $144 \pm 65$ & $9.2 \pm 1.6$ & $13.4 \pm 2.3$ & $5.4 \pm 1.3$ & $7.98^{+0.16}_{-0.18}$ & 1 & \textit{R23} & \\ 
CEERS\_00829 & $214.8615939$ & $52.8761590$ & $7.167$ & $-19.57^{+0.06}_{-0.10}$ & $7.58^{+0.23}_{-0.06}$ & \nodata & $212 \pm 96$ & $>3.7$ & \nodata & \nodata & $7.92^{+0.37}_{-0.37}$ & 1 & \textit{R3} & \\ 
CEERS\_00933 & $214.8337227$ & $52.8682822$ & $4.243$ & $-20.64^{+0.09}_{-0.30}$ & $8.13^{+0.54}_{-0.05}$ & $1.78^{+0.05}_{-0.06}$ & $171 \pm 25$ & $8.8 \pm 1.2$ & $12.8 \pm 1.8$ & $>6.7$ & $7.96^{+0.16}_{-0.17}$ & 1 & \textit{R23} & \\ 
CEERS\_01019 & $215.0353914$ & $52.8906618$ & $8.681$ & $-22.44^{+0.45}_{-0.17}$ & $10.12^{+0.12}_{-0.11}$ & $1.97^{+0.04}_{-0.05}$ & $81 \pm 11$ & $8.1 \pm 1.0$ & $11.0 \pm 1.3$ & $13.6 \pm 1.6$ & $8.04^{+0.17}_{-0.17}$ & 1 & \textit{R23} & $N^o=2$ \\ 
CEERS\_01023 & $215.1884129$ & $53.0336473$ & $7.779$ & $-21.06^{+0.17}_{-0.17}$ & $8.76^{+0.33}_{-0.33}$ & $1.49^{+0.15}_{-0.22}$ & \nodata & $3.1 \pm 0.8$ & $4.2 \pm 1.0$ & $>2.8$ & $7.61^{+0.22}_{-0.23}$ & 0 & \textit{R23} & $N^o=2$ \\ 
CEERS\_01025 & $214.9675468$ & $52.9329530$ & $8.714$ & $>-20.61$ & $<8.54$ & $<1.31$ & \nodata & $5.3 \pm 1.2$ & $7.4 \pm 1.7$ & $8.3 \pm 2.5$ & $7.64^{+0.18}_{-0.20}$ & 0 & \textit{R23} & \\ 
CEERS\_01027 & $214.8829941$ & $52.8404159$ & $7.825$ & $-20.73^{+0.12}_{-0.14}$ & $8.12^{+0.34}_{-0.05}$ & $1.23^{+0.03}_{-0.03}$ & $175 \pm 14$ & $7.4 \pm 0.5$ & $9.9 \pm 0.7$ & $21.5 \pm 5.5$ & $7.83^{+0.20}_{-0.15}$ & 1 & \textit{direct} & $N^o=4$ \\ 
CEERS\_01029 & $215.2187624$ & $53.0698619$ & $8.612$ & $-21.14^{+0.14}_{-0.14}$ & $8.80^{+0.33}_{-0.33}$ & \nodata & \nodata & $4.8 \pm 1.4$ & \nodata & \nodata & $8.19^{+0.25}_{-0.24}$ & 0 & \textit{R3} & $N^o=2$ \\ 
CEERS\_01038 & $215.0396971$ & $52.9015971$ & $7.197$ & $-19.24^{+0.18}_{-0.12}$ & $8.38^{+0.12}_{-0.05}$ & $0.74^{+0.08}_{-0.09}$ & $87 \pm 48$ & $5.1 \pm 1.0$ & $7.6 \pm 1.6$ & $>4.0$ & $7.84^{+0.25}_{-0.28}$ & 1 & \textit{R23} & $N^o=2$ \\ 
CEERS\_01064 & $215.1771672$ & $53.0489751$ & $6.794$ & $>-20.09$ & $<8.63$ & $<1.10$ & \nodata & $6.7 \pm 2.1$ & \nodata & \nodata & \nodata & 0 & \nodata & $N^o=2$ \\ 
CEERS\_01065 & $215.1168542$ & $53.0010814$ & $6.190$ & $>-20.31$ & $<8.91$ & $<2.00$ & \nodata & $10.8 \pm 3.2$ & $17.6 \pm 5.1$ & $2.5 \pm 0.3$ & $7.99^{+0.15}_{-0.15}$ & 0 & \textit{R23} & \\ 
CEERS\_01102 & $215.0910473$ & $52.9542849$ & $6.994$ & $>-20.32$ & $<8.75$ & \nodata & \nodata & $>4.9$ & \nodata & \nodata & $7.97^{+0.28}_{-0.28}$ & 0 & \textit{R3} & \\ 
CEERS\_01115 & $215.1628177$ & $53.0730973$ & $6.300$ & \nodata & \nodata & \nodata & \nodata & $5.6 \pm 0.7$ & $7.1 \pm 0.8$ & $>3.5$ & $7.62^{+0.15}_{-0.16}$ & \nodata & \textit{R23} & \\ 
CEERS\_01142 & $215.0607164$ & $52.9587084$ & $6.957$ & $-21.64^{+0.14}_{-0.14}$ & $9.41^{+0.17}_{-0.17}$ & \nodata & \nodata & $>2.7$ & \nodata & \nodata & $7.98^{+0.26}_{-0.26}$ & 0 & \textit{R3} & \\ 
CEERS\_01143 & $215.0770064$ & $52.9695042$ & $6.930$ & $>-20.24$ & $<8.71$ & $<1.66$ & \nodata & $8.4 \pm 1.3$ & $11.9 \pm 1.9$ & $8.3 \pm 2.1$ & $7.91^{+0.16}_{-0.17}$ & 0 & \textit{R23} & $N^o=2$ \\ 
CEERS\_01149 & $215.0897143$ & $52.9661828$ & $8.179$ & $-20.83^{+0.19}_{-0.19}$ & $8.65^{+0.33}_{-0.33}$ & $1.40^{+0.15}_{-0.22}$ & \nodata & $7.2 \pm 0.9$ & $10.1 \pm 1.3$ & $13.4 \pm 3.4$ & $7.82^{+0.16}_{-0.16}$ & 0 & \textit{R23} & $N^o=2$ \\ 
CEERS\_01160 & $214.8050474$ & $52.8458770$ & $6.566$ & $>-20.15$ & $<8.67$ & $<1.13$ & \nodata & $5.3 \pm 0.9$ & $8.5 \pm 1.4$ & $3.7 \pm 0.9$ & $8.26^{+0.18}_{-0.17}$ & 0 & \textit{R23} & \\ 
CEERS\_01163 & $214.9904678$ & $52.9719902$ & $7.451$ & $>-20.78$ & $<8.98$ & \nodata & \nodata & $>6.1$ & \nodata & \nodata & $7.99^{+0.24}_{-0.24}$ & 0 & \textit{R3} & $N^o=2$ \\ 
CEERS\_01173 & $215.1542076$ & $52.9558470$ & $4.997$ & $-20.98^{+0.11}_{-0.11}$ & $9.38^{+1.12}_{-1.12}$ & $2.26^{+0.14}_{-0.21}$ & \nodata & $7.0 \pm 1.5$ & $12.6 \pm 2.7$ & $1.7 \pm 0.2$ & $8.01^{+0.23}_{-0.20}$ & 0 & \textit{R23} & $N^o=2$ \\ 
CEERS\_01207 & $214.9600055$ & $52.8311713$ & $4.897$ & $-20.68^{+0.04}_{-0.08}$ & $8.81^{+0.05}_{-0.05}$ & \nodata & $16 \pm 14$ & $>4.7$ & \nodata & \nodata & $7.96^{+0.29}_{-0.29}$ & 1 & \textit{R3} & \\ 
CEERS\_01217 & $215.2422372$ & $53.0324421$ & $4.637$ & $-21.44^{+0.07}_{-0.07}$ & $9.61^{+1.13}_{-1.13}$ & \nodata & \nodata & $>7.9$ & \nodata & \nodata & \nodata & 0 & \nodata & \\ 
CEERS\_01236 & $215.1452982$ & $52.9672786$ & $4.480$ & $-19.61^{+0.27}_{-0.23}$ & $8.91^{+0.18}_{-0.18}$ & $0.24^{+0.11}_{-0.15}$ & $34 \pm 15$ & $3.3 \pm 1.0$ & $5.4 \pm 1.7$ & $>2.9$ & $7.55^{+0.29}_{-0.28}$ & 1 & \textit{R23} & $N^o=2$ , \textit{Broad} \\ 
CEERS\_01244 & $215.2406522$ & $53.0360412$ & $4.482$ & $-19.50^{+0.18}_{-0.18}$ & $8.62^{+0.88}_{-0.88}$ & $2.36^{+0.15}_{-0.22}$ & \nodata & $5.6 \pm 1.5$ & $7.8 \pm 2.1$ & $>3.7$ & $7.67^{+0.20}_{-0.23}$ & 0 & \textit{R23} & $N^o=2$ , \textit{Broad} \\ 
CEERS\_01267 & $215.2220383$ & $53.0273690$ & $5.152$ & $>-19.12$ & $<8.45$ & \nodata & \nodata & $>3.0$ & \nodata & \nodata & $7.98^{+0.26}_{-0.26}$ & 0 & \textit{R3} & $N^o=2$ \\ 
CEERS\_01289 & $214.9476341$ & $52.8370980$ & $4.880$ & $-20.12^{+0.04}_{-0.07}$ & $8.56^{+0.05}_{-0.05}$ & $0.79^{+0.07}_{-0.09}$ & $60 \pm 11$ & $5.1 \pm 1.0$ & $8.5 \pm 1.6$ & $3.4 \pm 0.6$ & $8.12^{+0.15}_{-0.37}$ & 1 & \textit{R23} & \\ 
CEERS\_01294 & $214.9552513$ & $52.8428999$ & $4.999$ & $-20.86^{+0.12}_{-0.08}$ & $8.89^{+0.14}_{-0.07}$ & $1.44^{+0.09}_{-0.11}$ & $43 \pm 11$ & $11.2 \pm 2.5$ & $17.0 \pm 3.9$ & $3.8 \pm 0.8$ & $7.98^{+0.14}_{-0.14}$ & 1 & \textit{R23} & \\ 
CEERS\_01324 & $215.0305739$ & $52.9026048$ & $5.007$ & $-19.94^{+0.11}_{-0.07}$ & $8.53^{+0.05}_{-0.05}$ & $0.66^{+0.10}_{-0.13}$ & $62 \pm 16$ & $4.5 \pm 1.2$ & $5.8 \pm 1.5$ & $>2.2$ & $7.61^{+0.28}_{-0.26}$ & 1 & \textit{R23} & \\ 
CEERS\_01334 & $214.7683564$ & $52.7176411$ & $5.503$ & $>-20.41$ & $<8.96$ & $<2.08$ & \nodata & \nodata & \nodata & \nodata & \nodata & 0 & \nodata & \\ 
CEERS\_01358 & $215.0909401$ & $52.9515310$ & $5.504$ & $-19.69^{+0.05}_{-0.05}$ & $8.39^{+0.05}_{-0.05}$ & $0.95^{+0.10}_{-0.14}$ & $39 \pm 11$ & $9.6 \pm 2.6$ & $14.3 \pm 4.0$ & $>3.7$ & $8.06^{+0.24}_{-0.24}$ & 1 & \textit{R3} & \\ 
CEERS\_01365 & $215.0268968$ & $52.9072017$ & $4.307$ & $-19.73^{+0.32}_{-0.35}$ & $9.05^{+0.08}_{-0.32}$ & $1.32^{+0.07}_{-0.08}$ & $238 \pm 56$ & $4.2 \pm 0.8$ & $7.5 \pm 1.4$ & $2.4 \pm 0.6$ & $8.34^{+0.19}_{-0.17}$ & 1 & \textit{R23} & \\ 
CEERS\_01374 & $214.9439110$ & $52.8500419$ & $4.999$ & $-21.48^{+0.07}_{-0.11}$ & $9.17^{+0.18}_{-0.06}$ & $1.81^{+0.04}_{-0.05}$ & $131 \pm 27$ & $7.2 \pm 1.1$ & $10.6 \pm 1.6$ & $5.2 \pm 0.6$ & $7.90^{+0.15}_{-0.16}$ & 1 & \textit{R23} & $N^o=2$ \\ 
CEERS\_01388 & $215.0001061$ & $52.8910612$ & $4.548$ & $>-19.70$ & $<8.74$ & $<1.36$ & \nodata & $3.4 \pm 1.0$ & $7.8 \pm 2.3$ & $0.9 \pm 0.2$ & $8.31^{+0.26}_{-0.21}$ & 0 & \textit{R23} & \\ 
CEERS\_01395 & $215.2349864$ & $53.0572423$ & $4.283$ & $-20.46^{+0.13}_{-0.13}$ & $9.14^{+0.90}_{-0.90}$ & $1.93^{+0.14}_{-0.21}$ & \nodata & $7.7 \pm 2.2$ & $11.1 \pm 3.1$ & $5.0 \pm 1.1$ & $7.87^{+0.20}_{-0.24}$ & 0 & \textit{R23} & \\ 
CEERS\_01400 & $215.1161048$ & $52.9741843$ & $4.492$ & $-19.52^{+0.18}_{-0.11}$ & $8.45^{+0.17}_{-0.05}$ & $0.88^{+0.11}_{-0.14}$ & $65 \pm 19$ & $8.8 \pm 2.5$ & $10.9 \pm 3.1$ & $>2.5$ & $8.11^{+0.25}_{-0.21}$ & 1 & \textit{R23} & $N^o=2$ \\ 
CEERS\_01401 & $215.2458005$ & $53.0652953$ & $5.375$ & $>-20.06$ & $<8.92$ & $<1.18$ & \nodata & $7.1 \pm 1.2$ & $11.1 \pm 1.9$ & $4.4 \pm 0.6$ & $7.92^{+0.20}_{-0.21}$ & 0 & \textit{R23} & $N^o=2$ \\ 
CEERS\_01410 & $215.2321183$ & $53.0573686$ & $5.262$ & $>-20.04$ & $<8.91$ & \nodata & \nodata & $>5.3$ & \nodata & \nodata & $7.97^{+0.27}_{-0.27}$ & 0 & \textit{R3} & \\ 
CEERS\_01420 & $215.0928636$ & $52.9606975$ & $5.296$ & $-20.44^{+0.21}_{-0.21}$ & $9.11^{+1.11}_{-1.11}$ & $1.47^{+0.15}_{-0.23}$ & \nodata & $3.7 \pm 0.9$ & $7.0 \pm 1.7$ & $1.8 \pm 0.4$ & $8.21^{+0.26}_{-0.25}$ & 0 & \textit{R23} & $N^o=2$ \\ 
CEERS\_01433 & $215.2248759$ & $53.0555927$ & $4.481$ & $-20.47^{+0.14}_{-0.14}$ & $9.15^{+0.90}_{-0.90}$ & \nodata & \nodata & $>4.9$ & \nodata & \nodata & $7.97^{+0.28}_{-0.28}$ & 0 & \textit{R3} & \\ 
CEERS\_01449 & $215.0800050$ & $52.9567860$ & $4.756$ & $-21.18^{+0.32}_{-0.12}$ & $9.24^{+0.20}_{-0.05}$ & $1.68^{+0.05}_{-0.06}$ & $71 \pm 13$ & $5.8 \pm 1.4$ & $9.2 \pm 2.2$ & $3.8 \pm 0.6$ & $8.22^{+0.22}_{-0.19}$ & 1 & \textit{R23} & $N^o=2$ \\ 
CEERS\_01452 & $215.2261665$ & $53.0602231$ & $4.471$ & $-19.84^{+0.19}_{-0.19}$ & $8.80^{+0.89}_{-0.89}$ & $1.00^{+0.15}_{-0.22}$ & \nodata & $5.9 \pm 1.2$ & $8.1 \pm 1.7$ & $>6.8$ & $7.69^{+0.18}_{-0.20}$ & 0 & \textit{R23} & \\ 
CEERS\_01465 & $214.8880049$ & $52.8882513$ & $5.274$ & $-19.72^{+1.81}_{-0.45}$ & $8.87^{+0.48}_{-0.14}$ & $0.84^{+0.10}_{-0.12}$ & $87 \pm 30$ & $4.4 \pm 1.3$ & $7.1 \pm 2.1$ & $3.5 \pm 1.1$ & $8.25^{+0.23}_{-0.22}$ & 1 & \textit{R23} & $N^o=2$ \\ 
CEERS\_01467 & $215.0029264$ & $52.9694972$ & $4.631$ & $>-19.37$ & $<8.58$ & \nodata & \nodata & $>3.9$ & \nodata & \nodata & $7.95^{+0.34}_{-0.34}$ & 0 & \textit{R3} & \\ 
CEERS\_01477 & $215.0034914$ & $52.9695391$ & $4.631$ & $>-19.92$ & $<8.85$ & $<1.46$ & \nodata & $6.1 \pm 0.5$ & $9.0 \pm 0.8$ & $5.1 \pm 0.7$ & $7.75^{+0.15}_{-0.15}$ & 0 & \textit{R23} & \\ 
CEERS\_01518 & $215.0068020$ & $52.9650411$ & $6.108$ & $-21.07^{+0.16}_{-0.16}$ & $9.29^{+0.86}_{-0.86}$ & $1.62^{+0.14}_{-0.22}$ & \nodata & $7.8 \pm 0.8$ & $11.0 \pm 1.1$ & $9.2 \pm 2.0$ & $7.87^{+0.15}_{-0.15}$ & 0 & \textit{R23} & \\ 
CEERS\_01534 & $214.8861409$ & $52.8769248$ & $4.588$ & $-20.02^{+0.46}_{-0.10}$ & $8.93^{+0.24}_{-0.05}$ & $1.27^{+0.08}_{-0.10}$ & $44 \pm 9$ & $8.0 \pm 1.7$ & $15.6 \pm 3.4$ & $1.5 \pm 0.3$ & $7.99^{+0.15}_{-0.15}$ & 1 & \textit{R23} & \\ 
CEERS\_01536 & $214.9772267$ & $52.9407821$ & $5.036$ & $-19.91^{+0.18}_{-0.18}$ & $8.85^{+1.09}_{-1.09}$ & $1.20^{+0.15}_{-0.22}$ & \nodata & $8.1 \pm 0.8$ & $11.5 \pm 1.1$ & $13.4 \pm 3.3$ & $7.61^{+0.18}_{-0.15}$ & 0 & \textit{direct} & $N^o=2$ \\ 
CEERS\_01539 & $214.9800779$ & $52.9426590$ & $4.888$ & $>-19.65$ & $<8.72$ & $<1.41$ & \nodata & $6.2 \pm 1.4$ & $9.0 \pm 1.9$ & $3.2 \pm 0.3$ & $7.96^{+0.19}_{-0.21}$ & 0 & \textit{R23} & $N^o=2$ \\ 
CEERS\_01544 & $215.0616558$ & $52.9995867$ & $4.192$ & $>-19.30$ & $<8.51$ & $<0.98$ & \nodata & $5.2 \pm 1.1$ & $6.5 \pm 1.4$ & $>3.2$ & $7.69^{+0.27}_{-0.28}$ & 0 & \textit{R23} & $N^o=2$ \\ 
CEERS\_01561 & $215.1660971$ & $53.0707553$ & $6.204$ & $-19.91^{+0.13}_{-0.13}$ & $8.70^{+0.83}_{-0.83}$ & $1.28^{+0.15}_{-0.22}$ & \nodata & $4.6 \pm 0.5$ & $6.4 \pm 0.7$ & $10.5 \pm 2.6$ & $7.62^{+0.15}_{-0.15}$ & 0 & \textit{R23} & $N^o=2$ \\ 
CEERS\_01565 & $215.0575020$ & $52.9937149$ & $4.792$ & $-20.28^{+0.14}_{-0.14}$ & $9.03^{+1.10}_{-1.10}$ & \nodata & \nodata & \nodata & \nodata & \nodata & \nodata & 0 & \nodata & \\ 
CEERS\_01605 & $215.0754073$ & $52.9975786$ & $4.631$ & $-20.45^{+0.13}_{-0.13}$ & $9.11^{+1.10}_{-1.10}$ & $1.55^{+0.14}_{-0.21}$ & \nodata & $5.3 \pm 0.8$ & $9.4 \pm 1.4$ & $1.9 \pm 0.3$ & $8.12^{+0.26}_{-0.26}$ & 0 & \textit{R23} & $N^o=2$ \\ 
CEERS\_01617 & $215.0713061$ & $52.9922950$ & $4.637$ & $-19.84^{+0.20}_{-0.20}$ & $8.81^{+1.09}_{-1.09}$ & $1.33^{+0.15}_{-0.23}$ & \nodata & $6.9 \pm 2.2$ & \nodata & \nodata & $7.98^{+0.26}_{-0.26}$ & 0 & \textit{R3} & $N^o=2$ \\ 
CEERS\_01620 & $215.0871730$ & $53.0028920$ & $5.303$ & $>-19.37$ & $<8.58$ & $<0.82$ & \nodata & $4.6 \pm 1.6$ & $6.3 \pm 2.1$ & $>2.9$ & $7.54^{+0.22}_{-0.28}$ & 0 & \textit{R23} & \\ 
CEERS\_01626 & $215.0710921$ & $52.9903228$ & $4.638$ & $>-19.33$ & $<8.55$ & \nodata & \nodata & $8.4 \pm 1.3$ & \nodata & \nodata & $7.90^{+0.28}_{-0.28}$ & 0 & \textit{R3} & $N^o=2$ \\ 
CEERS\_01634 & $215.0211100$ & $52.9569223$ & $5.076$ & $-22.01^{+0.06}_{-0.06}$ & $9.90^{+1.15}_{-1.15}$ & \nodata & \nodata & \nodata & \nodata & \nodata & \nodata & 0 & \nodata & \\ 
CEERS\_01651 & $215.1692172$ & $53.0547657$ & $4.382$ & $-19.93^{+0.13}_{-0.13}$ & $8.85^{+0.89}_{-0.89}$ & $1.04^{+0.14}_{-0.21}$ & \nodata & $6.5 \pm 1.2$ & $9.0 \pm 2.5$ & $>11.4$ & $7.75^{+0.20}_{-0.23}$ & 0 & \textit{R23} & $N^o=2$ \\ 
CEERS\_01658 & $214.9852372$ & $52.9242590$ & $4.610$ & $>-19.35$ & $<8.56$ & $<1.09$ & \nodata & $5.6 \pm 1.0$ & $9.1 \pm 1.2$ & $4.8 \pm 0.5$ & $7.76^{+0.16}_{-0.16}$ & 0 & \textit{R23} & $N^o=2$ \\ 
CEERS\_01665 & $215.1781971$ & $53.0593487$ & $4.488$ & $-21.66^{+0.07}_{-0.07}$ & $9.79^{+0.92}_{-0.92}$ & $2.42^{+0.14}_{-0.20}$ & \nodata & $6.8 \pm 0.3$ & $11.5 \pm 0.5$ & $2.4 \pm 0.1$ & $8.05^{+0.19}_{-0.19}$ & 0 & \textit{R23} & $N^o=2$ , \textit{Broad} \\ 
CEERS\_01672 & $214.7446370$ & $52.7503136$ & $5.206$ & $-22.00^{+0.28}_{-0.06}$ & $9.54^{+0.16}_{-0.05}$ & \nodata & $16 \pm 6$ & $>2.8$ & \nodata & \nodata & \nodata & 1 & \nodata & \\ 
CEERS\_01677 & $215.1887384$ & $53.0643782$ & $5.872$ & $-20.64^{+0.11}_{-0.11}$ & $9.07^{+0.84}_{-0.84}$ & $1.44^{+0.15}_{-0.24}$ & \nodata & $4.9 \pm 1.1$ & $8.1 \pm 1.7$ & $3.0 \pm 0.6$ & $8.29^{+0.21}_{-0.19}$ & 0 & \textit{R23} & $N^o=2$ \\ 
CEERS\_01691 & $215.1854288$ & $53.0593709$ & $5.038$ & $-21.29^{+0.12}_{-0.12}$ & $9.54^{+1.13}_{-1.13}$ & $1.95^{+0.16}_{-0.27}$ & \nodata & $6.4 \pm 1.5$ & $7.8 \pm 1.8$ & $>2.3$ & $7.67^{+0.18}_{-0.21}$ & 0 & \textit{R23} & $N^o=2$ \\ 
CEERS\_01699 & $215.0533477$ & $52.9648857$ & $5.078$ & $-19.88^{+0.18}_{-0.18}$ & $8.83^{+1.09}_{-1.09}$ & $1.16^{+0.15}_{-0.22}$ & \nodata & $6.5 \pm 0.7$ & $9.6 \pm 1.0$ & $6.0 \pm 1.0$ & $7.79^{+0.15}_{-0.15}$ & 0 & \textit{R23} & \\ 
CEERS\_01706 & $215.1195558$ & $53.0100982$ & $4.581$ & $>-19.70$ & $<8.74$ & \nodata & \nodata & $>3.8$ & \nodata & \nodata & $7.96^{+0.31}_{-0.31}$ & 0 & \textit{R3} & $N^o=2$ \\ 
CEERS\_01732 & $215.2128103$ & $53.0712112$ & $4.641$ & $-21.30^{+0.10}_{-0.10}$ & $9.54^{+1.13}_{-1.13}$ & $1.59^{+0.14}_{-0.21}$ & \nodata & $1.8 \pm 0.4$ & $4.8 \pm 0.9$ & $0.8 \pm 0.1$ & $8.62^{+0.19}_{-0.18}$ & 0 & \textit{R23} & $N^o=2$ \\ 
CEERS\_01746 & $215.0540062$ & $52.9568667$ & $4.560$ & $-19.40^{+0.20}_{-0.20}$ & $8.59^{+1.08}_{-1.08}$ & $0.83^{+0.15}_{-0.23}$ & \nodata & $6.6 \pm 0.4$ & $10.1 \pm 0.6$ & $4.8 \pm 0.4$ & $7.82^{+0.14}_{-0.14}$ & 0 & \textit{R23} & \\ 
CEERS\_01756 & $214.7533620$ & $52.7409890$ & $5.452$ & $-20.57^{+0.26}_{-0.08}$ & $8.87^{+0.22}_{-0.09}$ & \nodata & $17 \pm 10$ & $>3.9$ & \nodata & \nodata & $8.27^{+0.24}_{-0.24}$ & 1 & \textit{R3} & \\ 
CEERS\_01759 & $215.0364390$ & $52.9417510$ & $5.104$ & $-19.83^{+0.16}_{-0.16}$ & $8.81^{+1.09}_{-1.09}$ & \nodata & \nodata & $>5.0$ & \nodata & \nodata & $7.97^{+0.28}_{-0.28}$ & 0 & \textit{R3} & \\ 
CEERS\_01767 & $215.1727576$ & $53.0357884$ & $4.547$ & $-20.22^{+0.12}_{-0.12}$ & $9.00^{+1.10}_{-1.10}$ & $1.36^{+0.21}_{-0.41}$ & \nodata & $4.3 \pm 1.2$ & $5.3 \pm 1.4$ & $>3.1$ & $7.45^{+0.20}_{-0.23}$ & 0 & \textit{R23} & $N^o=2$ \\ 
CEERS\_01836 & $215.0871958$ & $52.9649047$ & $4.474$ & $-20.02^{+0.18}_{-0.18}$ & $8.90^{+0.89}_{-0.89}$ & \nodata & \nodata & $>3.2$ & \nodata & \nodata & $7.99^{+0.24}_{-0.24}$ & 0 & \textit{R3} & $N^o=2$ \\ 
CEERS\_01912 & $215.0108339$ & $53.0133270$ & $5.104$ & $-20.71^{+0.38}_{-0.66}$ & $8.67^{+0.64}_{-0.24}$ & \nodata & $65 \pm 40$ & $>5.1$ & \nodata & \nodata & $7.97^{+0.27}_{-0.27}$ & 1 & \textit{R3} & \\ 
CEERS\_01953 & $214.9982666$ & $52.9947428$ & $4.608$ & $-20.00^{+0.43}_{-0.07}$ & $8.46^{+0.29}_{-0.06}$ & \nodata & $106 \pm 42$ & $>5.4$ & \nodata & \nodata & $7.98^{+0.26}_{-0.26}$ & 1 & \textit{R3} & \\ 
CEERS\_02000 & $214.8596290$ & $52.8881300$ & $4.809$ & $-20.29^{+0.33}_{-0.67}$ & $8.97^{+0.24}_{-0.64}$ & $1.35^{+0.04}_{-0.05}$ & $152 \pm 67$ & $6.3 \pm 0.7$ & $9.5 \pm 1.0$ & $6.5 \pm 1.1$ & $7.99^{+0.15}_{-0.15}$ & 1 & \textit{R23} & \\ 
CEERS\_02036 & $215.0144487$ & $52.9939022$ & $4.645$ & $>-19.62$ & $<8.70$ & \nodata & \nodata & $>6.8$ & \nodata & \nodata & \nodata & 0 & \nodata & \\ 
CEERS\_02089 & $214.9991752$ & $52.9733008$ & $4.645$ & $-20.16^{+0.18}_{-0.18}$ & $8.97^{+1.10}_{-1.10}$ & $1.40^{+0.15}_{-0.22}$ & \nodata & $9.0 \pm 2.3$ & $11.7 \pm 3.0$ & $>9.5$ & $7.92^{+0.25}_{-0.27}$ & 0 & \textit{R23} & $N^o=2$ \\ 
CEERS\_02116 & $214.8116852$ & $52.8372405$ & $5.280$ & $-20.03^{+0.45}_{-0.63}$ & $8.64^{+0.38}_{-0.30}$ & $0.87^{+0.05}_{-0.05}$ & $98 \pm 50$ & $5.4 \pm 0.7$ & $7.6 \pm 1.0$ & $8.0 \pm 2.3$ & $7.91^{+0.30}_{-0.21}$ & 1 & \textit{R23} & \\ 
CEERS\_02123 & $214.8245795$ & $52.8457265$ & $5.289$ & $-21.24^{+0.11}_{-0.11}$ & $9.51^{+1.13}_{-1.13}$ & $1.65^{+0.14}_{-0.21}$ & \nodata & $6.8 \pm 0.8$ & $9.6 \pm 1.1$ & $9.0 \pm 1.9$ & $7.79^{+0.15}_{-0.15}$ & 0 & \textit{R23} & \\ 
CEERS\_02140 & $214.7960090$ & $52.7158781$ & $4.897$ & $>-20.24$ & $<9.01$ & $<1.16$ & \nodata & $6.1 \pm 1.2$ & \nodata & \nodata & \nodata & 0 & \nodata & \\ 
CEERS\_02168 & $215.1526021$ & $53.0570621$ & $5.658$ & $>-20.39$ & $<8.94$ & \nodata & \nodata & $>5.3$ & \nodata & \nodata & $7.97^{+0.27}_{-0.27}$ & 0 & \textit{R3} & $N^o=2$ \\ 
CEERS\_02174 & $215.0834205$ & $52.9915130$ & $5.308$ & $>-19.16$ & $<8.47$ & $<0.73$ & \nodata & $4.7 \pm 1.4$ & $8.7 \pm 2.7$ & $2.1 \pm 0.5$ & $8.12^{+0.28}_{-0.26}$ & 0 & \textit{R23} & $N^o=2$ \\ 
CEERS\_02362 & $214.8645053$ & $52.8709642$ & $5.322$ & $-18.02^{+0.66}_{-0.11}$ & $9.67^{+0.35}_{-0.58}$ & $1.80^{+0.12}_{-0.17}$ & $47 \pm 30$ & $5.8 \pm 1.9$ & $6.3 \pm 2.1$ & $>0.8$ & $8.45^{+0.28}_{-0.22}$ & 1 & \textit{R23} & \\ 
CEERS\_02782 & $214.8234525$ & $52.8302813$ & $5.251$ & $-19.50^{+0.12}_{0.00}$ & $8.53^{+0.42}_{-0.05}$ & $1.68^{+0.05}_{-0.05}$ & $157 \pm 17$ & $4.5 \pm 0.5$ & $6.0 \pm 0.6$ & $>10.0$ & $7.51^{+0.16}_{-0.16}$ & 1 & \textit{R23} & $N^o=2$ , \textit{Broad} \\ 
CEERS\_03584 & $214.9887523$ & $52.9980436$ & $4.642$ & $-20.94^{+0.96}_{-0.32}$ & $8.89^{+0.55}_{-0.19}$ & $1.39^{+0.04}_{-0.04}$ & $142 \pm 53$ & \nodata & \nodata & \nodata & \nodata & 1 & \nodata & \\ 
CEERS\_04144 & $215.1752272$ & $53.0632832$ & $4.673$ & $>-19.47$ & $<8.63$ & $<1.28$ & \nodata & $7.9 \pm 2.4$ & $12.7 \pm 3.8$ & $3.5 \pm 0.8$ & $8.10^{+0.24}_{-0.22}$ & 0 & \textit{R23} & $N^o=2$ \\ 
CEERS\_04196 & $215.1534142$ & $53.0163637$ & $5.065$ & $-20.03^{+0.14}_{-0.14}$ & $8.90^{+1.09}_{-1.09}$ & $1.50^{+0.14}_{-0.21}$ & \nodata & $5.2 \pm 1.5$ & $9.4 \pm 2.7$ & $1.9 \pm 0.4$ & $8.20^{+0.25}_{-0.21}$ & 0 & \textit{R23} & \\ 
CEERS\_04210 & $215.2372077$ & $53.0610871$ & $5.261$ & $>-20.32$ & $<9.05$ & $<1.67$ & \nodata & $7.3 \pm 1.0$ & $11.5 \pm 1.6$ & $3.5 \pm 0.5$ & $8.04^{+0.23}_{-0.22}$ & 0 & \textit{R23} & $N^o=2$ \\ 
CEERS\_08288 & $214.7703534$ & $52.7216670$ & $4.180$ & $-21.69^{+0.12}_{-0.12}$ & $9.80^{+0.92}_{-0.92}$ & $2.08^{+0.14}_{-0.21}$ & \nodata & $8.3 \pm 0.7$ & $11.2 \pm 0.9$ & $13.0 \pm 3.3$ & $7.88^{+0.15}_{-0.15}$ & 0 & \textit{R23} & \\ 
CEERS\_10995 & $215.1520322$ & $52.9740418$ & $4.482$ & $-22.29^{+0.08}_{-0.07}$ & $9.48^{+0.08}_{-0.05}$ & \nodata & $24 \pm 8$ & $>5.2$ & \nodata & \nodata & $7.97^{+0.27}_{-0.27}$ & 1 & \textit{R3} & \\ 
CEERS\_11117 & $215.2109830$ & $53.0238252$ & $4.105$ & $-19.84^{+0.13}_{-0.13}$ & $8.80^{+0.88}_{-0.88}$ & $1.00^{+0.14}_{-0.21}$ & \nodata & $6.1 \pm 1.8$ & \nodata & \nodata & \nodata & 0 & \nodata & \\ 
CEERS\_11383 & $215.0861440$ & $52.9522083$ & $5.072$ & $-22.74^{+0.44}_{-0.13}$ & $9.87^{+0.22}_{-0.07}$ & \nodata & $18 \pm 15$ & $>2.4$ & \nodata & \nodata & \nodata & 1 & \nodata & \\ 
CEERS\_11624 & $215.0724897$ & $52.9561377$ & $3.796$ & $-20.34^{+0.02}_{-0.08}$ & $8.52^{+0.08}_{-0.05}$ & $1.28^{+0.09}_{-0.12}$ & $105 \pm 25$ & $10.0 \pm 2.4$ & \nodata & \nodata & $8.05^{+0.24}_{-0.24}$ & 1 & \textit{R3} & \\ 
CEERS\_11676 & $215.0519465$ & $52.9447738$ & $4.269$ & $-19.86^{+0.13}_{-0.13}$ & $8.81^{+0.88}_{-0.88}$ & \nodata & \nodata & $>6.4$ & \nodata & \nodata & $7.99^{+0.24}_{-0.24}$ & 0 & \textit{R3} & \\ 
CEERS\_11728 & $215.0848700$ & $52.9707382$ & $3.887$ & $-20.20^{+0.11}_{-0.11}$ & $9.00^{+0.89}_{-0.89}$ & $1.15^{+0.14}_{-0.21}$ & \nodata & $6.1 \pm 0.7$ & \nodata & \nodata & \nodata & 0 & \nodata & \\ 
CEERS\_12221 & $214.7580014$ & $52.7664947$ & $4.158$ & $-20.05^{+0.21}_{-0.21}$ & $8.92^{+0.89}_{-0.89}$ & $1.28^{+0.15}_{-0.23}$ & \nodata & $4.7 \pm 1.5$ & \nodata & \nodata & \nodata & 0 & \nodata & \\ 
CEERS\_12496 & $215.0594623$ & $52.9955051$ & $4.192$ & $-19.71^{+0.22}_{-0.22}$ & $8.74^{+0.89}_{-0.89}$ & $0.95^{+0.15}_{-0.23}$ & \nodata & $2.0 \pm 0.4$ & \nodata & \nodata & $7.97^{+0.29}_{-0.29}$ & 0 & \textit{R3} & $N^o=2$ \\ 
CEERS\_14777 & $215.0228283$ & $52.9577664$ & $4.471$ & $-19.80^{+0.18}_{-0.18}$ & $8.78^{+0.89}_{-0.89}$ & \nodata & \nodata & \nodata & \nodata & \nodata & \nodata & 0 & \nodata & \\ 
CEERS\_31329 & $215.0551163$ & $53.0008499$ & $6.142$ & $-21.52^{+0.16}_{-0.16}$ & $9.51^{+0.86}_{-0.86}$ & $1.68^{+0.14}_{-0.22}$ & \nodata & \nodata & \nodata & \nodata & \nodata & 0 & \nodata & \\ 
CEERS\_37697 & $214.8914590$ & $52.8674580$ & $5.150$ & $-19.69^{+0.03}_{-0.08}$ & $8.41^{+0.12}_{-0.05}$ & \nodata & $10 \pm 10$ & $>2.2$ & \nodata & \nodata & \nodata & 1 & \nodata & \\ 
CEERS\_80072 & $214.8908500$ & $52.8139410$ & $5.268$ & $-18.21^{+0.11}_{-0.53}$ & $8.21^{+0.28}_{-0.25}$ & \nodata & $91 \pm 85$ & $>2.7$ & \nodata & \nodata & \nodata & 1 & \nodata & \\ 
CEERS\_80083 & $214.9612760$ & $52.8423640$ & $8.637$ & $-19.11^{+0.61}_{-0.02}$ & $8.94^{+0.08}_{-0.14}$ & $0.76^{+0.09}_{-0.11}$ & $105 \pm 30$ & $4.5 \pm 1.1$ & $6.1 \pm 1.5$ & $>6.9$ & $7.64^{+0.28}_{-0.25}$ & 1 & \textit{R23} & \\ 
CEERS\_80239 & $214.8960540$ & $52.8698530$ & $7.487$ & $-17.92^{+1.07}_{0.00}$ & $8.13^{+0.19}_{-0.56}$ & $0.36^{+0.11}_{-0.15}$ & $125 \pm 57$ & $4.6 \pm 1.4$ & $6.6 \pm 2.0$ & \nodata & $7.67^{+0.42}_{-0.31}$ & 1 & \textit{R23} & \\ 
CEERS\_80372 & $214.9277980$ & $52.8500030$ & $7.483$ & $-19.27^{+0.04}_{-0.03}$ & $8.19^{+0.05}_{-0.05}$ & $0.60^{+0.05}_{-0.06}$ & $104 \pm 14$ & $2.5 \pm 0.4$ & $3.4 \pm 0.5$ & $>6.6$ & $7.46^{+0.25}_{-0.25}$ & 1 & \textit{R3} & \\ 
CEERS\_80374 & $214.8980740$ & $52.8248950$ & $7.174$ & $-18.09^{+6.67}_{-2.38}$ & $9.00^{+0.07}_{-0.40}$ & \nodata & $125 \pm 36$ & $3.5 \pm 1.0$ & \nodata & \nodata & $7.61^{+0.30}_{-0.31}$ & 1 & \textit{R3} & \\ 
CEERS\_80432 & $214.8120560$ & $52.7467470$ & $7.477$ & $-20.05^{+0.04}_{-0.02}$ & $8.50^{+0.05}_{-0.05}$ & \nodata & $123 \pm 20$ & $7.0 \pm 1.2$ & \nodata & \nodata & $8.10^{+0.25}_{-0.25}$ & 1 & \textit{R3} & \\ 
CEERS\_80445 & $214.8431150$ & $52.7478860$ & $7.509$ & $-19.66^{+0.03}_{-0.02}$ & $8.34^{+0.05}_{-0.05}$ & \nodata & $8 \pm 18$ & $>3.1$ & \nodata & \nodata & \nodata & 1 & \nodata & \\ 
CEERS\_80573 & $214.7739240$ & $52.7805990$ & $5.428$ & $-19.70^{+0.02}_{-0.07}$ & $8.40^{+0.06}_{-0.05}$ & \nodata & $51 \pm 21$ & $>5.2$ & \nodata & \nodata & $7.97^{+0.27}_{-0.27}$ & 1 & \textit{R3} & \\ 
CEERS\_80576 & $214.7696270$ & $52.7729400$ & $5.450$ & $-19.13^{+0.11}_{-0.07}$ & $8.26^{+0.06}_{-0.05}$ & $0.88^{+0.07}_{-0.08}$ & $196 \pm 34$ & $6.1 \pm 1.1$ & $8.1 \pm 1.4$ & $>5.8$ & $7.69^{+0.39}_{-0.22}$ & 1 & \textit{R23} & \\ 
CEERS\_80596 & $214.7718650$ & $52.7781890$ & $6.542$ & $-18.54^{+0.35}_{-0.09}$ & $8.17^{+0.30}_{-0.12}$ & \nodata & $61 \pm 30$ & $>3.8$ & \nodata & \nodata & $8.23^{+0.23}_{-0.23}$ & 1 & \textit{R3} & \\ 
CEERS\_80710 & $214.8849850$ & $52.8360450$ & $6.545$ & $-17.85^{+0.34}_{-1.44}$ & $8.16^{+0.29}_{-0.28}$ & \nodata & $70 \pm 79$ & $>3.6$ & \nodata & \nodata & $8.19^{+0.23}_{-0.23}$ & 1 & \textit{R3} & \\ 
CEERS\_80916 & $214.8916300$ & $52.8159430$ & $5.674$ & $-17.88^{+0.53}_{-0.53}$ & $8.44^{+0.05}_{-0.23}$ & $0.78^{+0.08}_{-0.10}$ & $176 \pm 41$ & $6.3 \pm 1.3$ & $8.0 \pm 1.7$ & $>5.4$ & $7.73^{+0.35}_{-0.26}$ & 1 & \textit{R23} & \\ 
CEERS\_80954 & $214.9181830$ & $52.8118260$ & $5.930$ & $-17.72^{+0.13}_{-0.06}$ & $7.60^{+0.05}_{-0.05}$ & $0.14^{+0.11}_{-0.14}$ & $151 \pm 43$ & $5.0 \pm 1.5$ & \nodata & \nodata & $7.80^{+0.30}_{-0.37}$ & 1 & \textit{R3} & \\ 
CEERS\_81018 & $214.8054760$ & $52.7547300$ & $5.861$ & $-19.04^{+0.07}_{-0.06}$ & $8.14^{+0.10}_{-0.05}$ & \nodata & $13 \pm 16$ & $>1.6$ & \nodata & \nodata & $8.12^{+0.26}_{-0.26}$ & 1 & \textit{R3} & \\ 
CEERS\_81022 & $214.8004490$ & $52.7488890$ & $5.300$ & $-18.63^{+0.05}_{0.00}$ & $7.94^{+0.05}_{-0.05}$ & \nodata & $38 \pm 30$ & $>4.6$ & \nodata & \nodata & $7.96^{+0.29}_{-0.29}$ & 1 & \textit{R3} & \\ 
CEERS\_81026 & $214.8098410$ & $52.7542180$ & $5.433$ & $-18.28^{+0.05}_{-0.33}$ & $8.67^{+0.19}_{-0.09}$ & \nodata & $75 \pm 38$ & $>9.1$ & \nodata & \nodata & \nodata & 1 & \nodata & \\ 
CEERS\_81032 & $214.8126250$ & $52.7548490$ & $5.427$ & $-19.28^{+0.09}_{-0.10}$ & $8.25^{+0.13}_{-0.05}$ & \nodata & $45 \pm 16$ & $>7.0$ & \nodata & \nodata & \nodata & 1 & \nodata & \\ 
CEERS\_81049 & $214.7898220$ & $52.7307890$ & $6.738$ & $-19.61^{+0.11}_{-0.12}$ & $8.36^{+0.15}_{-0.05}$ & $1.21^{+0.04}_{-0.05}$ & $109 \pm 13$ & $8.5 \pm 0.9$ & $11.2 \pm 1.3$ & $>8.9$ & $7.88^{+0.15}_{-0.16}$ & 1 & \textit{R23} & \\ 
CEERS\_81063 & $214.7991100$ & $52.7251190$ & $6.086$ & $-19.04^{+0.12}_{-0.09}$ & $8.16^{+0.07}_{-0.07}$ & $0.98^{+0.09}_{-0.11}$ & $90 \pm 24$ & $9.2 \pm 2.1$ & $11.5 \pm 2.6$ & $>3.1$ & $8.08^{+0.21}_{-0.19}$ & 1 & \textit{R23} & \\ 
CEERS\_81068 & $214.8205070$ & $52.7371480$ & $6.272$ & $-18.30^{+0.18}_{-0.08}$ & $8.00^{+0.13}_{-0.05}$ & $0.60^{+0.06}_{-0.07}$ & $204 \pm 35$ & $6.0 \pm 1.0$ & $7.9 \pm 1.2$ & $>10.5$ & $7.65^{+0.22}_{-0.20}$ & 1 & \textit{R23} & \\ 
CEERS\_82043 & $214.7199860$ & $52.7502550$ & $4.324$ & $-18.85^{+0.34}_{-0.11}$ & $8.33^{+0.27}_{-0.07}$ & $1.00^{+0.10}_{-0.13}$ & $84 \pm 24$ & $7.3 \pm 1.9$ & \nodata & \nodata & $8.14^{+0.24}_{-0.24}$ & 1 & \textit{R3} & \\ 
CEERS\_82052 & $214.7665610$ & $52.7822700$ & $5.156$ & $-18.00^{+0.07}_{-0.12}$ & $8.43^{+0.05}_{-0.25}$ & \nodata & $80 \pm 33$ & $>5.6$ & \nodata & \nodata & $7.98^{+0.26}_{-0.26}$ & 1 & \textit{R3} & \\ 
CEERS\_82300 & $214.9005570$ & $52.8716120$ & $4.707$ & $-18.49^{+0.16}_{-0.49}$ & $8.47^{+0.05}_{-0.36}$ & \nodata & $57 \pm 19$ & $>3.8$ & \nodata & \nodata & $8.25^{+0.24}_{-0.24}$ & 1 & \textit{R3} & \\ 
CEERS\_82507 & $214.8532130$ & $52.8483650$ & $4.986$ & $-19.34^{+0.66}_{-0.81}$ & $8.36^{+0.85}_{-0.49}$ & \nodata & $87 \pm 74$ & $>8.1$ & \nodata & \nodata & \nodata & 1 & \nodata & \\ 
CEERS\_83398 & $214.9004090$ & $52.8268560$ & $4.554$ & $-19.79^{+0.05}_{-0.01}$ & $8.47^{+0.05}_{-0.05}$ & \nodata & $18 \pm 15$ & \nodata & \nodata & \nodata & \nodata & 1 & \nodata & \\ 
CEERS\_83439 & $214.9380700$ & $52.8471150$ & $4.824$ & $-16.98^{+0.05}_{-0.10}$ & $7.32^{+0.09}_{-0.05}$ & $-0.04^{+0.12}_{-0.16}$ & $193 \pm 59$ & $3.1 \pm 1.0$ & $4.1 \pm 1.3$ & $>1.7$ & $7.22^{+0.21}_{-0.23}$ & 1 & \textit{R23} & \\ 
CEERS\_83502 & $214.9058470$ & $52.8119060$ & $4.249$ & $-17.28^{+0.14}_{-0.12}$ & $7.46^{+0.26}_{-0.05}$ & $0.03^{+0.10}_{-0.13}$ & $171 \pm 44$ & $2.1 \pm 0.6$ & \nodata & \nodata & $7.14^{+0.27}_{-0.28}$ & 1 & \textit{R3} & \\ 
CEERS\_83592 & $214.9566930$ & $52.8337800$ & $5.736$ & $-18.43^{+0.38}_{-0.22}$ & $7.70^{+0.36}_{-0.08}$ & \nodata & $218 \pm 86$ & $>3.1$ & \nodata & \nodata & \nodata & 1 & \nodata & \\ 
CEERS\_83772 & $214.7857270$ & $52.7315530$ & $5.288$ & $-20.16^{+0.04}_{-0.04}$ & $8.63^{+0.05}_{-0.05}$ & $0.90^{+0.10}_{-0.13}$ & $88 \pm 23$ & $5.6 \pm 1.6$ & $7.1 \pm 2.0$ & $>2.6$ & $7.82^{+0.27}_{-0.33}$ & 1 & \textit{R23} & \\ 
CEERS\_83779 & $214.8214170$ & $52.7548380$ & $4.307$ & $-18.77^{+0.06}_{-0.23}$ & $8.23^{+0.08}_{-0.14}$ & \nodata & $103 \pm 42$ & $>6.9$ & \nodata & \nodata & \nodata & 1 & \nodata & \\ 
CEERS\_83856 & $214.8030220$ & $52.7265370$ & $4.556$ & $-18.63^{+0.11}_{-0.04}$ & $7.99^{+0.05}_{-0.05}$ & $0.68^{+0.06}_{-0.08}$ & $221 \pm 36$ & $2.7 \pm 0.5$ & \nodata & \nodata & $7.18^{+0.25}_{-0.25}$ & 1 & \textit{R3} & \\ 
CEERS\_83860 & $214.8283780$ & $52.7429470$ & $4.493$ & $-17.49^{+0.73}_{-0.04}$ & $8.30^{+0.34}_{-0.09}$ & \nodata & $35 \pm 48$ & \nodata & \nodata & \nodata & \nodata & 1 & \nodata & \\ 
CEERS\_85836 & $214.8312040$ & $52.7517900$ & $4.995$ & $-18.08^{+0.05}_{-0.03}$ & $7.76^{+0.05}_{-0.05}$ & \nodata & $82 \pm 44$ & $>4.9$ & \nodata & \nodata & $7.97^{+0.28}_{-0.28}$ & 1 & \textit{R3} & \\ 
\enddata
\tablecomments{%
($\star$) Lower-limit values at the $3\sigma$ level. The values are not shown (and hence the metallicities are not estimated) for those whose measured \OIII$\lambda\lambda 5007/4959$ line ratio is not consistent with the theoretical value of $2.98$ at the $2.5\sigma$ level. R23 and O32 are given if the dust-reddening is properly evaluated.
($\dag$) Photometry flag. Objects with flag=1 have the NIRCam photometry, and thus the stellar masses are reliably determined with the SED fit and SFRs are derived with the slit-loss corrected \Hb\ luminosities. For the objects with flag=0, their masses and SFRs are estimated with the UV luminosities of HST.
($\ddag$) Metallcitiy flag, specifying which method is adopted; the direct $T_e$ method or the empirical indicator of R23 or R3.
($\P$) For the gravitationally lensed objects, the magnification factor $\mu$ is noted. For the lensed objects, the values of \Muv, mass and SFR are already corrected for the magnification. For the CEERS objects that are multiply observed in several pointings/gratings, the number of pointings/gratings is noted, and their averaged values of EW(\Hb), line ratios, and metallicity are given. The objects with spectroscopic signatures of AGNs as indicated by \citep{harikane2023_jwst_blagn} are marked with \textit{Broad}. These \textit{Broad} objects are not used for deriving the MZ and SFR-MZ relations due to the unknown contribution of AGNs to the continuum level (i.e., uncetainty in the stellar mass estimation based on the SED fitting).
}
\end{deluxetable*}
\end{longrotatetable}




\end{document}